\documentclass[12pt, twocolumn, numberedappendix, tighten]{aastex62}
\usepackage{color} 
\usepackage{graphicx}

\newcommand{\civ}{C\,{\sc iv}}
\newcommand{\ciii}{C\,{\sc iii}}
\newcommand{\siiv}{Si\,{\sc iv}}
\newcommand{\aliii}{Al\,{\sc iii}}
\newcommand{\pv}{P\,{\sc v}}

\shortauthors{Hemler et al.}

\begin{document}

\title{The Sloan Digital Sky Survey Reverberation Mapping Project: Systematic Investigations of Short-Timescale \civ \ Broad Absorption Line Variability}


\author{Z.~S.~Hemler} 
\affiliation{Department of Astronomy and Astrophysics, Eberly College of Science, The Pennsylvania State University, 525 Davey Laboratory, University Park, PA 16802}

\author{C.~J.~Grier}
\affiliation{Department of Astronomy and Astrophysics, Eberly College of Science, The Pennsylvania State University, 525 Davey Laboratory, University Park, PA 16802}
\affiliation{Institute for Gravitation \& the Cosmos, The Pennsylvania State University, University Park, PA 16802}

\author{W.~N.~Brandt}
\affiliation{Department of Astronomy and Astrophysics, Eberly College of Science, The Pennsylvania State University, 525 Davey Laboratory, University Park, PA 16802}
\affiliation{Institute for Gravitation \& the Cosmos, The Pennsylvania State University, University Park, PA 16802}
\affiliation{Department of Physics, The Pennsylvania State University, University Park, PA 16802, USA}

\author{P.~B.~Hall}
\affiliation{Department of Physics and Astronomy, York University, Toronto, ON M3J 1P3, Canada}

\author{Keith~Horne}
\affiliation{SUPA Physics and Astronomy, University of St. Andrews, Fife, KY16 9SS, Scotland, UK} 

\author{Yue~Shen}
\altaffiliation{Alfred P. Sloan Research Fellow}
\affiliation{Department of Astronomy, University of Illinois at Urbana-Champaign, Urbana, IL 61801, USA} 
\affiliation{National Center for Supercomputing Applications, University of Illinois at Urbana-Champaign, Urbana, IL 61801, USA} 

\author{J.~R.~Trump}
\affiliation{Department of Physics, University of Connecticut, 2152 Hillside Road, Unit 3046, Storrs, CT 06269, USA}

\author{D.~P.~Schneider}
\affiliation{Department of Astronomy and Astrophysics, Eberly College of Science, The Pennsylvania State University, 525 Davey Laboratory, University Park, PA 16802}
\affiliation{Institute for Gravitation \& the Cosmos, The Pennsylvania State University, University Park, PA 16802}

\author{M. Vivek}
\affiliation{Department of Astronomy and Astrophysics, Eberly College of Science, The Pennsylvania State University, 525 Davey Laboratory, University Park, PA 16802}
\affiliation{Institute for Gravitation \& the Cosmos, The Pennsylvania State University, University Park, PA 16802}

\author{Dmitry Bizyaev}
\affiliation{Apache Point Observatory and New Mexico State University, P.O. Box 59, Sunspot, NM, 88349-0059, USA}
\affiliation{Sternberg Astronomical Institute, Moscow State University, Moscow, Russia}

\author{Audrey Oravetz}
\affiliation{Apache Point Observatory and New Mexico State University, P.O. Box 59, Sunspot, NM, 88349-0059, USA}

\author{Daniel Oravetz}
\affiliation{Apache Point Observatory and New Mexico State University, P.O. Box 59, Sunspot, NM, 88349-0059, USA}

\author{Kaike Pan}
\affiliation{Apache Point Observatory and New Mexico State University, P.O. Box 59, Sunspot, NM, 88349-0059, USA}

\begin{abstract}

We systematically investigate short-timescale ($<$10-day rest-frame) \civ \ broad absorption-line (BAL) variability to constrain quasar-wind properties and provide insights into BAL-variability mechanisms in quasars. We employ data taken by the Sloan Digital Sky Survey Reverberation Mapping (SDSS-RM) project, as the rapid cadence of these observations provides a novel opportunity to probe BAL variability on shorter rest-frame timescales than have previously been explored. In a sample of 27 quasars with a median of 58 spectral epochs per quasar, we have identified 15 quasars ($55^{+18}_{-14}$\%), 19 of 37 \civ \ BAL troughs ($51^{+15}_{-12}$\%), and 54 of 1460 epoch pairs ($3.7 \pm 0.5$\%) that exhibit significant \civ \ BAL equivalent-width variability on timescales of less than 10 days in the quasar rest frame. These frequencies indicate that such variability is common among quasars and BALs, though somewhat rare among epoch pairs. Thus, models describing BALs and their behavior must account for variability on timescales down to less than a day in the quasar rest frame.
We also examine a variety of spectral characteristics and find that in some cases, BAL variability is best described by ionization-state changes, while other cases are more consistent with changes in covering fraction or column density. We adopt a simple model to constrain the density and radial distance of two outflows appearing to vary by ionization-state changes, yielding outflow density lower limits consistent with previous work.

\end{abstract}

\keywords{galaxies: active --- galaxies: nuclei --- galaxies: kinematics and dynamics --- quasars: absorption lines)
}
\section{INTRODUCTION}
One of the major questions driving studies of galaxy evolution is the nature of the interaction between galaxies and their central supermassive black holes (SMBHs). A possible avenue for this interaction is through quasar outflows, or winds, generated near the SMBH (e.g., \citealt{Dimatteo05}; \citealt{Moll07}; \citealt{KingPounds15}). Quasar outflows are launched from the accretion disk around the SMBH (e.g., \citealt{Murray95}; \citealt{Proga00}; \citealt{Baskin14}; \citealt{Higginbottom14}), and are thought to regulate the growth of the host galaxy, and the SMBH itself, through the process of AGN feedback. Specifically, outflows provide a mechanism for a host galaxy to evacuate or compress gas, slowing SMBH accretion and suppressing or enhancing host-galaxy star formation. Understanding the physical properties of these outflows is thus integral to developing models of galaxy evolution. In the context of rest-frame UV quasar spectra, these outflows manifest themselves most prominently as broad absorption lines (BALs), features that exhibit observed velocity widths on the scale of several thousand km~s$^{-1}$ (\citealt{Weymann91}). BALs are an extremely useful probe of the physical properties of outflows as well as the environments from which they originate.

Previous studies have demonstrated that BALs are significantly variable, both in strength and in shape (e.g., \citealt{Barlow93}; \citealt{Lundgren07}; \citealt{Gibson08}; \citealt{Filizak13}; \citealt{Capellupo13}). Proposed mechanisms for BAL variability include inhomogeneous emission, ionization-state changes, and ``cloud crossing'', which refers to the drift of absorbing gas across the line of sight (e.g., \citealt{Hall07}; \citealt{Capellupo13}; 
\citealt{Filizak13}). Characterizing the BAL variations can yield constraints for many physical properties of quasar outflows; for example, ionization-state changes and cloud-crossing allow an inference of the density and/or radial distance of relevant gas from the SMBH \cite[e.g.,][]{Capellupo13}.

While BAL variability has been commonly studied on rest-frame timescales of years and months \cite[e.g.,][]{Gibson10}, observations probing timescales of days and hours are rare \cite[e.g.,][]{Capellupo13, Grier15}. Moreover, most studies have suffered a trade-off between the number of observed quasars and the number of epochs. For this reason, there has not yet been a study with both the sample size and rapid cadence required to make robust statistical statements about variability on timescales on the order of days to hours in the quasar rest frame. Table~\ref{Table:tbl1} presents various sample characteristics of several prior BAL-variability studies for the purposes of comparison to this investigation. 

\begin{deluxetable*}{l*{7}{c}}
\tablewidth{0pt}
\tabletypesize{\footnotesize}
\renewcommand{\arraystretch}{0.85}
\tablecaption{Parameters of Various BAL Variability Studies}
\tablehead{
\colhead{} &
\colhead{No. of} &
\colhead{$N_{\rm epochs}$\tablenotemark{a}} &
\colhead{Median} &
\colhead{$\Delta$$t$\tablenotemark{b}} &
\colhead{Median} &
\colhead{Redshift} &
\colhead{Median} \\
\colhead{Reference} &
\colhead{Quasars} &
\colhead{Range} &
\colhead{$N_{\rm epochs}$\tablenotemark{a}} &
\colhead{Range (days)} &
\colhead{$\Delta$$t$\tablenotemark{b} (days)} &
\colhead{Range} &
\colhead{Redshift} 
}

\startdata
\cite{Barlow93} & 23 & 2--6 & \nodata & 73--440 & \nodata & \nodata & \nodata \\ 
\cite{Lundgren07} & 29 & 2 & 2 & 18--121 & 81 & 1.70--3.26 & 2.00 \\ 
\cite{Gibson08} & 13 & 2 & 2 & 1100--2200 & 1500 & 1.72--2.81 & 2.02 \\ 
\cite{Gibson10} & 14 & 2--4 & 3 & 15--2500 & 1100 & 2.08--2.89 & 2.41 \\ 
\cite{Capellupo11} & 24 & 2--4 & 4 & 130--2800 & 1600 & 1.22--2.91 & 2.23 \\ 
\cite{Capellupo12} & 24 & 2--10 & 5 & 15--3000 & 270 & 1.22--2.91 & 2.23 \\ 
\cite{Vivek12}	& 5 & 4--14 & 5 & 3.7--1200 & 140 & 0.81--1.97 & 1.23 \\ 
\cite{Haggard12} & 17 & 6 & \nodata	 & 0.37--330 & \nodata & \nodata & \nodata \\ 
\cite{Filizak12} & 19 & 2--4 & 2 & 400--1400 & 730 & 1.71--2.57 & 2.00 \\ 
\cite{Capellupo13} & 24 & 2--13 & 7 & 7.3--3200 & 270 & 1.22--2.91 & 2.23 \\ 
\cite{Welling14} & 46 & 2--6 & 2 & 80--6000 & 500 & 0.24--3.46 & 2.21 \\ 
\cite{Filizak13} & 291 & 2--12 & 2 & 0.22--1350 & 770 & 2.00--3.93 & 2.55 \\ 
\cite{Filizak14} & 671 & 2 & 2 & 310-1500 & 930 & 1.90--3.83 & 2.28 \\ 
\cite{Vivek14} & 22 & 3--7 & 5 & 10--2800 & 280 & 0.20--2.10 & 1.28 \\
\cite{Vivek16} & 50 & 2--5 & 2 & 800--2000 & 1100 & 1.65--2.90 & 1.96 \\
\cite{Rogerson18} & 105 & 3--7 & 3 & 1.8--1210 & 800 & 1.68--4.13 & 2.51 \\
This work & 27 & 3--68 & 58 & 0.21--250 & 2.4 & 1.62--3.72 & 2.35 
\enddata
\label{Table:tbl1}
\tablenotetext{a}{$N_{\rm epochs}$ refers to the number of epochs per quasar. } 
\tablenotetext{b}{$\Delta t$ is measured in the quasar rest frame.} 
\end{deluxetable*} 

Spectroscopic data from the Sloan Digital Sky Survey Reverberation Mapping project \cite[SDSS-RM;][]{Shen15} provide a novel opportunity to investigate BAL variability in a representative sample of quasars using many epochs of spectra on shorter timescales than have previously been explored. The SDSS-RM project is an ongoing reverberation-mapping campaign that began in 2014 and has continued to obtain data in subsequent years during January -- July of each year. Many quasars in the SDSS-RM quasar sample exhibit BAL features. Prior to this program, the shortest BAL-variability timescales probed were on the order of $8-10$ days (\citealt{Capellupo13}).
However, with the rapid cadence of the SDSS-RM dataset, \cite{Grier15} were able to investigate even shorter timescales: They identified significant variability of a \civ \ BAL on timescales as short as 1.2 days in the quasar rest frame. On these short timescales, the equivalent width of this BAL varied by roughly 10\%. Because the BAL showed little variability in shape, \cite{Grier15} proposed that this feature was likely varying via ionization-state changes in the absorbing medium. This discovery of such short-timescale BAL variability motivates subsequent investigations to determine whether this variability is common among BAL quasars. 

In this work, we present the results of a search for short-timescale variability within a sample of BAL quasars using data from the SDSS-RM project. The goals of this study are to determine the frequency of short-timescale BAL variability, provide insights into quasar-wind variability mechanisms, and further constrain quasar-wind properties in BAL quasars exhibiting significant short-timescale variability. Section~\ref{sec:observations} presents the SDSS-RM observations and characterizes our quasar sample. Section~\ref{sec:balmeas} describes data processing, the measured BAL parameters such as equivalent widths, mean depths, and centroid velocities, and our criteria for selecting robust cases of BAL variability. Section~\ref{sec:discussion} discusses our results, our examination of coordinated variability, physical constraints calculated from our measurements, and the implications of our work. Section~\ref{sec:summary} summarizes our findings and elaborates on feasible future work in regards to short-timescale BAL variability studies. Where necessary, we adopt a cosmology with $H_0 = 70 \rm{\ km \ s^{-1} \ Mpc^{-1}}$, $\Omega_{M} = 0.3$, and $\Omega_\Lambda = 0.7$.

\section{OBSERVATIONS \& DATA PREPARATION}
\label{sec:observations}
\subsection{SDSS-RM Observations}
	  In 2014, SDSS-RM obtained 32 epochs of spectra of 849 quasars using the Baryon Oscillation Spectroscopic Survey \cite[BOSS;][]{Eisenstein11, Dawson13} spectrograph on the SDSS 2.5-meter telescope \citep{Gunn06, Smee13}; 12--13 additional epochs per year were taken subsequently in 2015, 2016, and 2017 as a part of the SDSS-IV eBOSS program (\citealt{Dawson16}; \citealt{Blanton17}), yielding 69 total epochs of spectroscopy over the four years. The BOSS spectrograph covers a wavelength range of 3650 -- 10,400\,\AA\ and has a spectral resolution of R = 2000 with a field of view that is 3 degrees in diameter. At the beginning of the SDSS-RM program, all 849 quasars were assigned numerical identifiers based upon their position on the spectrograph plate --- we hereafter refer to these identifiers as RM IDs and refer to all quasars by this designation. 
    
\subsection{Initial Quasar Sample}

\begin{figure}
\begin{center}
\includegraphics[scale = 0.41, trim = 0 0 0 0, clip]{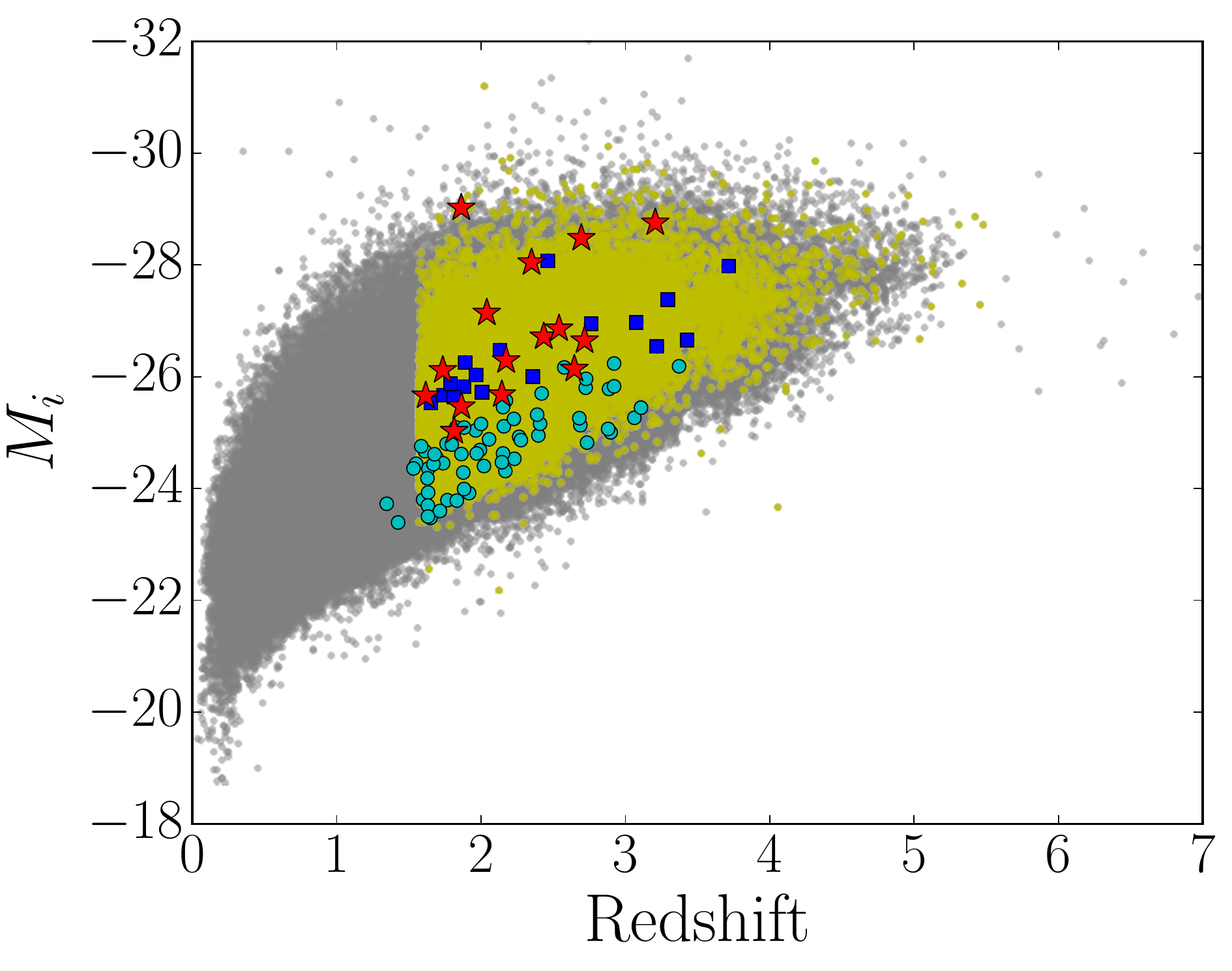}
\caption{Absolute $i$-magnitude as a function of redshift. Gray and yellow points show non-BAL and C\,{\sc iv}-BAL quasars from the DR14 quasar catalog (\citealt{Paris18}), respectively. Cyan circles represent quasars in our initial 94-object sample that were cut due to low SN; blue squares show the quasars in our final sample that show no significant short-timescale ($<$10-day) variability, and red stars indicate our 15 quasars that show significant short-timescale variability. Data for the SDSS-RM quasars were retrieved from Shen et al. (2018, in preparation). All absolute magnitudes have been corrected to $M_{i}[z = 2]$, following \cite{Richards06}.}
\label{fig:mag_vs_z}
\end{center}
\end{figure}

	We began with a parent sample of 94 quasars in the SDSS-RM sample that were identified to likely host prominent C\,{\sc iv} absorption features, each with 69 epochs of data from the four years of SDSS-RM monitoring. This initial sample was chosen by visual inspection, and thus may contain \civ \ absorbers that do not meet the formal requirements for BALs (e.g., \citealt{Weymann91}). We later refine our sample to include only formally detected BALs; see Section~\ref{sec:finalsample}. Figure~\ref{fig:mag_vs_z} shows the absolute magnitude-redshift distribution of our parent quasar sample compared to the general population of quasars from the DR14 quasar catalog (\citealt{Paris18}). The quasars in our parent sample are typical of BAL quasars within the magnitude/redshift plane. 

\subsection{Data Preparation}

The SDSS-RM data were first processed using the standard BOSS spectroscopic reduction pipeline; however, to further improve the spectrophotometric flux calibrations of the data, a custom flux calibration scheme was applied \citep{Shen15}. We then corrected the spectra for Galactic extinction using an $R_{\rm V}$ = 3.1 Milky Way extinction model (\citealt{Cardelli89}) and $A_{\rm V}$ values from \cite{SchaflyFinkbeiner11}. 

We searched for pixels with poor sky subtraction residuals using the SDSS bitmask flag ``BRIGHTSKY" and linearly interpolated over these pixels. This process suppressed much of the spectral mid-optical sky contamination, although significant sky contamination residuals are still visible at the far red and blue ends of the spectra. A few spectra contained additional bad pixels that were not flagged during the SDSS data reduction process. These pixels appeared as dramatic spikes or dips in flux -- we manually removed these points and linearly interpolated over the region for ease of spectra visualization. Before continuing our analysis, we translated the wavelengths of the spectra into the quasar rest frame using systemic redshifts from \cite{Shen16}. We adopt these redshift measurements for the remainder of this investigation. To remove the aforementioned sky contamination that was often present at the edges of the spectra, we cropped them to cover roughly the wavelength regions spanning 1200--2600 \AA\ in the quasar rest frame. 

During the course of our study, we identified an issue with the SDSS eBOSS DR14 pipeline reductions which produced wavelength-dependent residuals in data processed after 2014. The residuals resulted from incorrect background estimation in the presence of nearby standard stars. Spectra are grouped in bundles of 20 fibers on the detectors, and the DR14 background estimation used a new scheme to measure the background flux within these bundles. This issue has since been fixed in more recent versions of the pipeline reductions; however, it was not yet resolved while we were carrying out the bulk of our analysis. 
Empirically, we found that excluding fibers from bundles including two or more standard stars removed all fibers with obvious spectral residuals. Our team thus identified the spectra from fibers in these bundles at each epoch in our sample and removed these individual spectra from our sample to avoid introducing resultant systematic uncertainties.
Eighteen of the quasars in our sample had at least one epoch that was eliminated. For some of these quasars only a single epoch was removed (e.g., RM\,770), while in others, all of the post-2014 epochs were removed
(e.g., RM\,786).

\subsection{Magnitude and Signal-to-Noise Cuts} 
\label{sec:magcut}
\begin{figure}
\begin{center}
	\includegraphics[scale= 0.42, trim = 10 0 50 20, clip]{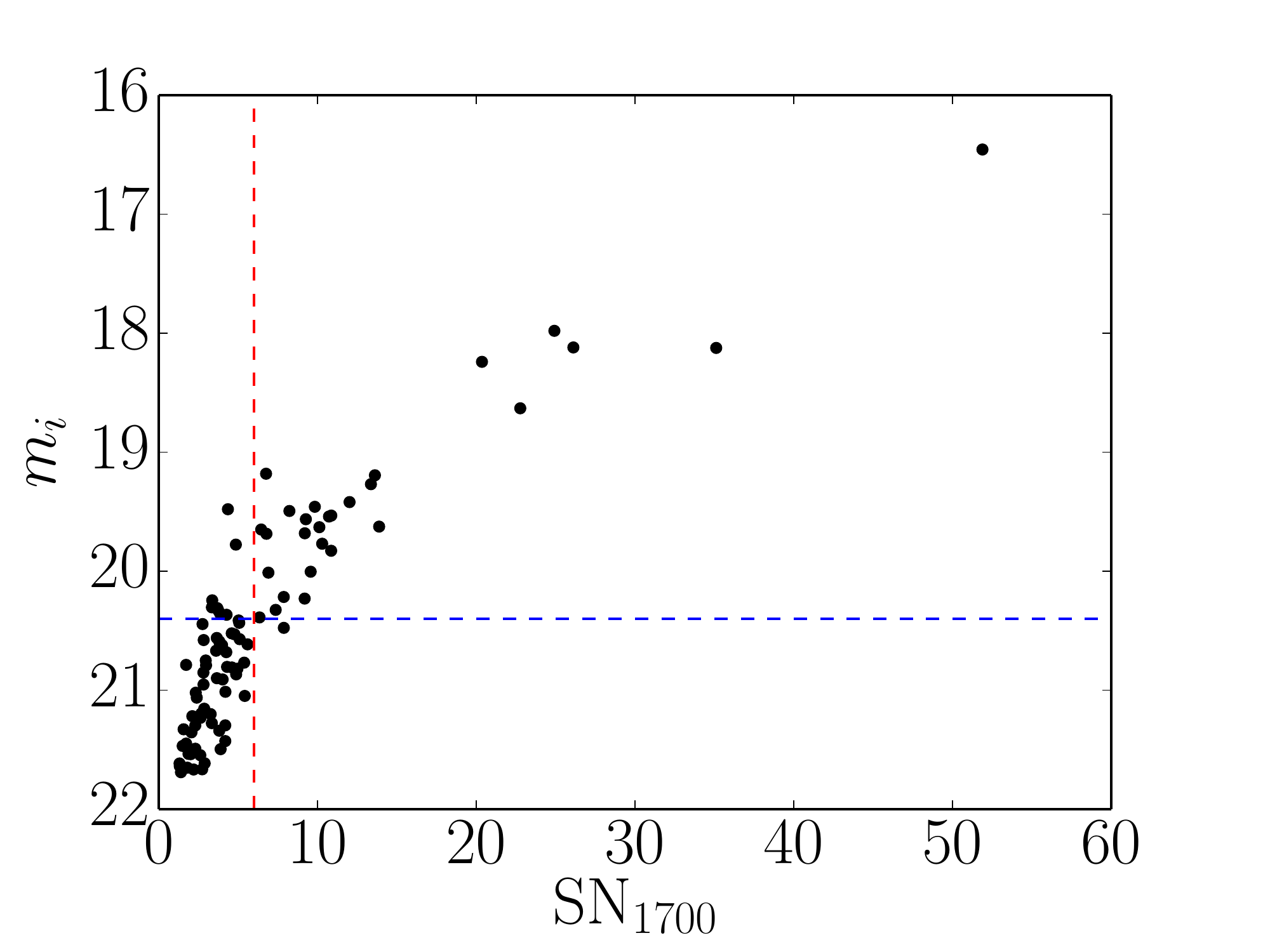}
    \caption{The quasar apparent $i$-magnitude ($m_i$) vs. median SN$_{1700}$ for the entire SDSS-RM BAL parent sample. 
The dashed blue line represents the location of our $i$-magnitude limit, and the dashed vertical red line shows the SN$_{1700}$ cut. We note, however, that the SN$_{1700}$ cut was made on individual spectral epochs, while the figure includes the median SN$_{1700}$ of all spectra for each quasar.}
    \label{fig:imagvsn17dat}
\end{center}
\end{figure}

Our goal is to construct a sample that is representative of the general BAL quasar population, so we make only two simple cuts to our sample. Because we are searching for small-amplitude BAL variability, we require our spectra to have reasonably high signal-to-noise ratios (SN) near the \civ \ region of the spectrum. To quantify the SN, we measure $\rm{SN_{1700}}$, defined as the median SN of the pixels in the 1650--1750~\AA\ region in the quasar rest frame. We follow previous work (e.g. \citealt{Gibson09}) and require our objects to have $\rm{SN_{1700}}$ of 6 or higher per pixel (reduced SDSS spectra have a resolution of 69~km~s$^{-1}$ per pixel). To achieve this, we selected only quasars with $i$-band magnitudes $m_i<$ 20.4, which mostly eliminated quasars with $\rm{SN_{1700}}$ lower than this threshold (see Figure~\ref{fig:imagvsn17dat}). This $m_i$ constraint restricted the sample to 36 objects. Figure~\ref{fig:imagvsn17dat} displays $m_i$ vs. the median $\rm{SN_{1700}}$ calculated from all observed epochs for each quasar. 

Even after this magnitude cut, individual epochs of some objects had fairly low SN due to variations in observing conditions. Often, a reliable continuum fit was difficult to obtain for these epochs, rendering further measurements difficult or impossible. Therefore, we impose a subsequent $\rm{SN_{1700}} \geq 6$ cut on each individual spectral epoch to remove epochs where reliable fits were unobtainable due to low SN. For some quasars, this removed an epoch or two from our sample, while in four cases, all observations of the given quasar were dropped. There is some scatter in the relationship between $m_i$ and SN$_{1700}$, and these four objects fell below our SN threshold despite our $m_i$ cut; see Figure~\ref{fig:imagvsn17dat}. Imposing this SN$_{1700}$ limit yielded 32 quasars that potentially host C\,{\sc iv} BALs and have sufficient SN for reliable analysis.

\subsection{Continuum and Emission-Line Fitting}

We follow previous BAL-variability studies \cite[e.g.][]{Filizak12, Filizak13, Grier15, Grier16} and use a least-squares fitting algorithm to fit reddened power laws to all spectra using the SMC-like reddening model of \cite{Pei92}. We began the fitting process by adopting the relatively line-free (RLF) regions determined by \cite{Gibson09b}, with a few revisions made after consulting the composite quasar spectra published by \cite{Vandenberk01} to identify regions relatively free of emission or absorption. Most quasars required additional deviations from the \cite{Gibson09b} regions for optimal continuum fits due to the presence of strong absorption or emission within these regions. For a given quasar, we adopted the same RLF fit regions for all epochs. Because we manually adjusted the line-free regions, we did not need to utilize the sigma clipping techniques commonly employed by previous investigations. (e.g., \citealt{Filizak13}). One epoch of one quasar (RM\,284, MJD 57451) was missing sections of the spectrum toward the red end, causing difficulties with the continuum fits; we exclude this epoch from our analysis, as the missing pixels left us unable to obtain a reasonable fit. 

\begin{figure*}
\begin{center}
\[\includegraphics[scale = 0.45, trim = 70 155 60 100, clip]{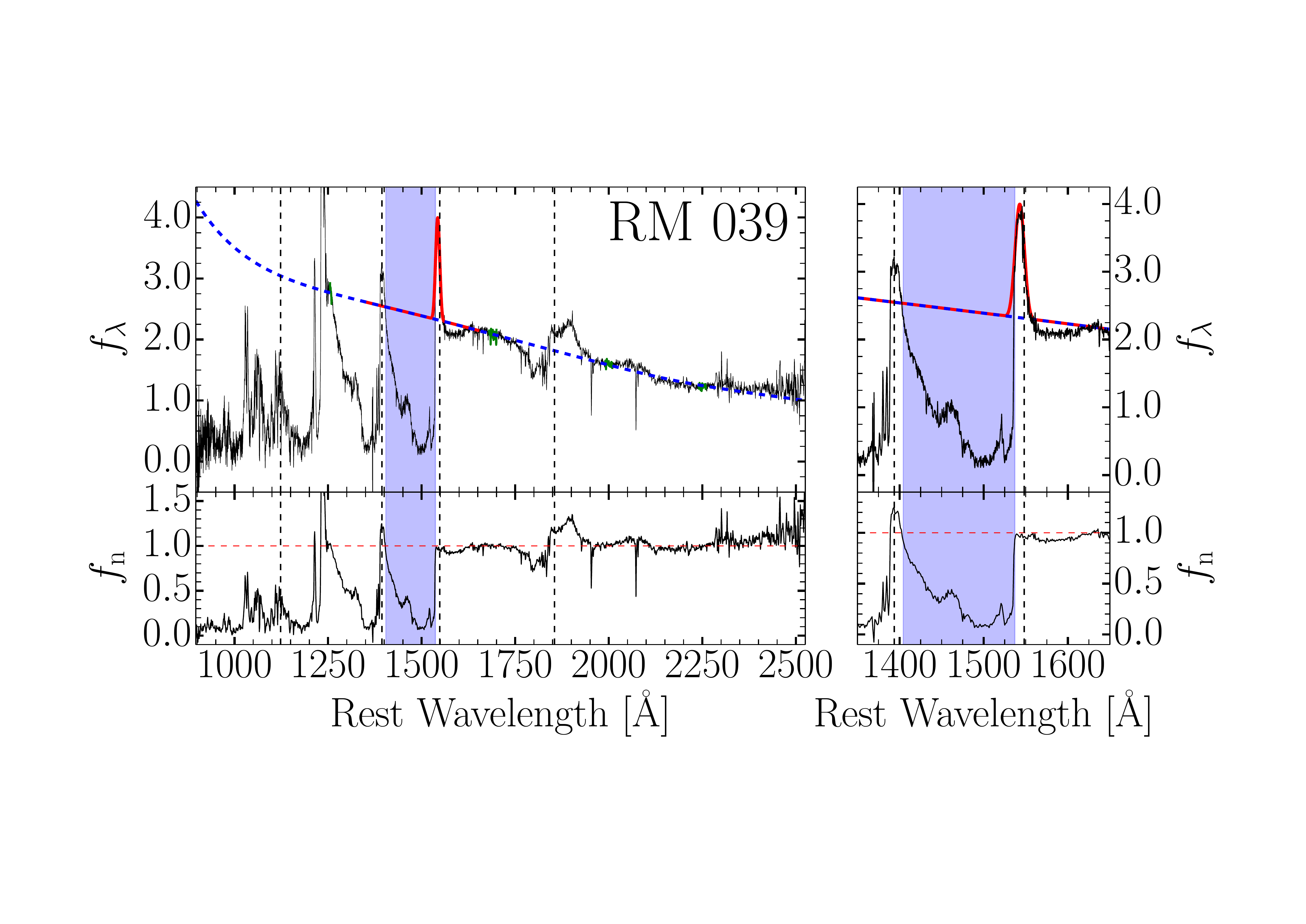}\]
\caption{The mean spectrum of quasar RM\,039, generated from data taken by the SDSS-RM project in 2014. The top subpanels show the quasar flux density, and the bottom subpanels show the continuum+emission-line-normalized fluxes $f_{\rm n}$. The left subpanels show the entire spectrum, while the right subpanels display only the \civ \ region. Thick dashed blue lines represent the continuum fits, solid red lines show our C\,{\sc iv} emission-line fits, and green regions show the adopted continuum regions used in the continuum fits. Shaded regions indicate the \civ \ BALs identified in each object. The horizontal red dashed lines in the bottom subpanels mark a normalized flux of 1.0 to guide the eye. Vertical black dotted lines indicate the rest-frame wavelengths of various species of interest: \pv \ (1115 \AA), \siiv \ (1393.8 \AA), \civ \ (1548.2 \AA), and \aliii \ (1854.7 \AA). $f_\lambda$ has units of $10^{-17}$ erg~s$^{-1}$~cm$^{-2}$\AA$^{-1}$. Figures for all 27 quasars in our final sample are provided in the online version of the article.}
\label{fig:meanspec}
\end{center}
\end{figure*}

For some spectra, the best-fit power law  was consistent with no reddening, so we adopted a simple power law for our continuum --- our choices of reddened or unreddened fits are indicated for each quasar in Table~\ref{Table:sample}. Uncertainties in the continuum fits were determined using Monte Carlo methods; the flux values of each pixel within the fitting region were altered by a random Gaussian deviate scaled to the pixel uncertainties, and the spectra were subsequently refit. Our continuum fits to the mean spectra created from the 2014 observations are presented in Figure~\ref{fig:meanspec}. 

Several quasars exhibited BALs at velocities insufficient to separate them cleanly from the C\,{\sc iv} emission line. To remove emission-line contamination from the BAL features, we follow \cite{Grier15} and fit Gauss-Hermite \citep{vanderMarel93}, double-Gaussian \citep{Park13}, and Voigt profile (Gibson et al. 2009b) models to the C\,{\sc iv} emission line. To do this, we constructed mean spectra for each individual year of the campaign for each object from all epochs passing our SN$_{1700}$ threshold criterion defined in Section~\ref{sec:magcut}. For each object, we choose the best-fit model (Gauss-Hermite, double-Gaussian, or Voigt) based on visual inspection; the chosen models for each quasar are listed in Table \ref{Table:sample}. We adopt the fit parameters from the mean spectrum (except the amplitude parameter) for all epochs in a given year (i.e., the 2014 mean spectrum fit parameters were used for all 2014 SDSS-RM epochs, etc.). We then fit the emission lines to each individual epoch while treating only the amplitude as a free parameter. This process assumes that only the amplitude of the emission lines varied within the same observing season and all other characteristics (e.g., widths, centers, etc) remained constant. Our emission-line fits are displayed in Figure~\ref{fig:meanspec}. We add the continuum and emission-line fits together, hereafter referred to as the ``continuum+emission-line" fits. 

\section{Final Sample Selection and BAL Measurements}
\label{sec:balmeas} 

\subsection{Final BAL Quasar Sample} 
\label{sec:finalsample}
We are primarily interested in quasars that host \civ \ BALs at the beginning of the SDSS-RM campaign. To identify such BALs, we first binned all of the 2014 mean normalized spectra by three pixels to account for SDSS oversampling of the spectral line function. Binning ensures that neighboring pixels are almost entirely uncorrelated. We then smoothed the binned spectra by three pixels to reduce noise. We note that smoothing was used only for the BAL search; we perform all additional measurements on the un-smoothed, binned spectra. We then searched each of the 2014 mean spectra for regions in which the normalized flux density drops below 0.9 for velocity widths greater than 2000~km~s$^{-1}$ (i.e., a BAL exists that is consistent with the formal definition of \citealt{Weymann91}). We restricted our search to within 30,000 km~s$^{-1}$ blueward of the adopted \civ \ rest wavelength (1548.202 \AA) and include BALs that extend redward of the \civ \ rest wavelength by up to $1000$~km~s$^{-1}$. Quasars without formal C\,{\sc iv} BALs in their 2014 mean spectra, such as those containing only mini-BALs or those for which a BAL did not appear until later on in the campaign, were dropped from our sample. 

After identifying the BALs, we visually inspected all spectra to ensure no BALs were contaminated by residual sky flux or bad pixels. Three BALs were found to either have bad pixels within the BAL trough  (in quasars RM 528 and RM 729) or to contain residual sky lines within the BAL (RM 195) that could potentially affect our analysis. These features were not detected by the SDSS pipeline, and affect BALs with an unfortunate combination of velocity and quasar redshift. We excluded these BALs from our sample. This resulted in the removal of RM\,528 from our sample, as this quasar possessed only the contaminated BAL. The other two quasars harbor additional uncontaminated BALs, and so remain in our final sample.

The final sample contains 27 quasars that have 37 uncontaminated C\,{\sc iv} BALs between them. We present various characteristics of the quasars in our final BAL quasar sample in Table~\ref{Table:sample} and provide information on each individual identified BAL in Table~\ref{Table:baltbl}. To distinguish between \civ \ BALs when a spectrum has more than one present, we assign identifiers [A] and [B] to the BALs as wavelength increases (and as velocity decreases); therefore, in cases with multiple BALs, [A] refers to the higher-velocity BAL, and [B] to the lower-velocity BAL. None of our quasars contained more than two \civ \ BALs. The regions of the spectra containing C\,{\sc iv} BALs in each quasar are highlighted in Figure~\ref{fig:meanspec}. 

\begin{deluxetable*}{ccccccrcrcc} 
\tabletypesize{\footnotesize}
\renewcommand{\arraystretch}{0.85}
\tablewidth{0pt} 
\tablecaption{Final Quasar Sample Information} 
\tablehead{ 
\colhead{} & 
\colhead{} & 
\colhead{RA\tablenotemark{a}} & 
\colhead{DEC\tablenotemark{a}} & 
\colhead{} & 
\colhead{} & 
\colhead{} & 
\colhead{} & 
\colhead{$\Delta t_{\rm rest}$} & 
\colhead{Continuum} & 
\colhead{Emission-} \\ 
\colhead{SDSS ID} & 
\colhead{RM ID} & 
\colhead{(deg)} & 
\colhead{(deg)} & 
\colhead{$z$\tablenotemark{a}} & 
\colhead{$m_i$\tablenotemark{a}} & 
\colhead{SN$_{1700}$} & 
\colhead{$N_{\rm epochs}$} &
\colhead{(days)} & 
\colhead{Fit\tablenotemark{b}} & 
\colhead{Line Fit\tablenotemark{c}}
}
\startdata 
J141607.12+531904.8    &    RM 039    &     214.02967    &      53.31800    &          3.08    &         19.77    &         10.24    &    58    &          2.88    &     R    &    V    \\ 
J141741.72+530519.0    &    RM 073    &     214.42385    &      53.08863    &          3.43    &         20.37    &          6.76    &    3    &          7.34    &     R    &    DG    \\ 
J141432.46+523154.5    &    RM 116    &     213.63526    &      52.53182    &          1.88    &         19.68    &          9.50    &    64    &          3.99    &     R    &    DG    \\ 
J141103.17+531551.3    &    RM 128    &     212.76324    &      53.26426    &          1.86    &         20.01    &          7.25    &    50    &          3.60    &     R    &    V    \\ 
J141123.68+532845.7    &    RM 155    &     212.84871    &      53.47936    &          1.65    &         19.65    &          7.14    &    41    &          2.61    &     R    &    DG    \\ 
J141935.58+525710.7    &    RM 195    &     214.89825    &      52.95297    &          3.22    &         20.33    &          7.66    &    59    &          2.75    &     R    &    DG    \\ 
J141000.68+532156.1    &    RM 217    &     212.50287    &      53.36560    &          1.81    &         20.39    &          7.12    &    41    &          2.46    &     R    &    V    \\ 
J140931.90+532302.2    &    RM 257    &     212.38295    &      53.38395    &          2.43    &         19.54    &         10.88    &    65    &          3.02    &     U    &    V    \\ 
J141927.35+533727.7    &    RM 284    &     214.86398    &      53.62436    &          2.36    &         20.22    &          8.16    &    63    &          3.46    &     U    &    GH    \\ 
J142014.84+533609.0    &    RM 339    &     215.06184    &      53.60252    &          2.01    &         20.00    &          9.68    &    64    &          3.47    &     R    &    DG    \\ 
J141955.27+522741.1    &    RM 357    &     214.98029    &      52.46144    &          2.14    &         20.23    &          9.25    &    65    &          3.49    &     R    &    DG    \\ 
J142100.22+524342.3    &    RM 361    &     215.25092    &      52.72844    &          1.62    &         19.46    &          9.92    &    66    &          3.94    &     R    &    V    \\ 
J141409.85+520137.2    &    RM 408    &     213.54107    &      52.02702    &          1.74    &         19.63    &         10.27    &    65    &          4.00    &     R    &    DG    \\ 
J142129.40+522752.0    &    RM 508    &     215.37253    &      52.46445    &          3.21    &         18.12    &         35.09    &    69    &          2.21    &     R    &    DG    \\ 
J142233.74+525219.8    &    RM 509    &     215.64060    &      52.87219    &          2.65    &         20.30    &          6.51    &    10    &          3.84    &     R    &    V    \\ 
J142306.05+531529.0    &    RM 564    &     215.77521    &      53.25807    &          2.46    &         18.24    &         20.36    &    67    &          2.69    &     U    &    DG    \\ 
J140634.14+525407.8    &    RM 565    &     211.64227    &      52.90219    &          1.79    &         19.48    &          6.25    &    3    &         13.78    &     U    &    V    \\ 
J141007.73+541203.4    &    RM 613    &     212.53224    &      54.20095    &          2.35    &         18.12    &         26.18    &    68    &          2.87    &     R    &    V    \\ 
J140554.87+530323.5    &    RM 631    &     211.47865    &      53.05655    &          2.72    &         19.83    &         11.28    &    65    &          2.79    &     R    &    GH    \\ 
J141648.26+542900.9    &    RM 717    &     214.20109    &      54.48360    &          2.17    &         19.69    &          7.18    &    49    &          3.74    &     R    &    DG    \\ 
J142419.18+531750.6    &    RM 722    &     216.07996    &      53.29739    &          2.54    &         19.49    &          8.65    &    59    &          3.10    &     U    &    DG    \\ 
J142404.67+532949.3    &    RM 729    &     216.01946    &      53.49704    &          2.76    &         19.56    &          9.66    &    63    &          2.91    &     R    &    V    \\ 
J142225.03+535901.7    &    RM 730    &     215.60432    &      53.98381    &          2.69    &         17.98    &         24.74    &    65    &          2.81    &     R    &    GH    \\ 
J142405.10+533206.3    &    RM 743    &     216.02126    &      53.53509    &          1.74    &         19.18    &          6.97    &    49    &          3.47    &     R    &    DG    \\ 
J142106.86+533745.2    &    RM 770    &     215.27862    &      53.62923    &          1.86    &         16.46    &         51.89    &    69    &          3.25    &     R    &    DG    \\ 
J141322.43+523249.7    &    RM 785    &     213.34349    &      52.54715    &          3.72    &         19.19    &         13.81    &    67    &          2.11    &     U    &    GH    \\ 
J141421.53+522940.1    &    RM 786    &     213.58974    &      52.49447    &          2.04    &         18.63    &         22.96    &    68    &          3.17    &     R    &    None 
\enddata
\tablenotetext{a}{J2000 Position, redshift, and magnitude measurements were obtained from SDSS Data Release 10 (\citealt{Ahn14}). The $i$-magnitudes provided are point-spread function magnitudes that have not been corrected for Galactic extinction.}
\tablenotetext{b}{A designation of ``R" indicates that a reddened power law continuum fit was used for this quasar, and a designation of ``U" indicates the use of an unreddened power law.}
\tablenotetext{c}{A designation of ``V" indicates that a Voigt profile to was used to represent the \civ \ emission line,  ``DG" indicates the use of a double-Gaussian profile, ``GH" indicates a Gauss-Hermite profile was used, and ``None" indicates that we were unable to obtain any acceptable emission-line fit, so the continuum-only fit was adopted.}
\label{Table:sample} 
\end{deluxetable*}  

Our final sample of BAL quasars does not generally have properties different from those of the larger BAL-quasar population (see Figure~\ref{fig:mag_vs_z}). Most (26 out of 27) of the quasars in our sample are radio quiet and are undetected by the FIRST radio survey (\citealt{White97}; compiled by \citealt{Shen18}). The one exception is RM\,155, which has a measured radio-loudness parameter of $R$~=~42.89; $R$ is defined as the ratio of fluxes at rest-frame 6~cm and 2500~\AA \ (e.g., \citealt{Shen18}).

\begin{figure}
\begin{center}
\includegraphics[scale = 0.28, trim = 50 0 0 0, clip]{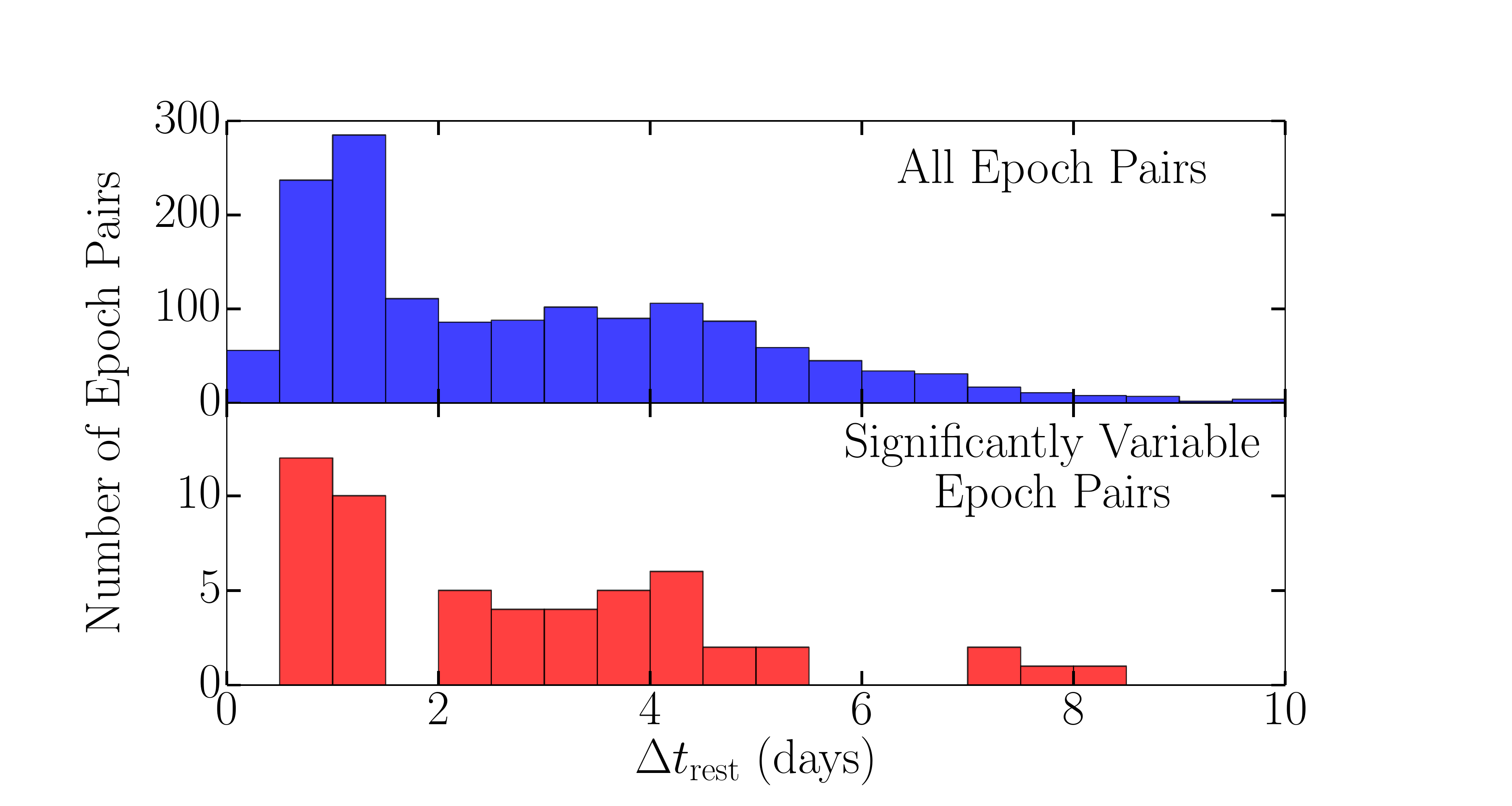}
\caption{Histogram of the rest-frame time difference $\Delta t_{\rm rest}$ distribution of our sample. the top subpanel displays the distribution of our entire 27-object sample; the bottom subpanel displays only that of epoch pairs exhibiting significant short-
timescale variability (see Section~\ref{sec:criteria}). The region $\Delta t_{\rm rest} < 2$ days is well-populated.}
\label{fig:fa4em}
\end{center}
\end{figure}

Our final sample of spectra has a median of 58 epochs per object; this is the highest of any BAL variability study by almost a factor of 10 (see Table~\ref{Table:tbl1}). Epochs separated by rest-frame timescales (hereafter referred to as $\Delta t_{\rm rest}$) on the order of a single day are frequent. Hereafter, we refer to all variability on timescales of less than 10 days in the quasar rest frame  as ``short-timescale", or ``rapid" variability. The median rest-frame time resolution of the SDSS-RM observations among our sample of quasars is 2.4 days, nearly a factor of 10 lower than even the shortest-timescale studies previously reported (see Table~\ref{Table:tbl1}). Our sample spans a redshift range of 1.62 to 3.72, which is comparable to previous studies. This is by necessity, as the the C\,{\sc iv} region must be redshifted to within the range of spectral coverage of the instruments used in each study. Figure~\ref{fig:fa4em} displays the distribution of $\Delta t_{\rm rest}$ between all pairs of subsequent epochs for the sample, including only sequential pairs of epochs for each quasar that are separated by less than 10 days in the quasar rest frame. 
   
\begin{deluxetable*}{ccrrrrrcc} 
\tablewidth{0pt} 
\tabletypesize{\footnotesize}
\renewcommand{\arraystretch}{0.85}
\tablecaption{C{\sc iv} BAL Measurements from 2014 Mean Spectra} 
\tablehead{ 
\colhead{} & 
\colhead{BAL} & 
\colhead{} & 
\colhead{} & 
\colhead{} & 
\colhead{} & 
\colhead{${\rm{EW}}$} & 
\colhead{FA14} & 
\colhead{} \\ 
\colhead{RM ID} & 
\colhead{ID} & 
\colhead{${v_{\rm max}}$\tablenotemark{a}} & 
\colhead{${v_{\rm min}}$\tablenotemark{a}} & 
\colhead{${v_{\rm cent}}$\tablenotemark{a}} & 
\colhead{$\langle d \rangle$} & 
\colhead{(\AA)} & 
\colhead{ID\tablenotemark{b}} & 
\colhead{P\,{\sc v}\tablenotemark{c}} 
}  
\startdata 
RM 039    &    A    &         29148    &          2190    &         13445    &          0.67    &          88.5$\pm$0.7    &    C{\sc iv}$_{\rm SA}$    &    Y    \\ 
RM 073    &    A    &         10142    &          5768    &          7821    &          0.23    &           5.1$\pm$0.4    &    C{\sc iv}$_{\rm 00}$    &    N    \\ 
RM 073    &    B    &          4586    &          1456    &          2680    &          0.51    &           8.0$\pm$0.2    &    C{\sc iv}$_{\rm 00}$    &    N    \\ 
RM 116    &    A    &         11599    &          5467    &          7946    &          0.26    &           8.2$\pm$0.3    &    C{\sc iv}$_{\rm S0}$    &    \nodata    \\ 
RM 116    &    B    &          4860    &          2746    &          3650    &          0.36    &           4.0$\pm$0.2    &    C{\sc iv}$_{\rm 00}$    &    \nodata    \\ 
RM 128    &    A    &         21360    &          4814    &         10569    &          0.38    &          31.5$\pm$0.7    &    C{\sc iv}$_{\rm S0}$    &    \nodata    \\ 
RM 128    &    B    &          2691    &           583    &          1486    &          0.22    &           2.5$\pm$0.2    &    C{\sc iv}$_{\rm 00}$    &    \nodata    \\ 
RM 155    &    A    &          7670    &           447    &          3249    &          0.54    &          19.9$\pm$0.4    &    C{\sc iv}$_{\rm SA}$    &    \nodata    \\ 
RM 195    &    A    &          3243    &           378    &          1788    &          0.41    &           5.9$\pm$0.1    &    C{\sc iv}$_{\rm 00}$    &    N    \\ 
RM 217    &    A    &         19323    &         14820    &         17014    &          0.19    &           4.2$\pm$0.4    &    C{\sc iv}$_{\rm S0}$    &    \nodata    \\ 
RM 217    &    B    &         11640    &          2901    &          6379    &          0.47    &          20.8$\pm$0.5    &    C{\sc iv}$_{\rm S0}$    &    \nodata    \\ 
RM 257    &    A    &         24762    &          6659    &         13586    &          0.48    &          42.6$\pm$0.4    &    C{\sc iv}$_{\rm SA}$    &    Y    \\ 
RM 257    &    B    &          4117    &          2098    &          3067    &          0.50    &           5.2$\pm$0.1    &    C{\sc iv}$_{\rm 00}$    &    N    \\ 
RM 284    &    A    &         24141    &         19496    &         21736    &          0.12    &           2.8$\pm$0.2    &    C{\sc iv}$_{\rm N0}$    &    \nodata    \\ 
RM 339    &    A    &         15188    &          7789    &         10850    &          0.26    &           9.7$\pm$0.3    &    C{\sc iv}$_{\rm SA}$    &    \nodata    \\ 
RM 357    &    A    &         14782    &         11935    &         13284    &          0.15    &           2.2$\pm$0.3    &    C{\sc iv}$_{\rm 00}$    &    \nodata    \\ 
RM 357    &    B    &          7152    &          1891    &          3934    &          0.53    &          14.1$\pm$0.3    &    C{\sc iv}$_{\rm S0}$    &    \nodata    \\ 
RM 361    &    A    &          7122    &          1463    &          3980    &          0.54    &          15.3$\pm$0.3    &    C{\sc iv}$_{\rm N0}$    &    \nodata    \\ 
RM 408    &    A    &          4451    &           656    &          2201    &          0.50    &           9.8$\pm$0.2    &    C{\sc iv}$_{\rm 00}$    &    \nodata    \\ 
RM 508    &    A    &         21344    &         18101    &         19590    &          0.36    &           5.7$\pm$0.1    &    C{\sc iv}$_{\rm 00}$    &    N    \\ 
RM 509    &    A    &         24553    &         13814    &         18176    &          0.26    &          13.6$\pm$0.5    &    C{\sc iv}$_{\rm S0}$    &    N    \\ 
RM 509    &    B    &          2464    &           337    &          1371    &          0.46    &           5.1$\pm$0.1    &    C{\sc iv}$_{\rm 00}$    &    N    \\ 
RM 564    &    A    &         25282    &         21819    &         23559    &          0.17    &           2.8$\pm$0.1    &    C{\sc iv}$_{\rm S0}$    &    \nodata    \\ 
RM 565    &    A    &         27794    &         14010    &         20647    &          0.21    &          13.7$\pm$1.3    &    C{\sc iv}$_{\rm N0}$    &    \nodata    \\ 
RM 565    &    B    &         13688    &          3878    &          8127    &          0.24    &          11.7$\pm$1.0    &    C{\sc iv}$_{\rm S0}$    &    \nodata    \\ 
RM 613    &    A    &         20046    &         15713    &         17872    &          0.22    &           4.8$\pm$0.2    &    C{\sc iv}$_{\rm 00}$    &    \nodata    \\ 
RM 631    &    A    &          8931    &          5858    &          7295    &          0.47    &           7.2$\pm$0.2    &    C{\sc iv}$_{\rm S0}$    &    Y    \\ 
RM 717    &    A    &         23343    &          8567    &         15064    &          0.30    &          21.8$\pm$0.7    &    C{\sc iv}$_{\rm S0}$    &    \nodata    \\ 
RM 722    &    A    &          6238    &          2063    &          3955    &          0.58    &          12.5$\pm$0.2    &    C{\sc iv}$_{\rm 00}$    &    N    \\ 
RM 729    &    A    &          9754    &          3465    &          6306    &          0.37    &          11.8$\pm$0.3    &    C{\sc iv}$_{\rm S0}$    &    Y    \\ 
RM 730    &    A    &         27969    &         15310    &         22035    &          0.40    &          24.5$\pm$0.2    &    C{\sc iv}$_{\rm S0}$    &    Y    \\ 
RM 730    &    B    &         14823    &         11675    &         12893    &          0.15    &           2.5$\pm$0.1    &    C{\sc iv}$_{\rm 00}$    &    N    \\ 
RM 743    &    A    &         11177    &          4423    &          7402    &          0.31    &          10.6$\pm$0.6    &    C{\sc iv}$_{\rm SA}$    &    \nodata    \\ 
RM 770    &    A    &         20663    &         15904    &         18231    &          0.15    &           3.6$\pm$0.1    &    C{\sc iv}$_{\rm 00}$    &    \nodata    \\ 
RM 785    &    A    &         26858    &         12732    &         19713    &          0.21    &          14.3$\pm$0.3    &    C{\sc iv}$_{\rm S0}$    &    Y    \\ 
RM 786    &    A    &         14967    &         11437    &         13062    &          0.15    &           2.7$\pm$0.2    &    C{\sc iv}$_{\rm 00}$    &    \nodata    \\ 
RM 786    &    B    &          8814    &          $-338$    &          3456    &          0.75    &          34.4$\pm$0.2    &    C{\sc iv}$_{\rm SA}$    &    \nodata     
\enddata 
\tablenotetext{a}{Velocities are in units of km~s$^{-1}$.} 
\tablenotetext{b}{BAL designation according to whether or not BALs or mini-BALs are present for Si\,{\sc iv} and Al\,{\sc iii}, as defined by \cite{Filizak14}; see Section~\ref{sec:otherspecies}. A designation of C\,{\sc iv}$_{\rm N0}$ indicates that the Si\,{\sc iv} line is not completely covered by our spectrum at those velocities or we are unable to otherwise determine whether or not it is present, so we are unsure of the designation.} 
\tablenotetext{c}{This column indicates whether or not there is a suspected corresponding P\,{\sc v} BAL or mini-BAL (see Section~\ref{sec:otherspecies}): ``Y" indicates that P\,{\sc v} is suspected to be present; ``N" indicates that it is not. Entries with no data indicate that the P\,{\sc v} region is not covered by the SDSS spectra.} 
\label{Table:baltbl} 
\end{deluxetable*} 

\subsection{BAL Measurements} 
\begin{figure}
\begin{center}
\includegraphics[scale = 0.26, trim = 15 0 0 0, clip]{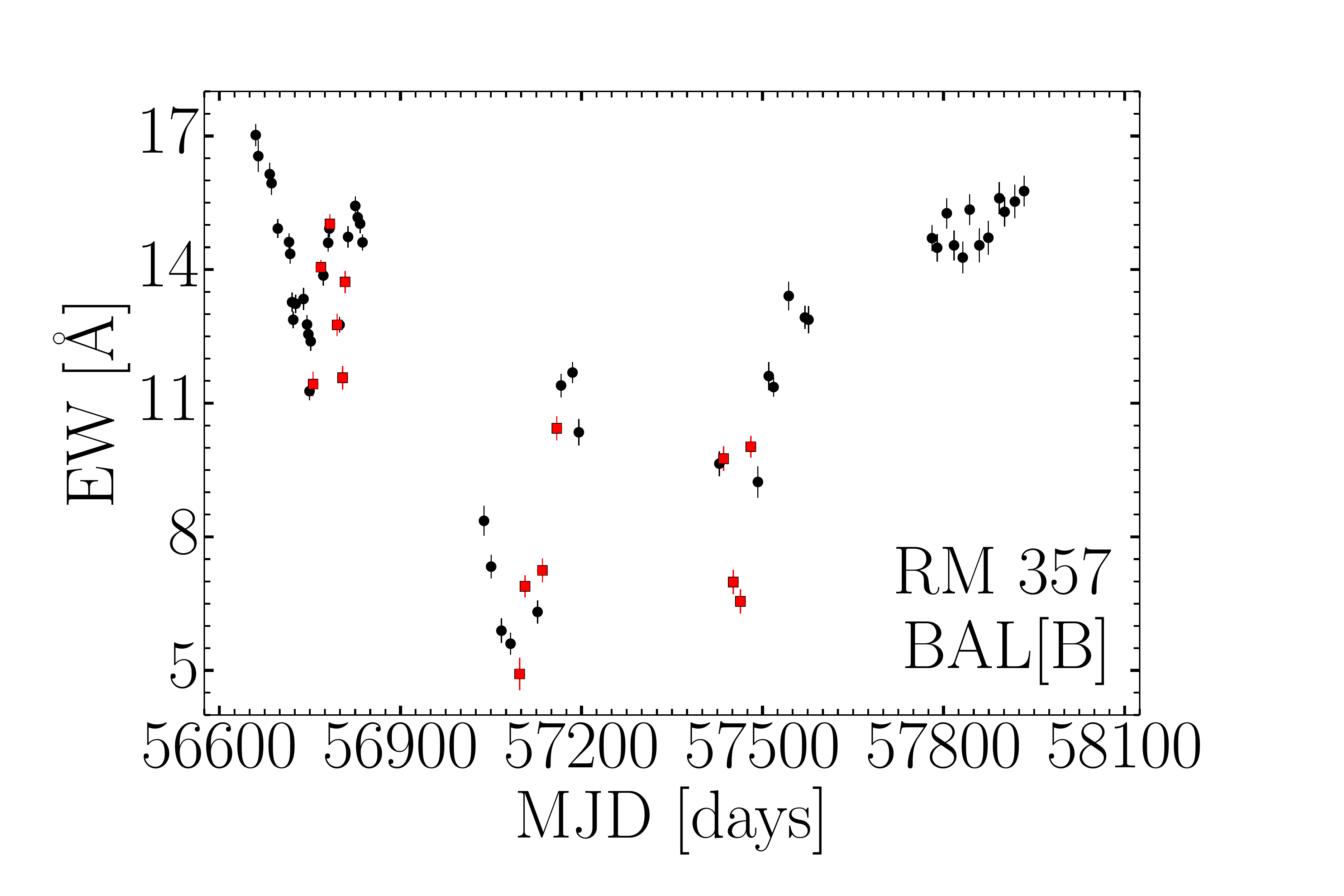}
\caption{EW vs MJD for RM\,357 BAL[B]. Epoch pairs that satisfy our criteria for significant short-term variability (see Section~\ref{sec:criteria}) are marked by red squares. Figures for all 37 individual BALs in our sample are provided in the online version of the article.}
\label{fig:ew}
\end{center}
\end{figure}

After identifying the BALs and refining the BAL sample, we calculated the BAL velocity bounds ($v_{\rm min}$ and $v_{\rm max}$) from the pixels on either end of the trough that recovered to normalized flux densities of 0.9 or higher. We measure these velocity bounds from the mean spectrum created from the data taken in 2014 for each quasar and adopt them for all individual epochs. We define our velocity ranges as positive and increasing at larger blueshifts. We computed the rest-frame equivalent widths (EWs) for each BAL, along with rest-frame mean fractional depth $\langle d \rangle$ and absorbed-flux weighted centroid velocity ($v_{\rm cent}$). The EW and $v_{\rm cent}$ are defined as follows:

\begin{equation}
	\rm{EW} = \int_{\lambda_{\rm min}}^{\lambda_{\rm max}} [1 - f_{n}(\lambda)]\ \rm{d}\lambda
\end{equation}

\begin{equation}
	v_{\rm {cent}} = \frac{\int_{v_{\rm min}}^{v_{\rm max}} v [1 - f_{\rm n}(v)]\ \rm{d}\textit{v}}{\int_{v_{\rm min}}^{v_{\rm max}}{[1-f_{\rm n}(v)]}\ \rm{d}\textit{v}}
\end{equation}
where $f_{\rm n}$ is the continuum+emission-line-normalized flux across the BAL trough. We propagate the normalized spectral uncertainties to determine the uncertainty in our EW measurements. 
We provide these measurements from the 2014 mean spectra for our final sample in Table~\ref{Table:baltbl}, and provide measurements for each individual epoch in Table~\ref{Table:monster}. The EW measurements of each BAL as a function of the Modified Julian Date (MJD) are presented in Figure~\ref{fig:ew}. Nearly all of our BALs show significant variability over the four years of monitoring --- most of the EW light curves reveal visible trends across even single observing seasons. 

\begin{deluxetable}{ccccrr} 
\tablewidth{0.45\textwidth} 
\tabletypesize{\footnotesize}
\renewcommand{\arraystretch}{0.85}
\tablecaption{\civ \ BAL Measurements per Epoch} 
\tablehead{ 
\colhead{} & 
\colhead{BAL} & 
\colhead{MJD} &
\colhead{$v_{\rm cent}$} & 
\colhead{} & 
\colhead{$\rm{EW}$} \\ 
\colhead{RM ID} & 
\colhead{ID} & 
\colhead{(days)} & 
\colhead{($\rm{km \ s^{-1}}$)} & 
\colhead{$\langle d \rangle$} & 
\colhead{(\AA)} 
}  
\startdata 
RM 039    &    A    &      56660.21    &         13250    &          0.65    &          85.7 $\pm$        0.8    \\ 
RM 039    &    A    &      56664.51    &         13320    &          0.63    &          83.8 $\pm$        1.0    \\ 
RM 039    &    A    &      56683.48    &         13260    &          0.66    &          88.0 $\pm$        0.8    \\ 
RM 039    &    A    &      56686.47    &         13230    &          0.66    &          87.6 $\pm$        0.8    \\ 
RM 039    &    A    &      56696.78    &         13390    &          0.68    &          89.5 $\pm$        0.9    \\ 
RM 039    &    A    &      56715.39    &         13200    &          0.67    &          88.6 $\pm$        0.7    \\ 
RM 039    &    A    &      56717.33    &         13440    &          0.65    &          86.1 $\pm$        0.6    \\ 
RM 039    &    A    &      56720.45    &         13380    &          0.65    &          86.6 $\pm$        0.7    \\ 
RM 039    &    A    &      56722.39    &         13310    &          0.66    &          87.2 $\pm$        0.5    \\ 
RM 039    &    A    &      56726.46    &         13270    &          0.63    &          83.6 $\pm$        0.6    \\ 
RM 039    &    A    &      56739.41    &         13470    &          0.67    &          89.0 $\pm$        0.6    \\ 
RM 039    &    A    &      56745.28    &         13530    &          0.67    &          89.4 $\pm$        0.6    \\ 
RM 039    &    A    &      56747.42    &         13850    &          0.65    &          86.4 $\pm$        0.8    \\ 
RM 039    &    A    &      56749.37    &         13480    &          0.67    &          88.9 $\pm$        0.7    \\ 
RM 039    &    A    &      56751.34    &         13520    &          0.65    &          86.0 $\pm$        0.7    \\ 
RM 039    &    A    &      56755.34    &         13560    &          0.67    &          89.1 $\pm$        0.8    \\ 
RM 039    &    A    &      56768.23    &         13590    &          0.67    &          89.0 $\pm$        0.6    \\ 
RM 039    &    A    &      56772.23    &         13470    &          0.67    &          88.9 $\pm$        0.5  
\enddata 
\tablecomments{This table is available in its entirety in the online version of the article. A portion is shown here for guidance on formatting.}
\label{Table:monster} 
\end{deluxetable} 
\subsection{Detection of Significant Variability}
\label{sec:criteria}
\begin{figure}
\begin{center}
	\includegraphics[scale = 0.25, trim = 20 0 0 120, clip]{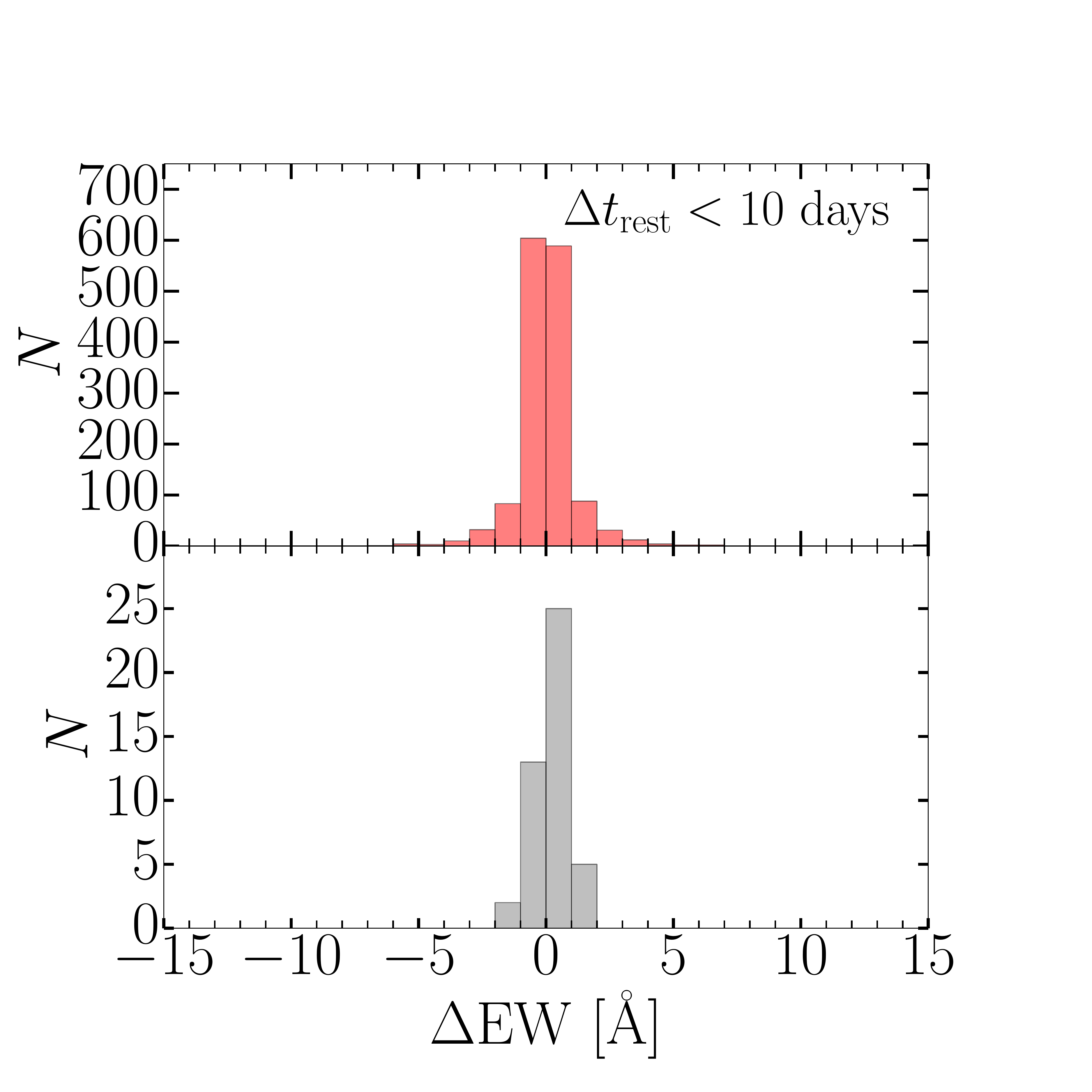}
	\includegraphics[scale = 0.25, trim = 20 40 0 120, clip]{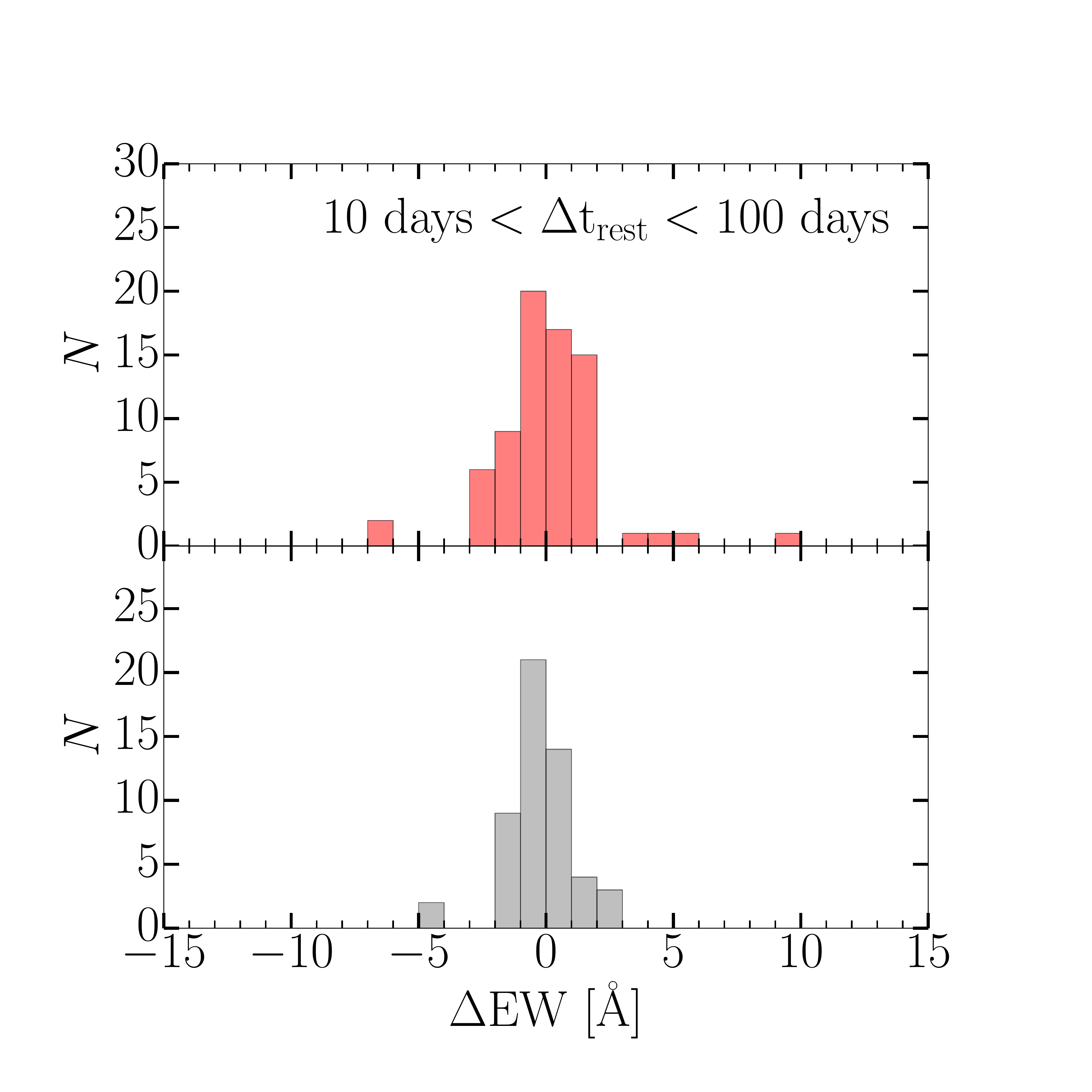}
    \caption{The distribution of $\Delta$EW among BAL quasars. Panels with red histograms show measurements from our work using all pairs of sequential epochs for each quasar in our sample. Panels with gray histograms represent data from \cite{Barlow93}, \cite{Lundgren07}, \cite{Gibson08}, and \cite{Filizak13}. The top two panels show the distributions for epochs that are separated by 0--10 days in the quasar rest frame and the bottom two panels show the distributions for epochs separated by 10--100 days in the quasar rest frame. See Section~\ref{sec:samplechar} for a comparison between our work and previous studies.}
\label{fig:ewdist}
\end{center}
\end{figure}

We began our investigation into BAL variability by measuring $\Delta$EW, the change in equivalent width of the BAL between two subsequent epochs. We calculated the uncertainty in the $\Delta$EW parameter using a quadrature sum of equivalent width uncertainties from each epoch of interest. As discussed above, we removed some epochs from our sample due to various issues (low SN, bad sky subtraction, etc); therefore, from object to object, the cadence deviates from the SDSS-RM observation cadence. Figure~\ref{fig:ewdist} shows the distribution of $\Delta \rm{EW}$ for all sequential epoch pairs in our sample. The distribution of $\Delta$EW is centered on zero and contains mostly small-amplitude variations, though there are rare cases of more dramatic variability, particularly on timescales ranging from $10~-~100$~days in the quasar rest frame.

To identify significant variability between epoch pairs, we follow previous works (e.g., \citealt{Grier15}) and require that the ratio of $\Delta$EW to its uncertainty, $\sigma_{\Delta \rm EW}$, be at least four (i.e., we impose a 4$\sigma$ significance requirement on $\Delta$EW). However, solely relying on the aforementioned significance criteria returned several cases of supposed BAL variability for which nearby regions outside the BAL also varied considerably. This phenomenon suggests that the variability observed within the BAL may be due to spectral variability unassociated with the BAL itself (whether real deviations from a power-law, or spurious due to spectrophotometric residuals on scales of tens of \AA), or inconsistencies in continuum fits. These inconsistencies are likely due to nearby emission-line variability or a lack of line-free regions close to the BAL, which, combined with noise, causes the continuum fits to vary somewhat from epoch to epoch. 
Such cases motivated the implementation of additional criteria which could isolate true BAL variability from variations in continuum and emission-line fits. For these purposes, we define a quantity hereafter referred to as $G$: 
\begin{equation}
	G = \frac{\chi^2 - (N - 1)}{\sqrt{2(N - 1)}}
\end{equation}
where the $\chi^2$ quantity is the square of the difference between fluxes at two spectral epochs divided by their combined uncertainty, summed over the $N$ pixels within a specified region. The expected value and standard deviation of the distribution of the $\chi^2$ statistic for a region of $N$ pixels are $N - 1$ and $\sqrt{2(N - 1)}$, respectively; thus the $G$ statistic should have a mean of zero and standard deviation of one if there are no variations between the two spectra. Larger absolute values of $G$ indicate inconsistencies between the two spectra, suggesting variability at some level of significance. 

For each BAL, we located nearby regions that contained very few features, hereafter referred to as ``continuum" regions. We use these regions to search for cases where there is variability within the BAL region of the normalized spectra but {\it not} within the continuum regions. Because we did not fit the nearby \siiv \  emission line, this region (and all regions blueward) were not included in the continuum regions. 
Continuum regions redward of \siiv \ were selected via visual inspection, referencing the composite quasar spectra of \cite{Vandenberk01} to search for wavelength regions with minimal contaminants. We then calculated the $G$ values in the identified continuum regions and within the BAL region. Hereafter, we refer to these $G$ values as $G_{\rm C}$ and $G_{\rm B}$, respectively.

\begin{figure}
\begin{center}
	\includegraphics[scale = 0.23, trim = 15 0 0 0, clip]{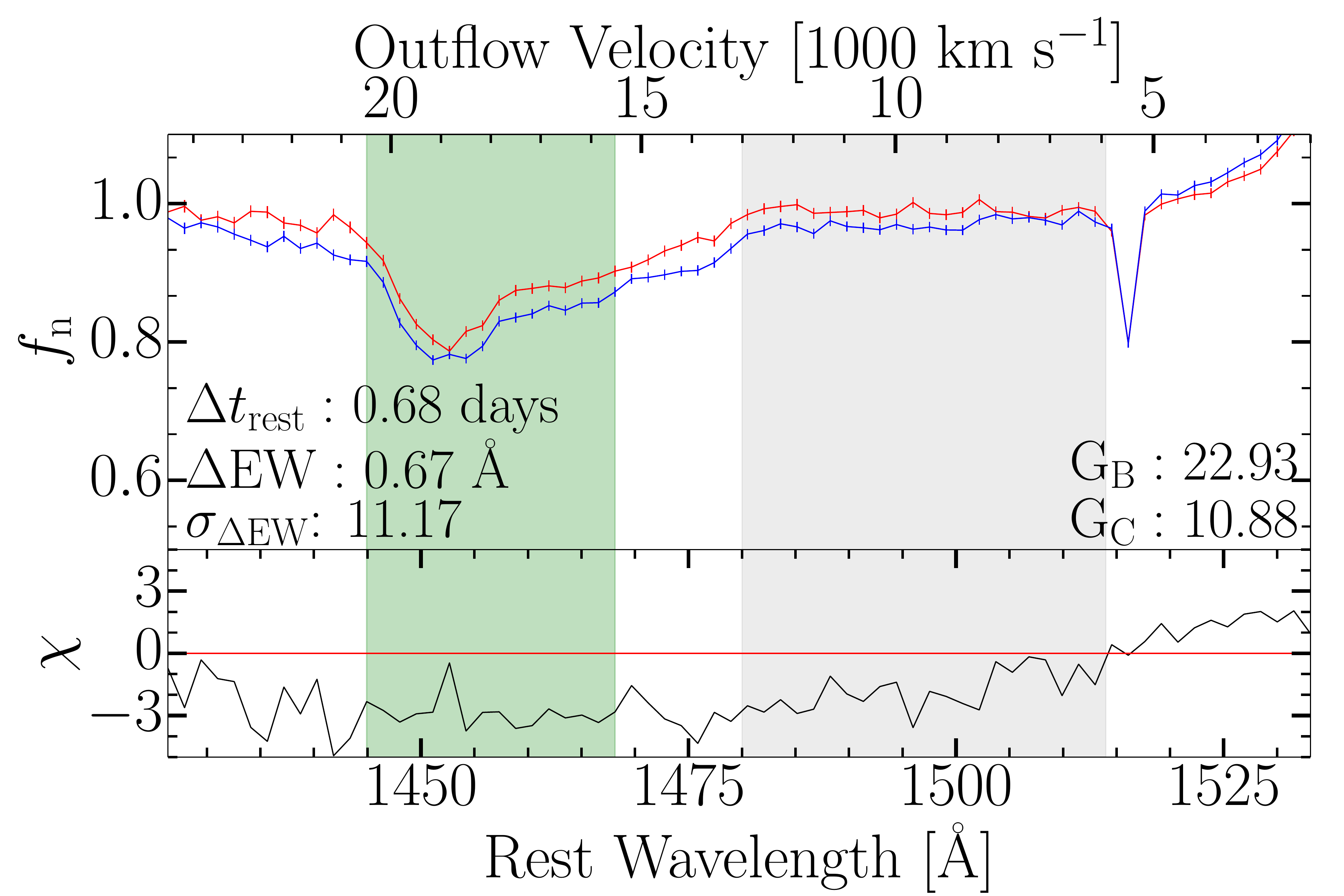}
    \caption{A demonstration of the utility of our additional significance criteria. The spectra shown are from RM\,770, MJDs 56715--56717. The top subpanel shows the normalized flux, $f_{\rm n}$. The red line and error bars correspond to the first epoch listed, and blue corresponds to the second epoch. The gray shaded region displays the continuum region used to calculate G$_{C}$. The green shaded region highlights the region containing the BAL being examined --- this region was used to calculate $G_{\rm B}$. Relevant measurements ($\Delta t_{\rm rest}$, $\Delta$EW, $\sigma_{\Delta \rm{EW}}$, $G_{\rm B}$, and $G_{\rm C}$) are provided. The bottom subpanel shows the $\chi$ statistic (showing the difference between the two spectra in units of their uncertainties) as a function of wavelength.}
\label{fig:falsepos}
\end{center}
\end{figure}

For a pair of spectra to exhibit ``significant" variability, we require that $G_{\rm B} > 4$ and $G_{\rm C} < 2$. 
Figure~\ref{fig:falsepos} demonstrates the use of these criteria in excluding cases of spurious detections from our sample. 
These criteria may exclude cases of true BAL variability due to random fluctuations in the continuum regions; however, the use of these criteria allows us to create a conservative sample of spectra showing significant short-timescale BAL variability. 

\begin{figure*}
\begin{center}
\includegraphics[scale = 0.45, trim = 0 0 0 0, clip]{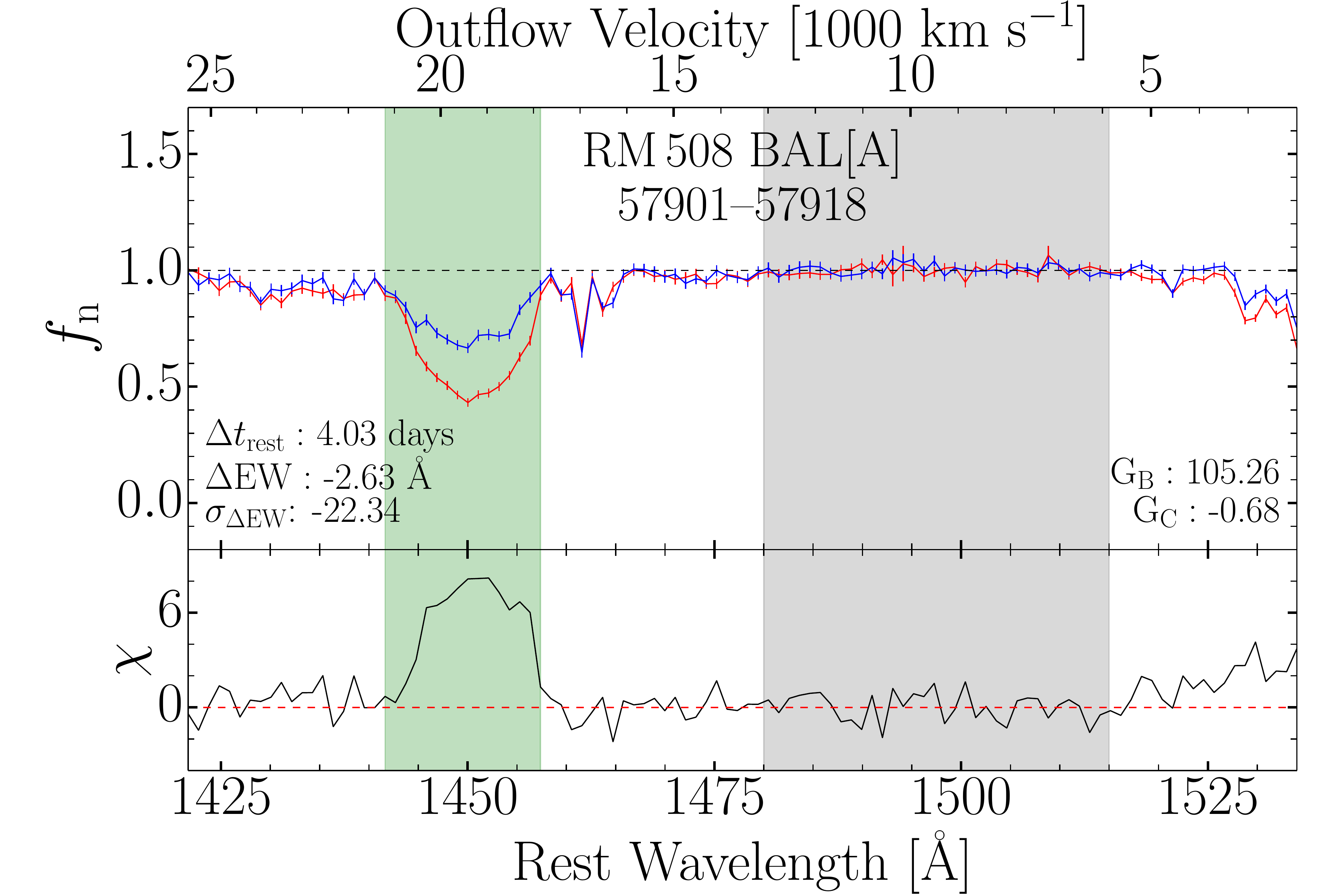}    
\caption{A pair of spectra showing ``significant" variability for one epoch pair of RM\,508. The top subpanels show the normalized flux, $f_{\rm n}$, with the MJDs of the two epochs provided below the RM identifier in the title of each plot. The red line and error bars correspond to the first epoch listed, and blue corresponds to the second epoch. The gray shaded regions display the continuum regions used to calculate $G_{\rm C}$. The green shaded region highlights the region containing the BAL being examined --- this region was used to calculate $G_{\rm B}$. Yellow shaded regions indicate additional C\,{\sc iv} BALs, when present. Relevant measurements ($\Delta t_{\rm rest}$, $\Delta$EW, $\sigma_{\Delta \rm{EW}}$, $G_{\rm B}$, and $G_{\rm C}$) are provided for each pair of spectra. The bottom subpanels for each pair show the $\chi$ statistic (quantifying the difference between the two spectra in units of their uncertainties) as a function of wavelength. Figures for all 54 instances of significant short-timescale variability are provided in the online version of the article.} 
\label{fig:var}
\end{center}
\end{figure*}

\section{RESULTS \& DISCUSSION}
\label{sec:discussion}

\subsection{BALs Exhibiting Short-Timescale Variability}
\label{sec:var_overview}
	We searched for short-timescale variability in our sample by identifying pairs of epochs with significant variability (as defined above) on rest-frame timescales of less than 10 days. Figure~\ref{fig:var} shows all pairs of spectra in which we identified significant short-timescale variability, and Table~\ref{Table:shortvartbl} provides information on which epochs show significant variability and relevant measurements. We identify 54 epoch pairs in 19 unique BALs in 15 different quasars that show significant variability on these short timescales (four quasars had two C\,{\sc iv} BALs present that both varied significantly). This indicates that short-timescale BAL variability occurs in at least $55^{+18}_{-14}$\% of BAL quasars and in $51^{+15}_{-12}$\% of BAL troughs (upper and lower confidence limits, here and henceforth, are all calculated via \citealt{Gehrels86}, unless noted otherwise).
    
We also examine the frequency of rapid variability among all epoch pairs. In our sample, we have a total 1460 pairs of sequential epochs separated by less than 10 days in the quasar rest frame. As noted above, 54 of these epoch pairs exhibit significant short-timescale variability, which corresponds to a frequency of $3.7 \pm 0.5\%$.

The observed high fraction of BAL troughs exhibiting short-timescale variability raises the possibility that all BAL troughs would exhibit short-timescale variability if observed a sufficient number of times.  To test this possibility, we consider the following simple model (in which ``epoch pairs" refers to subsequent epoch pairs and ``variability" to significant short-timescale variability).  First, we calculate two probabilities: the probability that an epoch pair will show variability if the immediately previous epoch pair did not show such variability (3.0\%) and if the immediately preceding epoch pair did (22\%). These probabilities show that significant short-timescale variability is not distributed completely at random in time in a given BAL trough.

We assume the probability of {\sl not} seeing a variability event in a given trough and short-timescale epoch pair is $P_{\rm no}$=0.97, regardless of the exact time separation between subsequent epochs. For a BAL trough observed $N_i$ times, the probability of not seeing a variability event in that BAL is $P_i = 0.97^{(N_i-1)}$. Summing over all troughs in a
sample ($i$=1 to $N_{\rm troughs}$), the expected number of BAL troughs without variability events is $N_{\rm no} = \sum_i P_i$, with a variance of $\sigma^2 = N_{\rm troughs} \bar{P_i} (1-\bar{P_i})$, where $\bar{P_i}$ is the average of $P_i$ over all $i$ troughs. This toy model for short-timescale variability occurring in all BAL troughs predicts that we should see $13.5 \pm 2.9$ troughs without significant variability in our sample of 37 troughs with an average of 39.5 epoch pairs per trough. This prediction is likely a small underestimate because it ignores cases of three or more epoch pairs in a row showing variability; accounting for that effect would increase $P_{\rm no}$ and $N_{\rm no}$. In reality, we observed 18 such troughs, a +1.5$\sigma$ deviation. Thus, our data are consistent with a toy model in which short-timescale variability occurs in all BAL troughs.

Our observed frequency of rapid variability is significantly higher than that reported by \cite{Capellupo13}; they found that only $12^{+15}_{-7.6}$\% of their quasar sample (2/17 quasars) showed significant variability on timescales of less than 72 days (they report 29\% for a less-conservative estimate). As in our study, several of the \cite{Capellupo13} quasars host multiple BALs --- \cite{Capellupo11} identify 25 unique BALs in the sample of 17 quasars with short-timescale data. Two of these BALs varied on timescales of less than 10 days, corresponding to a BAL variability frequency of $8^{+10}_{-5}$\% (2/25 BALs). Applying the same arguments above to the \cite{Capellupo13} sample and assuming $P_{\rm no}$~=~0.97 regardless of exact time separation, we find that their expected number of BAL troughs without observed variability events is $N_{\rm no} = 22.5 \pm 1.5$. This agrees very well with their true observed number of troughs without variability events, 23. Thus, although our aforementioned frequency of observed short-timescale variability is significantly higher than that of \cite{Capellupo13}, this discrepancy could be explained by their comparative lack of spectral epochs.

\begin{deluxetable*}{lccccrrrrrrccr} 
\tablewidth{0pt} 
\tabletypesize{\footnotesize}
\renewcommand{\arraystretch}{0.85}
\tablecaption{Short-Timescale Variability Among Epoch Pairs} 
\tablehead{ 
\colhead{} & 
\colhead{BAL} & 
\colhead{$\rm MJD_1$} & 
\colhead{$\rm MJD_2$} & 
\colhead{$\Delta t_{\rm rest}$} & 
\colhead{$\rm EW_1$} & 
\colhead{$\rm EW_2$} & 
\colhead{$\Delta \rm{EW}$} & 
\colhead{} & 
\colhead{} & 
\colhead{} & 
\colhead{$f$} & 
\colhead{$f$} & 
\colhead{} \\
\colhead{RM ID} & 
\colhead{ID} & 
\colhead{(days)} & 
\colhead{(days)} & 
\colhead{(days)} & 
\colhead{(\AA)} & 
\colhead{(\AA)} & 
\colhead{(\AA)} & 
\colhead{$\sigma_{\Delta \rm{EW}}$} & 
\colhead{$G_{\rm B}$} & 
\colhead{$G_{\rm C}$} & 
\colhead{($\chi$$>$1)} & 
\colhead{($\chi$$<$$-1$)} & 
\colhead{$\Delta v_{\rm co}$\tablenotemark{a}} 
}  
\startdata 
RM 128    &    A    &    $  56768.23$    &    $  56772.23$    &    $      1.40$    &    $      34.9 \pm        0.5$    &    $      31.6 \pm        0.5$    &    $      -3.3 \pm        0.8$    &   $     -4.26$    &    $      5.15$    &    $     -0.70$    &    $      0.32$    &    $      0.08$    &    $       620$    \\ 
RM 128    &    A    &    $  56783.25$    &    $  56795.18$    &    $      4.16$    &    $      31.7 \pm        0.7$    &    $      26.4 \pm        0.7$    &    $      -5.3 \pm        1.0$    &   $     -5.08$    &    $      4.46$    &    $      0.39$    &    $      0.32$    &    $      0.09$    &    $       610$    \\ 
RM 128    &    B    &    $  56720.45$    &    $  56722.39$    &    $      0.68$    &    $       1.7 \pm        0.1$    &    $       2.6 \pm        0.1$    &    $       0.9 \pm        0.2$    &   $      5.23$    &    $      8.87$    &    $      0.29$    &    $      0.00$    &    $      0.57$    &    $       900$    \\ 
RM 128    &    B    &    $  56825.19$    &    $  56829.21$    &    $      1.41$    &    $       2.0 \pm        0.1$    &    $       3.1 \pm        0.1$    &    $       1.1 \pm        0.2$    &   $      5.50$    &    $      9.41$    &    $      1.29$    &    $      0.00$    &    $      0.86$    &    $      1500$    \\ 
RM 217    &    B    &    $  56660.21$    &    $  56683.48$    &    $      8.27$    &    $      15.7 \pm        0.5$    &    $      19.2 \pm        0.5$    &    $       3.5 \pm        0.7$    &   $      4.93$    &    $      4.13$    &    $     -1.27$    &    $      0.00$    &    $      0.41$    &    $       310$    \\ 
RM 217    &    B    &    $  56683.48$    &    $  56686.47$    &    $      1.06$    &    $      19.2 \pm        0.5$    &    $      16.3 \pm        0.5$    &    $      -2.9 \pm        0.7$    &   $     -4.30$    &    $      4.33$    &    $      0.28$    &    $      0.56$    &    $      0.15$    &    $       940$    \\ 
RM 217    &    B    &    $  56686.47$    &    $  56696.78$    &    $      3.66$    &    $      16.3 \pm        0.5$    &    $      19.3 \pm        0.4$    &    $       3.1 \pm        0.6$    &   $      4.76$    &    $      4.70$    &    $     -0.77$    &    $      0.07$    &    $      0.48$    &    $       630$    \\ 
RM 257    &    A    &    $  57510.32$    &    $  57518.31$    &    $      2.33$    &    $      40.6 \pm        0.4$    &    $      47.1 \pm        0.5$    &    $       6.4 \pm        0.6$    &   $     10.10$    &    $     27.03$    &    $      0.38$    &    $      0.13$    &    $      0.54$    &    $      7250$    \\ 
RM 257    &    A    &    $  57518.31$    &    $  57543.45$    &    $      7.32$    &    $      47.1 \pm        0.5$    &    $      42.1 \pm        0.5$    &    $      -4.9 \pm        0.7$    &   $     -7.11$    &    $      9.46$    &    $      1.13$    &    $      0.49$    &    $      0.06$    &    $      3650$    \\ 
RM 257    &    B    &    $  56715.39$    &    $  56717.33$    &    $      0.57$    &    $       4.9 \pm        0.1$    &    $       5.3 \pm        0.1$    &    $       0.4 \pm        0.1$    &   $      4.60$    &    $      5.26$    &    $     -0.66$    &    $      0.00$    &    $      0.75$    &    $       510$    \\ 
RM 257    &    B    &    $  56717.33$    &    $  56720.45$    &    $      0.91$    &    $       5.3 \pm        0.1$    &    $       4.9 \pm        0.1$    &    $      -0.4 \pm        0.1$    &   $     -4.64$    &    $      6.24$    &    $     -0.42$    &    $      0.75$    &    $      0.00$    &    $       510$    \\ 
RM 257    &    B    &    $  57510.32$    &    $  57518.31$    &    $      2.33$    &    $       4.9 \pm        0.1$    &    $       5.6 \pm        0.1$    &    $       0.7 \pm        0.1$    &   $      5.90$    &    $     14.09$    &    $      0.38$    &    $      0.12$    &    $      0.62$    &    $      1010$    \\ 
RM 257    &    B    &    $  57518.31$    &    $  57543.45$    &    $      7.32$    &    $       5.6 \pm        0.1$    &    $       5.1 \pm        0.1$    &    $      -0.6 \pm        0.1$    &   $     -4.32$    &    $      5.41$    &    $      1.13$    &    $      0.62$    &    $      0.00$    &    $       760$    \\ 
RM 257    &    B    &    $  57918.16$    &    $  57933.42$    &    $      4.44$    &    $       4.6 \pm        0.1$    &    $       5.3 \pm        0.1$    &    $       0.6 \pm        0.1$    &   $      5.06$    &    $      7.20$    &    $     -0.40$    &    $      0.00$    &    $      0.75$    &    $      1010$    \\ 
RM 357    &    A    &    $  56755.34$    &    $  56768.23$    &    $      4.10$    &    $       0.6 \pm        0.2$    &    $       2.1 \pm        0.2$    &    $       1.5 \pm        0.3$    &   $      5.09$    &    $      5.98$    &    $     -0.80$    &    $      0.00$    &    $      0.80$    &    $      1140$    \\ 
RM 357    &    A    &    $  56783.25$    &    $  56795.18$    &    $      3.79$    &    $       2.5 \pm        0.2$    &    $       0.6 \pm        0.2$    &    $      -1.9 \pm        0.3$    &   $     -6.08$    &    $      8.68$    &    $      1.59$    &    $      0.90$    &    $      0.00$    &    $      2280$    \\ 
RM 357    &    B    &    $  56755.34$    &    $  56768.23$    &    $      4.10$    &    $      11.4 \pm        0.3$    &    $      14.1 \pm        0.2$    &    $       2.6 \pm        0.3$    &   $      8.07$    &    $     17.98$    &    $     -0.80$    &    $      0.11$    &    $      0.53$    &    $      2220$    \\ 
RM 357    &    B    &    $  56783.25$    &    $  56795.18$    &    $      3.79$    &    $      15.0 \pm        0.2$    &    $      12.8 \pm        0.3$    &    $      -2.3 \pm        0.3$    &   $     -6.69$    &    $     10.46$    &    $      1.59$    &    $      0.63$    &    $      0.11$    &    $      1950$    \\ 
RM 357    &    B    &    $  56804.19$    &    $  56808.26$    &    $      1.29$    &    $      11.6 \pm        0.3$    &    $      13.7 \pm        0.2$    &    $       2.2 \pm        0.4$    &   $      5.88$    &    $     10.66$    &    $      0.31$    &    $      0.05$    &    $      0.47$    &    $      1110$    \\ 
RM 357    &    B    &    $  57097.48$    &    $  57106.47$    &    $      2.86$    &    $       4.9 \pm        0.4$    &    $       6.9 \pm        0.3$    &    $       2.0 \pm        0.4$    &   $      4.45$    &    $      4.21$    &    $     -0.98$    &    $      0.00$    &    $      0.53$    &    $       550$    \\ 
RM 357    &    B    &    $  57135.17$    &    $  57159.16$    &    $      7.63$    &    $       7.2 \pm        0.3$    &    $      10.4 \pm        0.3$    &    $       3.2 \pm        0.4$    &   $      8.30$    &    $     16.19$    &    $     -1.33$    &    $      0.00$    &    $      0.68$    &    $      1380$    \\ 
RM 357    &    B    &    $  57435.40$    &    $  57451.46$    &    $      5.11$    &    $       9.8 \pm        0.3$    &    $       7.0 \pm        0.3$    &    $      -2.8 \pm        0.4$    &   $     -7.10$    &    $     15.12$    &    $      0.48$    &    $      0.68$    &    $      0.05$    &    $      1390$    \\ 
RM 357    &    B    &    $  57463.40$    &    $  57480.60$    &    $      5.47$    &    $       6.6 \pm        0.3$    &    $      10.0 \pm        0.2$    &    $       3.5 \pm        0.4$    &   $      9.49$    &    $     19.82$    &    $      1.89$    &    $      0.00$    &    $      0.79$    &    $      2490$    \\ 
RM 361    &    A    &    $  56813.23$    &    $  56825.19$    &    $      4.57$    &    $      14.5 \pm        0.2$    &    $      16.7 \pm        0.2$    &    $       2.2 \pm        0.3$    &   $      6.29$    &    $      7.54$    &    $      1.80$    &    $      0.00$    &    $      0.65$    &    $      2010$    \\ 
RM 508    &    A    &    $  56686.47$    &    $  56696.78$    &    $      2.45$    &    $       6.0 \pm        0.0$    &    $       5.4 \pm        0.1$    &    $      -0.6 \pm        0.1$    &   $     -8.52$    &    $     16.87$    &    $      1.72$    &    $      0.80$    &    $      0.00$    &    $      2160$    \\ 
RM 508    &    A    &    $  57185.17$    &    $  57195.60$    &    $      2.48$    &    $       6.7 \pm        0.1$    &    $       6.3 \pm        0.1$    &    $      -0.5 \pm        0.1$    &   $     -5.25$    &    $      4.93$    &    $     -0.68$    &    $      0.73$    &    $      0.07$    &    $      1300$    \\ 
RM 508    &    A    &    $  57789.44$    &    $  57805.35$    &    $      3.78$    &    $       7.1 \pm        0.1$    &    $       6.2 \pm        0.1$    &    $      -0.8 \pm        0.1$    &   $     -9.61$    &    $     20.33$    &    $     -0.62$    &    $      0.87$    &    $      0.00$    &    $      2380$    \\ 
RM 508    &    A    &    $  57805.35$    &    $  57817.32$    &    $      2.84$    &    $       6.2 \pm        0.1$    &    $       6.6 \pm        0.1$    &    $       0.4 \pm        0.1$    &   $      4.29$    &    $      4.27$    &    $     -0.15$    &    $      0.07$    &    $      0.53$    &    $       860$    \\ 
RM 508    &    A    &    $  57874.21$    &    $  57892.29$    &    $      4.30$    &    $       7.1 \pm        0.1$    &    $       6.5 \pm        0.1$    &    $      -0.6 \pm        0.1$    &   $     -6.03$    &    $      4.95$    &    $      0.45$    &    $      0.87$    &    $      0.00$    &    $      1730$    \\ 
RM 508    &    A    &    $  57901.21$    &    $  57918.16$    &    $      4.03$    &    $       6.1 \pm        0.1$    &    $       3.5 \pm        0.1$    &    $      -2.6 \pm        0.1$    &   $    -22.34$    &    $    105.26$    &    $     -0.68$    &    $      0.93$    &    $      0.00$    &    $      2810$    \\ 
RM 508    &    A    &    $  57918.16$    &    $  57933.42$    &    $      3.63$    &    $       3.5 \pm        0.1$    &    $       2.8 \pm        0.1$    &    $      -0.7 \pm        0.1$    &   $     -6.25$    &    $      9.12$    &    $      0.08$    &    $      0.73$    &    $      0.00$    &    $      1510$    \\ 
RM 509    &    B    &    $  56780.23$    &    $  56782.25$    &    $      0.55$    &    $       4.4 \pm        0.1$    &    $       5.2 \pm        0.1$    &    $       0.8 \pm        0.2$    &   $      4.88$    &    $      6.78$    &    $      0.08$    &    $      0.00$    &    $      0.78$    &    $       710$    \\ 
RM 613    &    A    &    $  56686.47$    &    $  56696.78$    &    $      3.08$    &    $       5.0 \pm        0.1$    &    $       5.4 \pm        0.1$    &    $       0.4 \pm        0.1$    &   $      4.26$    &    $      4.57$    &    $      0.15$    &    $      0.00$    &    $      0.44$    &    $       540$    \\ 
RM 613    &    A    &    $  56751.34$    &    $  56755.34$    &    $      1.19$    &    $       5.1 \pm        0.1$    &    $       4.6 \pm        0.1$    &    $      -0.5 \pm        0.1$    &   $     -4.96$    &    $      5.13$    &    $      1.24$    &    $      0.56$    &    $      0.00$    &    $       810$    \\ 
RM 613    &    A    &    $  56772.23$    &    $  56780.23$    &    $      2.39$    &    $       4.7 \pm        0.1$    &    $       4.1 \pm        0.1$    &    $      -0.6 \pm        0.1$    &   $     -7.28$    &    $      8.76$    &    $      1.00$    &    $      0.88$    &    $      0.00$    &    $      1630$    \\ 
RM 631    &    A    &    $  56783.25$    &    $  56795.18$    &    $      3.21$    &    $       6.5 \pm        0.1$    &    $       7.2 \pm        0.1$    &    $       0.7 \pm        0.2$    &   $      4.37$    &    $      6.73$    &    $      1.28$    &    $      0.15$    &    $      0.46$    &    $       470$    \\ 
RM 717    &    A    &    $  56715.39$    &    $  56717.33$    &    $      0.61$    &    $      17.3 \pm        0.5$    &    $      23.6 \pm        0.7$    &    $       6.2 \pm        0.9$    &   $      7.14$    &    $      6.65$    &    $      1.07$    &    $      0.02$    &    $      0.50$    &    $      1690$    \\ 
RM 717    &    A    &    $  56717.33$    &    $  56720.45$    &    $      0.98$    &    $      23.6 \pm        0.7$    &    $      17.7 \pm        0.6$    &    $      -5.8 \pm        0.9$    &   $     -6.44$    &    $      5.17$    &    $      1.84$    &    $      0.56$    &    $      0.06$    &    $      1690$    \\ 
RM 722    &    A    &    $  57185.17$    &    $  57195.60$    &    $      2.94$    &    $      10.5 \pm        0.1$    &    $       8.9 \pm        0.2$    &    $      -1.5 \pm        0.2$    &   $     -7.28$    &    $     11.79$    &    $      1.34$    &    $      0.82$    &    $      0.06$    &    $       740$    \\ 
RM 730    &    A    &    $  56751.34$    &    $  56755.34$    &    $      1.08$    &    $      24.5 \pm        0.1$    &    $      23.5 \pm        0.1$    &    $      -0.9 \pm        0.2$    &   $     -5.08$    &    $      5.17$    &    $     -0.56$    &    $      0.41$    &    $      0.06$    &    $       750$    \\ 
RM 730    &    A    &    $  56755.34$    &    $  56768.23$    &    $      3.49$    &    $      23.5 \pm        0.1$    &    $      24.6 \pm        0.1$    &    $       1.1 \pm        0.2$    &   $      6.11$    &    $      4.20$    &    $      0.41$    &    $      0.04$    &    $      0.49$    &    $      1250$    \\ 
RM 743    &    A    &    $  56715.39$    &    $  56717.33$    &    $      0.71$    &    $       9.3 \pm        0.5$    &    $      12.7 \pm        0.6$    &    $       3.4 \pm        0.7$    &   $      4.52$    &    $      4.92$    &    $      0.54$    &    $      0.10$    &    $      0.43$    &    $       320$    \\ 
RM 743    &    A    &    $  56780.23$    &    $  56782.25$    &    $      0.74$    &    $      12.7 \pm        0.4$    &    $      10.0 \pm        0.4$    &    $      -2.6 \pm        0.6$    &   $     -4.69$    &    $      7.45$    &    $      0.16$    &    $      0.52$    &    $      0.10$    &    $      1610$    \\ 
RM 770    &    A    &    $  56825.19$    &    $  56829.21$    &    $      1.41$    &    $       3.2 \pm        0.0$    &    $       3.5 \pm        0.0$    &    $       0.3 \pm        0.1$    &   $      5.54$    &    $      4.72$    &    $      0.67$    &    $      0.00$    &    $      0.73$    &    $      1580$    \\ 
RM 786    &    A    &    $  56683.48$    &    $  56686.47$    &    $      0.98$    &    $       2.2 \pm        0.1$    &    $       1.6 \pm        0.1$    &    $      -0.6 \pm        0.1$    &   $     -5.03$    &    $      6.42$    &    $      1.56$    &    $      0.67$    &    $      0.00$    &    $       590$    \\ 
RM 786    &    B    &    $  56660.21$    &    $  56664.51$    &    $      1.42$    &    $      31.9 \pm        0.1$    &    $      33.6 \pm        0.2$    &    $       1.8 \pm        0.2$    &   $      7.89$    &    $     12.27$    &    $      1.21$    &    $      0.03$    &    $      0.74$    &    $      2290$    \\ 
RM 786    &    B    &    $  56669.50$    &    $  56683.48$    &    $      4.60$    &    $      34.3 \pm        0.3$    &    $      32.2 \pm        0.1$    &    $      -2.1 \pm        0.3$    &   $     -6.35$    &    $      7.92$    &    $      1.22$    &    $      0.61$    &    $      0.03$    &    $      2590$    \\ 
RM 786    &    B    &    $  56683.48$    &    $  56686.47$    &    $      0.98$    &    $      32.2 \pm        0.1$    &    $      31.4 \pm        0.1$    &    $      -0.8 \pm        0.2$    &   $     -4.92$    &    $      4.48$    &    $      1.56$    &    $      0.45$    &    $      0.06$    &    $       290$    \\ 
RM 786    &    B    &    $  56686.47$    &    $  56696.78$    &    $      3.39$    &    $      31.4 \pm        0.1$    &    $      32.8 \pm        0.1$    &    $       1.4 \pm        0.2$    &   $      9.17$    &    $     13.20$    &    $      0.38$    &    $      0.00$    &    $      0.71$    &    $      3140$    \\ 
RM 786    &    B    &    $  56717.33$    &    $  56720.45$    &    $      1.02$    &    $      33.3 \pm        0.1$    &    $      32.6 \pm        0.1$    &    $      -0.7 \pm        0.2$    &   $     -4.58$    &    $      4.99$    &    $      0.84$    &    $      0.52$    &    $      0.10$    &    $      1430$    \\ 
RM 786    &    B    &    $  56745.28$    &    $  56747.42$    &    $      0.70$    &    $      32.0 \pm        0.1$    &    $      33.3 \pm        0.1$    &    $       1.3 \pm        0.1$    &   $      8.73$    &    $     15.19$    &    $     -0.49$    &    $      0.00$    &    $      0.74$    &    $      2290$    \\ 
RM 786    &    B    &    $  56749.37$    &    $  56751.34$    &    $      0.65$    &    $      33.7 \pm        0.1$    &    $      32.1 \pm        0.1$    &    $      -1.5 \pm        0.2$    &   $     -9.39$    &    $     18.63$    &    $     -0.89$    &    $      0.77$    &    $      0.03$    &    $      3970$    \\ 
RM 786    &    B    &    $  56772.23$    &    $  56780.23$    &    $      2.63$    &    $      32.8 \pm        0.1$    &    $      33.5 \pm        0.1$    &    $       0.7 \pm        0.1$    &   $      5.66$    &    $      7.68$    &    $     -0.41$    &    $      0.06$    &    $      0.58$    &    $      1140$    \\ 
RM 786    &    B    &    $  56804.19$    &    $  56808.26$    &    $      1.34$    &    $      32.6 \pm        0.1$    &    $      33.7 \pm        0.1$    &    $       1.2 \pm        0.2$    &    $      6.48$    &    $      7.88$    &    $      1.53$    &    $      0.03$    &    $      0.48$    &    $      1710$    
\enddata 
\tablenotetext{a}{Coordinated velocity (See Section~\ref{sec:coord}), in units of km~s$^{-1}$.} 
\label{Table:shortvartbl} 
\end{deluxetable*} 

We analyzed several properties of the 15 quasars exhibiting significant short-term variability ($M_{i}$, $z$, radio-loudness) and find that none of them is an outlier compared to the general population of BAL quasars (see Figure~\ref{fig:mag_vs_z}). None of the varying BAL quasars is radio-loud. Additionally, we investigated whether the BAL troughs that exhibit significant short-timescale variability are distinct from BAL troughs that do not exhibit such variability in terms of velocity width, mean depth, EW, and centroid velocity ($\Delta v$, $\langle d \rangle$, EW, and $v_{\rm cent}$). The parameter distributions of the rapidly varying and non-varying BALs were compared using a two-sample Kolmogorov-Smirnov (K-S) test, returning the K-S statistic ($D$) and the corresponding probability $P$, which represents the probability that the two samples are drawn from the same parent distribution. We measure $D$~=~0.17 and $P$~=~0.94 for $\Delta v$, $D$~=~0.17 and $P$~=~0.94 for $\langle d \rangle$, $D$~=~0.21 and $P$~=~0.80 for EW, and $D$~=~0.22 and $P$~=~0.72 for $v_{\rm cent}$. Thus, we conclude that varying and non-varying BALs do not fundamentally differ in terms of the aforementioned parameters.

In this sample, we find significant variability on timescales down to 0.57 days, nearly as short as our data probe; the shortest rest-frame timescale probed for each object ranges between 0.1 and 0.3 days, depending on their redshifts. This is the first detection of such short-term variability, to our knowledge, likely because our study intensively explores timescales much shorter than previous works (see Table~\ref{Table:tbl1}). BAL variability has thus far been observed on all timescales that have been examined. 

An assortment of variability amplitudes appear in our rapidly varying pairs of spectra; however, in most cases, the fractional change in EW is fairly small. This is expected, as previous studies have established that longer timescales correspond to larger fractional changes in EW (e.g., \citealt{Filizak13}). We do not observe BAL variations to be preferentially strengthening or weakening on these short timescales --- specifically, 26 out of 54 epoch pairs show a decrease in EW, while the remainder exhibit an increase. 
In some cases, however, rather dramatic variability occurs. For example, between MJDs 57901~--~57918, BAL[A] in RM\,508 underwent a very significant ($\Delta$EW $>$ 20$\sigma$) EW change. The EW of the BAL dropped from 6.12 \AA \ to 3.49 \AA \ (a factor of 1.8) within a time span of about 4~days in the quasar rest frame (see Figure~\ref{fig:var}). 

Another example of dramatic variability occurs in RM\,357, which has a high-velocity BAL (denoted BAL[A] in our tables and figures) that is present at the beginning of the campaign. This BAL disappears and reappears at various points during the campaign. 
The EW in the absorption region containing BAL[A] steadily weakens during the first few months of the campaign (see Figure~\ref{fig:ew}), and decreases to within 3$\sigma$ of zero by MJD 56755 (at this point it is no longer formally considered a BAL). The absorption feature then reappeared as a formal BAL between MJD 56755 and 56768 ($\Delta t_{\rm rest} = 4.1$~days), and formally disappeared again between MJD 56783 and 56795 ($\Delta t_{\rm rest}$ = 3.79 days). Both of these epoch pairs are indicated in Figure~\ref{fig:ew} by red coloring. For each epoch where the BAL is not present (MJDs 56755 and 56795), we searched for residual low-level absorption. 
We consider both of these epochs to be ``pristine'' cases of BAL disappearance (see, e.g., \citealt{Filizak12} or \citealt{Decicco18} for further examples of ``pristine'' BAL disappearance) because the flux at all pixels within the trough region deviates by less than 3$\sigma$ from the continuum. 
 

\subsection{Variability Characteristics of the Entire Quasar Sample} 
\label{sec:samplechar}
\begin{figure}
\begin{center}
	\includegraphics[scale = 0.25, trim = 0 0 0 0, clip]{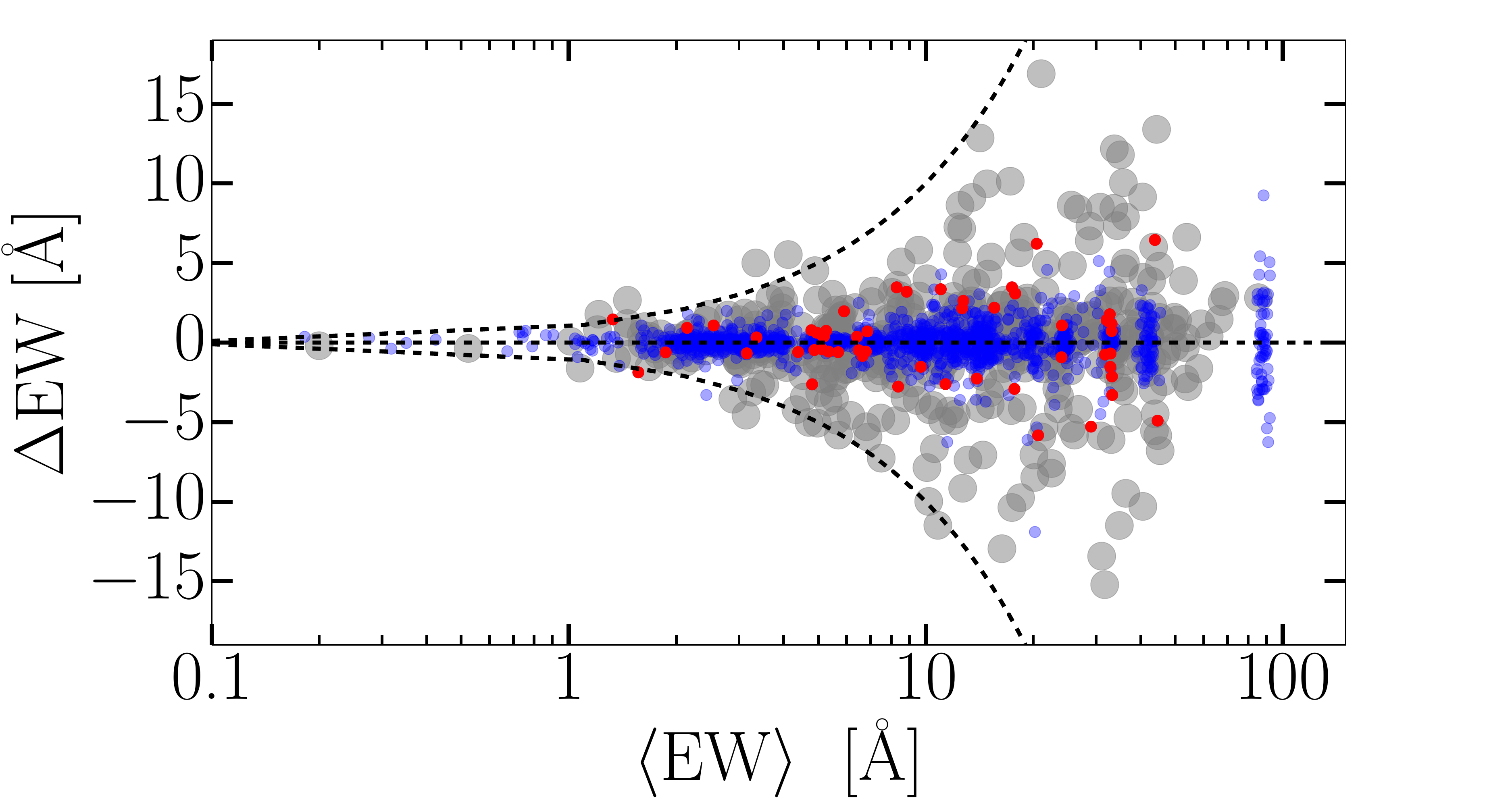}
    \caption{Sequential epoch $\Delta$EW versus average $\rm{EW}$ for all pairs of spectra in our sample (blue dots) and our rapidly-varying pairs (red dots). Gray dots represent data from \cite{Filizak13}, \cite{Barlow93}, \cite{Lundgren07}, and \cite{Gibson08}. The horizontal dashed line shows $\Delta \rm{EW} = 0$, and the curved dashed lines show $\langle \rm{EW} \rangle = \Delta \rm{EW}$. Our sample shows the same general distribution as previous work.}
	\label{fig:fa15}
\end{center}
\end{figure}

\begin{figure*}
\begin{center}
	\includegraphics[scale = 0.48, trim = 10 40 65 0, clip]{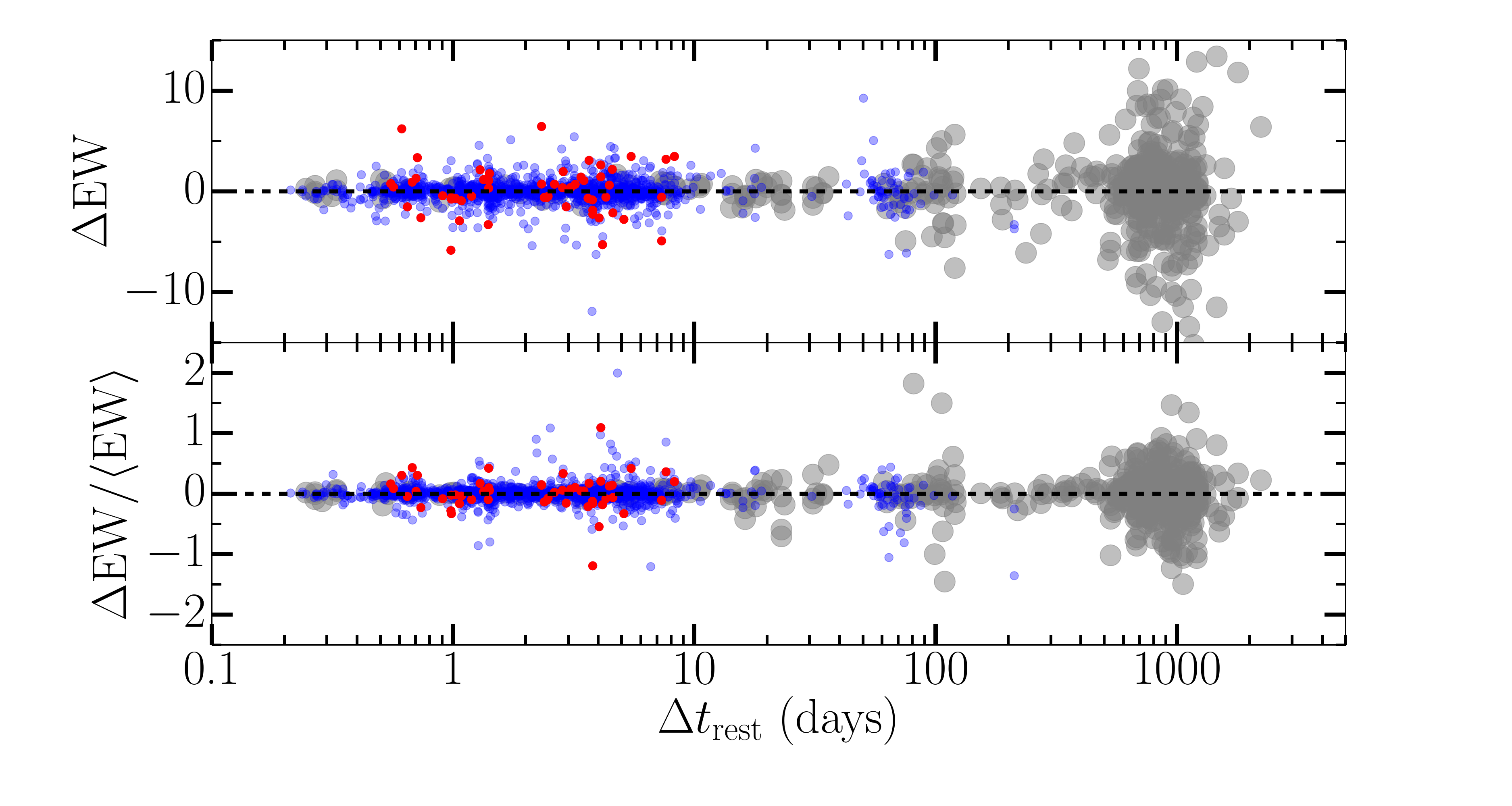}
    \caption{$\Delta \rm{EW}$ and $\Delta \rm{EW} / \langle \rm{EW} \rangle$ versus rest-frame time difference for every sequential epoch pair of every object (blue dots) and for our pairs of epochs that are identified as rapidly varying (red dots). Gray circles indicate data from \cite{Filizak13}, \cite{Barlow93}, \cite{Lundgren07}, and \cite{Gibson08}. Our sample shows the same general distribution as previous work.}
    \label{fig:g14}
\end{center}
\end{figure*}

We now compare the distributions of some variability parameters within our sample to those from previous studies. 
Figure~\ref{fig:ewdist} shows the distribution of $\Delta$EW from every sequential epoch pair in our sample compared with previous work, Figure \ref{fig:fa15} displays $\Delta$EW as a function of $\langle$EW$\rangle$ and Figure \ref{fig:g14} shows the distribution of $\Delta \rm{EW}$ and $\Delta \rm{EW}/\langle \rm{EW} \rangle$ as functions of $\Delta$t$_{\rm rest}$. Two BALs (RM\,217 BAL[A] and RM\,357 BAL[A]) disappeared at least once over the course of the campaign; in cases where these BALs were absent in both epochs, fluctuations in EW are dominated by noise. Thus, for our sample investigations (Figures~\ref{fig:fa15} and \ref{fig:g14}), we exclude epoch pairs where both of the EW values are consistent with zero to within 3$\sigma$.  

Together, Figures \ref{fig:ewdist}, \ref{fig:fa15}, and \ref{fig:g14} reveal that the $\Delta$EW distributions of our sample contain no significant outliers in comparison to the distributions of other samples. While Figure~\ref{fig:g14} appears to show excessive variability at short timescales, Figure~\ref{fig:ewdist} demonstrates that such variability represents the rare tails of the $\Delta$EW distribution, and that the majority of our measurements are, as expected, small-amplitude variations.

\subsection{Causes of Variability}
\label{sec:causes}
There are many different models of the processes that drive BAL variability. Previous studies have found that in some cases, the variability is consistent with models involving changes in the amount of ionizing radiation received by the gas (e.g., \citealt{Misawa10}; \citealt{Filizak12, Filizak13}; \citealt{Capellupo12}; \citealt{Grier15}). Other studies of BAL variability suggest that the variability is due to the transverse movement of outflow material across the line of sight of the observer (often referred to as ``cloud-crossing'', e.g., \citealt{Lundgren07}; \citealt{Gibson08}; \citealt{Hall11}; \citealt{Vivek12}; \citealt{Capellupo13}). Changes in ionizing flux can be difficult or impossible to observe, as ion species like \civ \ are ionized by radiation in the extreme ultraviolet, which has been reported to be more variable than the optical or near ultraviolet (e.g., \citealt{Marshall97}). Thus the variability in the ionizing continuum may not be reflected by observations of optical or near-ultraviolet continuum variability (see Section~4.2.1 of \citealt{Grier15} for a more detailed discussion). 

It is thus advantageous to use other evidence, such as certain characteristics of the variability, or other properties of the quasar spectra, to identify the most likely cause of variability. 
For example, the presence of coordinated variability across the full profile of a BAL trough and between distant \civ \ troughs provides evidence in support of variability driven by changes in the ionization state of the outflow gas. The presence of BAL troughs of other species can provide information on whether the \civ \ BAL is highly saturated (e.g., \citealt{Capellupo17}); a single-$\tau$ saturated BAL trough will not change its EW in response to changes in ionizing continuum in the same manner as unsaturated troughs. Thus, observed variability in saturated BALs would favor cloud-crossing models. We below examine several characteristics of the observed short-timescale BAL variability and the corresponding quasar spectra to determine the physical models responsible. 

\subsubsection{Coordinated Variability Across the Width of Individual BAL Troughs} 
\label{sec:coord}
We first search for coordinated variability across individual BAL troughs. At each (binned) pixel in a normalized trough, we calculate $\chi$, the error-normalized difference in flux between the two spectra. We assume that all pixels with measured $|\chi|<1$ exhibit only noise-dominated fluctuations consistent with zero variability. For each rapidly-varying BAL, we measure the fraction of pixels in the BAL trough with $\chi>1$, denoted $f(\chi>1)$, and the fraction with $f(\chi<-1)$. 
The various possible combinations of these parameters can be interpreted as follows: 
\begin{enumerate}
\item High values of either $f(\chi>1)$ or $f(\chi<-1)$ correspond to variability consistent with being coordinated in the same direction across a trough. Such variability is consistent with a model in which ionization-state changes drive the observed BAL variations. These cases cannot be explained solely by crossing clouds because such clouds are unlikely to cause coordinated variability over thousands of km~s$^{-1}$ on timescales of days or less. 
There may be some cases where part(s) of the trough vary in the same direction, but unequally (e.g., RM128 BAL[A] MJDs 56768--56772; see Figure Set~\ref{fig:var}). Such variability may be explained by ionization-state changes with a range of densities across the outflow (see \citealt{Arav12}, Appendix A), or with ionization-state variability {\it and} crossing clouds. 

\item If $\chi$ is nonzero in parts of the trough and effectively zero in other parts, we are seeing variable and stable regions of the trough, respectively (e.g., RM\,357 BAL[B] MJDs 56755--56768, RM\,717 BAL[A] MJDs 56717--56720, or RM\,786 BAL[B] MJDs 56745--56747). Such variability may be consistent with crossing clouds if the velocity widths of the variable regions are sufficiently small, but this behavior can also be explained by ionization variability combined with saturation or different densities within the outflow.

\item If $f(\chi>1)$ and $f(\chi<-1)$ are both significantly nonzero, we are seeing variability in both directions simultaneously (i.e., the BAL is weakening in some regions and strengthening in some regions). Such variability may be consistent with crossing clouds or ionization variability if the density is nonuniform within the outflow. 

\end{enumerate} 

\begin{figure}
\begin{center}
\includegraphics[scale = 0.35, trim = 0 0 0 0, clip]{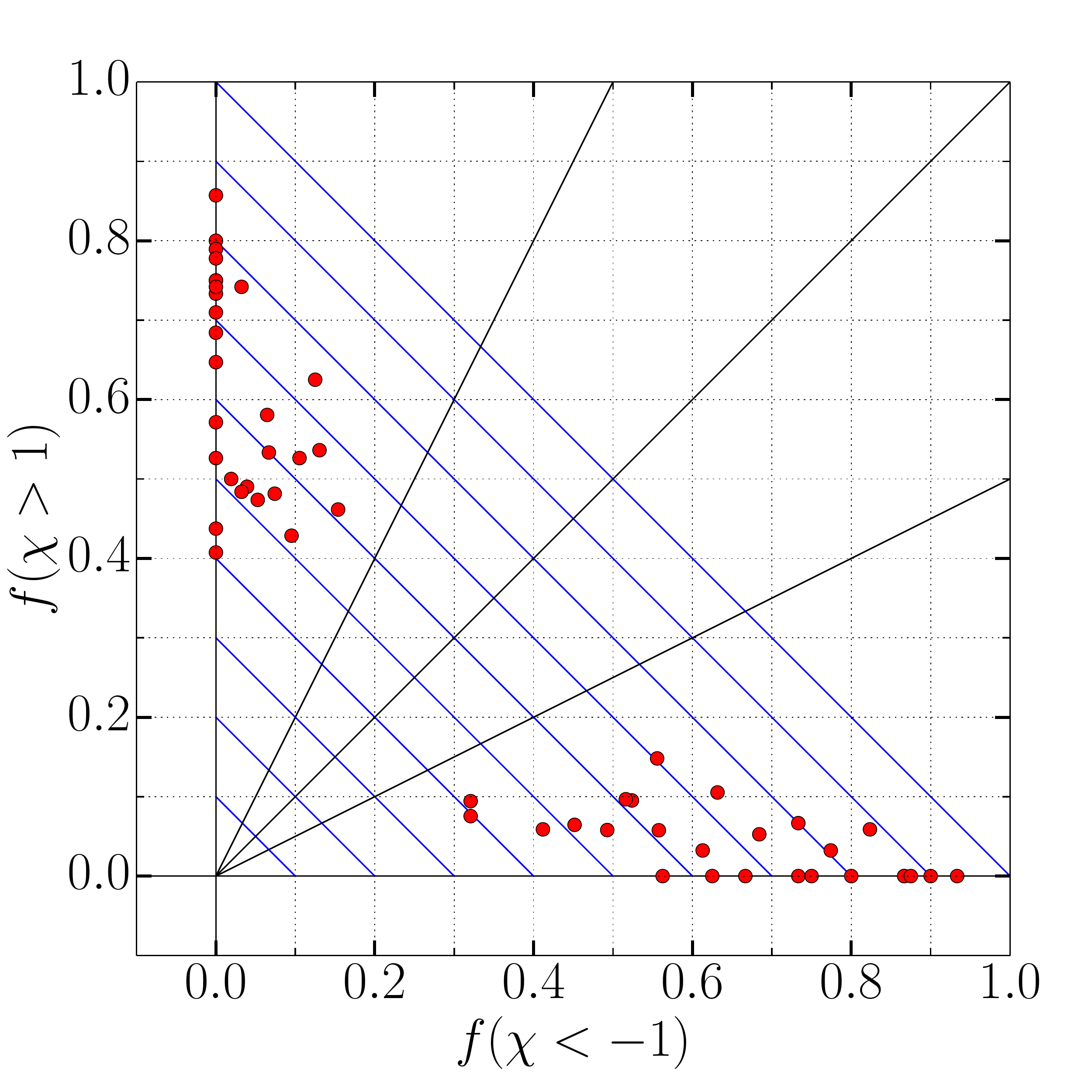}    
\caption{The fraction of pixels with $\chi$ $>$ 1 ($f(\chi>1)$) as a function of the fraction of pixels with $\chi < -1$ ($f(\chi<-1)$) for the epoch pairs showing rapid BAL variability (red circles). Black solid lines indicate ratios of 2:1, 1:1, and 1:2 to guide the eye, and the blue diagonals represent lines of constant $f(-1 < \chi < 1)$. } 
\label{fig:fchi}
\end{center}
\end{figure}

Figure~\ref{fig:fchi} shows $f(\chi>1)$ vs. $f(\chi<-1)$ for our rapidly-varying pairs of spectra, and we provide $f(\chi>1)$ and $f(\chi<-1)$ values in Table~\ref{Table:shortvartbl}. BALs experiencing coordinated variability across the trough (the first situation described above) cluster along either the horizontal or vertical axis of Figure~\ref{fig:fchi}. Cases showing coordinated variability in the same direction across the majority of the trough (e.g., RM\,508) appear farthest from the origin, and objects for which only part of the trough varies lie closer to the origin along the axes. BALs with significant variability in both directions (the third scenario described above) will lie close to a 45-degree line.  
All of our cases lie on or between the vertical or horizontal axes and the lines indicating a 2:1 ratio between the two measured quantities. Most of the sets of measurements are consistent with situations (1) and (2) above, which suggests that ionization variability likely plays a role. 

We see no cases representing the scenario (3) described above, where there is significant amounts of variability in both directions (i.e., part of the trough strengthens while another part weakens). However, we note that our aforementioned variability significance criteria select against such cases. Consider, for example, a scenario in which half of a BAL strengthens by some large amount, and the other half weakens by a similar amount. The $\Delta$EW of such a case would approach zero. Thus, this case would not be flagged as significant by our $\sigma_{\Delta \rm{EW}} > 4$ criterion, although significant variability could exist in the individual regions within the BAL. 
Such preferential selection could help explain the lack of variability in both directions in our sample (see Figure~\ref{fig:fchi}). 
However, we searched our data for potential cases of variability in opposite directions and find no clear examples of this behavior in our sample. 

We also measured the largest number of contiguous pixels in each trough with $\chi>1$ and $\chi<-1$ and denote this quantity as $\Delta v_{\rm co}$ --- this value is the largest velocity width with ``coordinated" variability-state changes. These numbers are provided in Table~\ref{Table:shortvartbl} for each pair of epochs. There is a large variety of $\Delta v_{\rm co}$ within our sample, ranging from just under 300 km~s$^{-1}$ to just under 4000~km~s$^{-1}$, indicating a wide variety of coordinated velocity widths. More than half of epoch pairs show coordinated variability in regions $>$1000~km~s$^{-1}$. 
Large regions of coordinated variability can be explained by ionization variability, but models that rely on cloud-crossing scenarios must be able to explain coordinated variability across such velocity widths, e.g., through time-dependent studies of accretion-disk winds (\citealt{Proga12}). 

\subsubsection{Coordinated Variability Between Additional \civ \ BALs}  
\label{sec:other_civ}
\begin{figure}
\begin{center}
\includegraphics[scale = 0.35, trim = 0 40 60 70, clip]{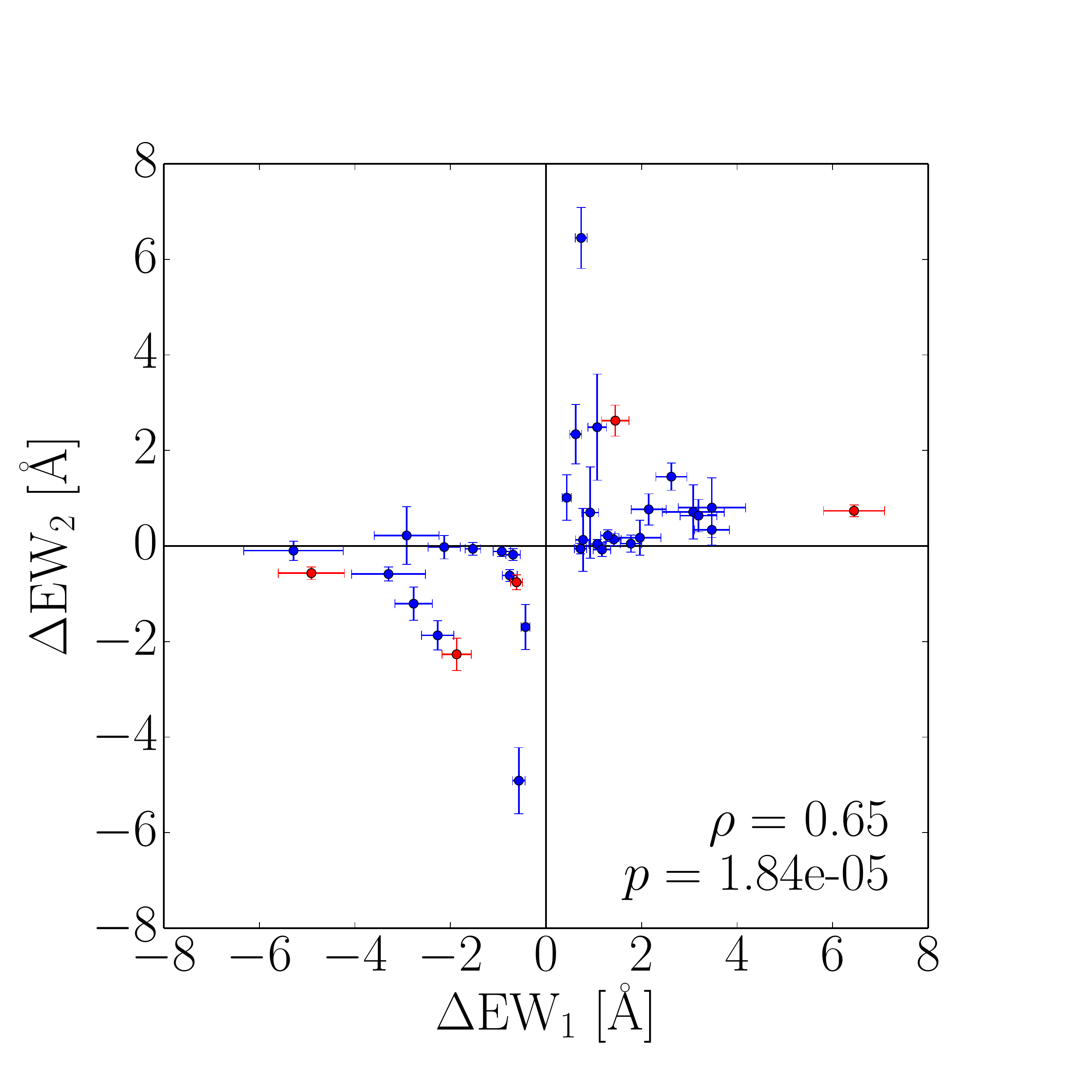}    
\includegraphics[scale = 0.35, trim = 0 10 60 70, clip]{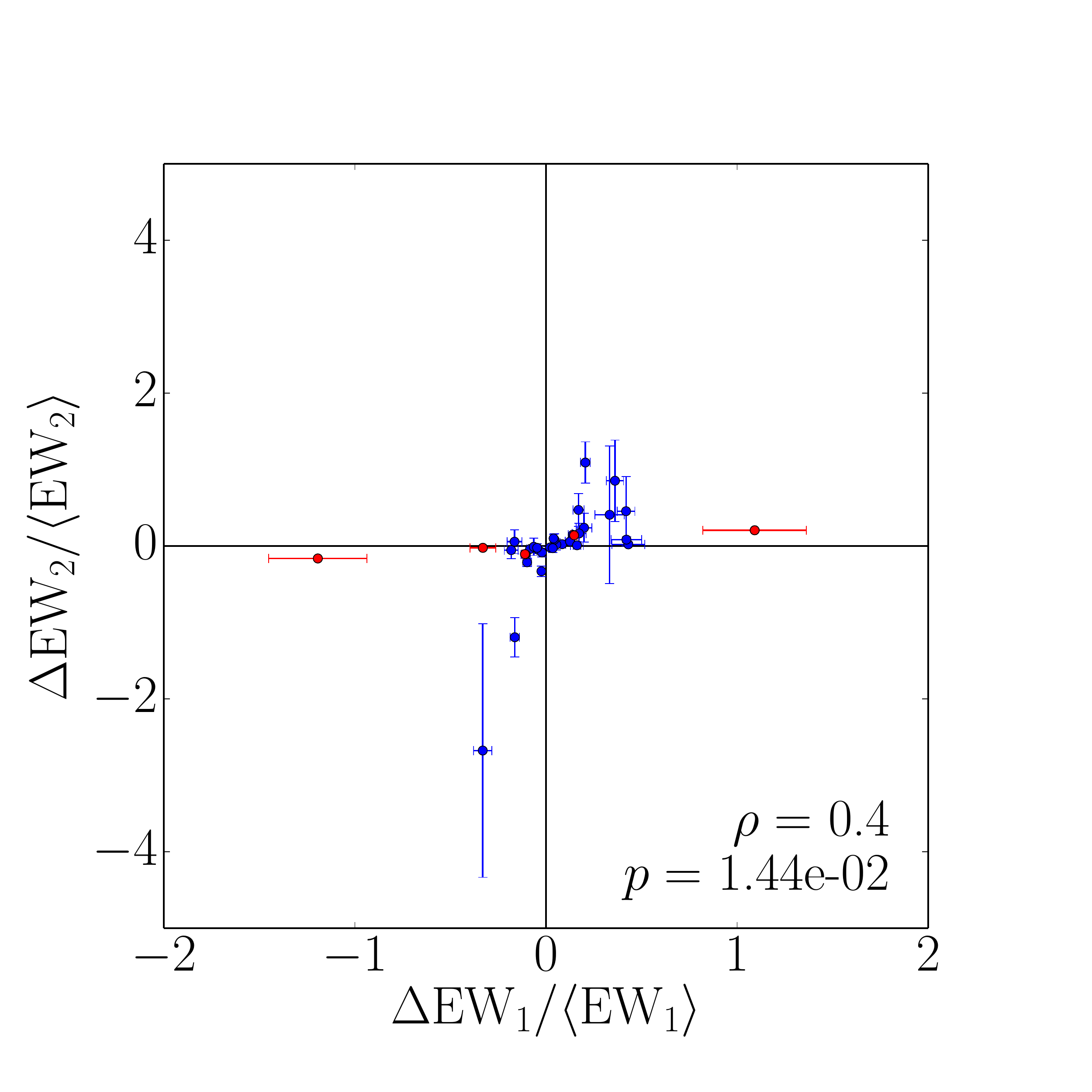}    
\caption{$\Delta$EW (top panel) and $\Delta$EW/$\langle$EW$\rangle$ (bottom panel) of accompanying \civ \ BALs as a function of the $\Delta$EW of the rapidly-varying BALs. In all cases, $\Delta$EW$_{1}$ refers to the measurement of the BAL that is identified as rapidly-variable, and $\Delta$EW$_{2}$ refers to the accompanying BAL during the same pair of epochs. Blue points represent pairs where only one of the BALs ($\Delta$EW$_{1}$) passed our criteria for rapid variability; in these cases, the second BAL varied with a lower significance (or is sometimes statistically consistent with zero variability). Red points represent pairs of epochs where both BALs varied significantly. The results of a Pearson correlation test are provided in the lower-right corner of each subplot; $r$ is the measured Pearson coefficient and $p$ is the corresponding probability that the measured $r$ would be obtained randomly if the two quantities were actually uncorrelated.} 
\label{fig:civcoord}
\end{center}
\end{figure}

Seven of our 15 quasars hosting rapidly varying BALs (RM 128, 217, 257, 357, 509, 730, and 786) have additional C\,{\sc iv} BALs present in their spectra at different velocities. These quasars provide an opportunity to search for coordination between rapidly varying C\,{\sc iv} troughs and those at different velocities --- such coordination was reported by \cite{Capellupo12}, \cite{Filizak12, Filizak13}, and \cite{Decicco18} on longer time baselines. For all pairs of epochs where at least one of the BALs  experienced significant rapid variations, we also explore the behavior of the other BAL during that same time period. Figure~\ref{fig:civcoord} shows $\Delta$EW for each pair of significantly-varying \civ \ BALs. In addition, the behavior of the additional \civ \ BALs can be compared to the rapidly-varying BALs using Figure Sets~\ref{fig:ew} and~\ref{fig:var}. 

We have 36 pairs of spectra with at least one BAL identified as rapidly varying in quasars hosting a second BAL. A detailed discussion of each individual object and its additional \civ \ BAL is provided in the Appendix. 
Figure~\ref{fig:civcoord} shows a potential correlation between $\Delta$EW of \civ \ BAL pairs, providing tentative evidence for coordinated variability between the two troughs in many cases. We also performed a Spearman rank correlation test to determine the strength of the correlation between the two quantities, computing the Spearman rank-order correlation coefficient $\rho$ and the corresponding probability $p$ that describes the likelihood of obtaining the measured $\rho$ value if the two quantities are actually uncorrelated. We obtain $\rho~=~0.65$ and $p= 1.84 \times 10^{-5}$ for $\Delta$EW changes and $\rho~=~0.4$ and $p= 1.44 \times 10^{-2}$ for $\Delta$EW/$\langle$EW$\rangle$, indicating that the EW changes of the two \civ \ troughs are  correlated. We interpret this as evidence for coordinated variability between \civ \ BAL troughs in our sample. 



\subsubsection{Si\ {\sc iv} and Al\ {\sc iii}}
\label{sec:otherspecies}
The presence of BALs at velocities corresponding to the \civ \ BAL but arising from other absorption species can provide clues to the physical properties of BAL outflows (e.g., \citealt{Filizak14}; \citealt{Capellupo17}). 
To investigate this, \cite{Filizak14} divided their sample of C\,{\sc iv} BALs into categories according to the presence of other species: If neither \siiv $\lambda\lambda1393,1403$ nor Al\,{\sc iii}$\lambda\lambda1855,1863$ is present at velocities corresponding to an identified C\,{\sc iv} BAL, said BAL is denoted a C\,{\sc iv}$_{\rm 00}$ trough. If a \siiv \ BAL is present but Al\,{\sc iii} is not detected, it is denoted as a C\,{\sc iv}$_{\rm S0}$ trough\footnote{The troughs are designated C\,{\sc iv}$_{\rm s0}$ if \siiv \ is a mini-BAL rather than a formal BAL, C\,{\sc iv}$_{\rm sa}$ if \siiv \ and \aliii \ are both present as mini-BALs, and C\,{\sc iv}$_{\rm Sa}$ if the \siiv \ feature is a formal BAL but the \aliii \ feature is a mini-BAL.}, and if both \siiv \ and Al\,{\sc iii} BALs are present at corresponding velocities, it is denoted as a C\,{\sc iv}$_{\rm SA}$ trough. In a sample of 851 BAL troughs, they find that 13$\pm$1\% of their BAL troughs fall into the C\,{\sc iv}$_{\rm 00}$ category, 61$\pm$3\% into the C\,{\sc iv}$_{\rm S0}$ or C\,{\sc iv}$_{\rm s0}$ category, and 26$\pm$1\% into the C\,{\sc iv}$_{\rm SA}$, C\,{\sc iv}$_{\rm Sa}$, or C\,{\sc iv}$_{\rm sa}$ category. 

We visually inspected the mean spectra of our quasars to determine whether or not these additional species are present in our sample, and assigned each BAL a category following \cite{Filizak14}, with one notable change: we do not differentiate between BALs and mini-BALs when considering \siiv \ and \aliii. We thus place our BAL troughs into only three categories (C\,{\sc iv}$_{\rm 00}$, C\,{\sc iv}$_{\rm S0}$, and C\,{\sc iv}$_{\rm SA}$) as long as a BAL or mini-BAL of the other species is present. In some cases, there was evidence for absorption of \siiv \ or \aliii \ that was not deep enough to meet the normalized flux threshold of 0.9 used in BAL searches, and thus the absorption is considered undetected. Our classifications for all identified BALs are given in Table~\ref{Table:baltbl}, with details on individual objects given in the Appendix. There are three cases in our BAL trough sample where there is not sufficient coverage of the \siiv \ region to determine a class --- we denote these as C\,{\sc iv}$_{\rm N0}$ troughs in Table~\ref{Table:baltbl}. 

In our sample of 34 troughs with coverage of the \siiv \ region, we find 15 C\,{\sc iv}$_{\rm 00}$ troughs (44$^{+15}_{-11}$\% of our sample), 13 C\,{\sc iv}$_{\rm S0}$ troughs (38$^{+14}_{-11}$\%), and six C\,{\sc iv}$_{\rm SA}$ troughs (18$^{+11}_{-7}$\%). This is a substantially higher fraction of C\,{\sc iv}$_{\rm 00}$ BALs than is present in the \cite{Filizak14} sample. This is primarily due to the different velocity ranges used in our BAL searches --- \cite{Filizak14} search only outflow velocities ranging from 3000--20,000~km~s$^{-1}$ in their work, while we extend ours from an apparent inflow velocity of $1000$~km~s$^{-1}$ to an apparent outflow velocity of 30,000~km~s$^{-1}$. Many of our BALs, primarily C\,{\sc iv}$_{\rm 00}$ troughs, extend into the ranges excluded by \cite{Filizak14} and would be excluded if we imposed similar velocity restrictions --- imposing these restrictions results in much better consistency between the observed percentages of each category. Another likely contributor to the different percentages of each kind of trough is that we are using coadded spectra to search for BALs --- sometimes including more than 60 spectra, whereas \cite{Filizak14} used only single-epoch spectra. Our mean spectra are thus generally much higher-SN, which allows us to detect weaker BAL troughs (which are often missed due to noise). Since C\,{\sc iv}$_{\rm 00}$ troughs are generally weaker than BALs in the other two categories (\citealt{Filizak14}), we expect that the use of coadded/mean spectra also contributes to the higher incidence of C\,{\sc iv}$_{\rm 00}$ troughs in our sample. 

Among only our 19 rapidly varying BAL troughs, we observe a distribution that is consistent with our full sample: nine (50$^{+23}_{-16}$\%) of the 18 BALs with short-term variability and \siiv \ coverage fall into the C\,{\sc iv}$_{\rm 00}$ category,  six (33$^{+20}_{-13}$\%) fall into the C\,{\sc iv}$_{\rm S0}$ category, and three (17$^{+16}_{-9}$\%) fall into the C\,{\sc iv}$_{\rm SA}$ category (the remaining rapidly varying BAL, RM\,361 BAL[A], does not have wavelength coverage of the \siiv \ region).  

When a deep \siiv \ trough accompanies the observed \civ \ trough,
the \civ \ trough is likely to be heavily saturated given the relative cosmic
abundances of Si and C and the relative ionic abundances expected in BAL
outflows (e.g., Figure~4 of \citealt{Hamann00}). Half of our rapidly varying BAL troughs show accompanying \siiv, which indicates saturation and suggests that cloud-crossing may play a part in the observed variability in many of our BALs, as saturated troughs will not respond to changes in the incident ionizing flux in the same manner as unsaturated troughs. However, the presence of saturated \civ \ troughs is not necessarily solid evidence for simple cloud-crossing scenarios. 

For example, consider RM\,257 BAL [A], a C\,{\sc iv}$_{\rm SA}$ trough with two pairs of epochs identified as rapidly variable. The BAL shows coordinated variability across velocities of $\sim$3600~km~s$^{-1}$ and $\sim$7250~km~s$^{-1}$ (about 20\% and 40\% of the entire trough width). This coordinated variability is suggestive of ionization-state changes as the driving mechanism. In addition, this quasar also contains a second, lower-velocity BAL that varies significantly in the same direction (also naturally explained by ionization-state changes). However, the presence of \siiv \ and \aliii \ BALs indicates that the \civ \ BAL is likely highly saturated; as discussed above, variability in saturated troughs is more easily explained by cloud-crossing scenarios. At face value, the evidence presented by this BAL seems contradictory. 

However, coordinated variability over such a wide velocity range can be explained by a model including inhomogenous partial covering of some
type; that is, different parts of the quasar emission regions are covered
by absorbers of different optical depths along our sight-line
(e.g., \citealt{dekool02}; \citealt{Arav05}; \citealt{Sabra05}).
For example, the optical depth values might follow a power-law distribution
(e.g., \citealt{Arav08}),
with a large fraction of the emitting area
covered by gas optically thick in \civ \ and a small fraction
covered by optically thin gas.  Variations in the ionizing continuum
leading to variations in the optical depth values would yield a more
detectable change in residual flux from the emitting area covered by
optically thin gas than from the emitting area covered by optically thick gas.
A BAL outflow exhibiting inhomogeneous partial covering therefore can
exhibit strong \siiv \ and even \pv \ absorption in a trough
which nonetheless exhibits variability in the \civ \ equivalent width that is caused by ionization-state changes.
Simultaneously modeling the variability of \civ \ and \siiv \
in such cases may constrain inhomogeneous partial convering models; however, we defer such detailed analysis to a future work.

\subsubsection{\pv} 
\label{sec:pv}
We also searched for  P\,{\sc v}$\lambda\lambda$1112,1118 absorption in our significant-variability sample. P\,{\sc v} is a lower-abundance ion that has an ionization energy similar to that of C\,{\sc iv}, and its presence, like that of \siiv, indicates that the corresponding component of C\,{\sc iv} absorption is highly saturated (e.g., \citealt{Capellupo17}). There have been differing reports on the frequency of \pv \ BALs; \cite{Capellupo17} report that only 3--6\% of BAL quasars examined show detectable \pv, while \cite{Filizak14} estimate that about half of their sample contains visible signs of \pv \ absorption. The incidence of detectable \pv \ rises when other species such as \siiv \ or \aliii \ are present --- \cite{Filizak14} report that 88\% of their C\,{\sc iv}$_{\rm SA}$ troughs show detectable \pv, about half of the C\,{\sc iv}$_{\rm S0}$ troughs show detectable \pv, and only 12\% of C\,{\sc iv}$_{\rm 00}$ troughs are accompanied by \pv \ absorption. The discrepancy in the \pv \ detection percentages between \cite{Capellupo17} and \cite{Filizak14} is likely due to differing identification criteria: \cite{Capellupo13} note that they are extremely conservative in identifying \pv \ and that their algorithm is biased against weaker and narrower features, whereas \cite{Filizak14} use visual inspection to identify \pv, and may thus identify many features that are excluded by the strict criteria used in the former work.

In the cases in which the P\,{\sc v} region is covered by our spectra, it falls toward the very blue end of the spectrum, which often contains significant noise, causing difficulties in its reliable detection, and this difficulty is compounded by the fact that it falls within the Ly$\alpha$ forest. A detailed analysis is thus beyond the scope of this work --- however, we visually inspected the spectra for signs of broad P\,{\sc v} absorption that corresponds to the velocities of our 37 identified C{\sc iv} BALs. The results for each identified BAL in our sample are presented in Table~\ref{Table:baltbl}. Only 15 of our 37 BALs have coverage of the \pv \ region at corresponding velocities; for these 15 \civ \ BALs, we find tentative signs of accompanying \pv \ in six cases (40$^{+23}_{-15}$\%). 

Three of the six likely \pv \ BALs accompany rapidly-varying \civ \ BALs --- in quasars RM\,257, RM\,631, and RM\,730. All three also contain \siiv \ absorption (RM\,631 and RM\,730 are both \civ$_{\rm S0}$ troughs and RM 257 BAL[A] is a \civ$_{\rm SA}$ trough), further supporting saturation. One would thus expect that the observed variability in these BALs is most likely due to cloud-crossing or, as discussed above (Section~\ref{sec:otherspecies}), by inhomogeneous partial covering scenarios. 

\subsection{Physical Constraints from Ionization Variability}  
\label{sec:physicalconstraints}
The evidence presented above suggests that the variability that we observe is due to a variety of causes; in some cases (e.g., RM\,508), we find substantial evidence in support of the ionization-change scenario, while in others (e.g., RM\,730; see Section~\ref{sec:pv}), the evidence may point to to cloud-crossing as the more likely scenario. We here carry out one example calculation of some of the physical constraints that can be obtained in the case of ionization changes for one such BAL. To do so, we adopt a model for ionization variability in optically thin \civ \ gas (details are provided in the Appendix). This model is only plausible for BAL troughs with a \civ$_{\rm 00}$ designation in Table~\ref{Table:baltbl}, indicating that neither \siiv \ nor \aliii \ accompany the \civ \ absorption; we do not expect ionization variability in optically thin \civ \ gas to explain variability in troughs with \civ$_{\rm SA}$ or \civ$_{\rm S0}$ designations, though ionization variability may play a role if the optical depth is inhomogenous (see Section~\ref{sec:otherspecies}).

We choose our most significant case, RM\,508, as our case study. BAL[A] in this quasar shows variability at a significance of $>20\sigma$ over a period of just four days in the quasar rest frame (see Figure~\ref{fig:var}).   
Following the procedure outlined in the Appendix, we find that the observed variability in RM\,508 is consistent with a model of an absorber that fully covers the continuum source, is optically thin in \civ \ (with a maximum optical depth $\tau_{\rm max}=0.847\pm 0.024$), has constant electron density over its velocity range, and responds to a variable ionizing flux by a drop in optical depth of a factor of $\bar{B}=2.04 \pm 0.07$ in 4.03 rest-frame days. Of course, consistency with a model is not proof the model is correct, 
but it is encouraging that simple absorber models may be useful in 
constraining the physical parameters of absorbers and of quasar ionizing 
flux variability (though we leave the latter as a topic for future work).


Our observations can further be employed to constrain the density and radial distance of quasar outflows (e.g., \citealt{Grier15}). For ionization state $i$ of some element, outflow density can be constrained given the equivalent widths of two epochs ($\rm{EW_1}$ and $\rm{EW_2}$), the rest-frame time difference between these epochs ($\Delta t_{\rm rest}$), the ion's recombination coefficient to ionization state $i-1$ ($\alpha_{i-1}$), and several simplifying assumptions \cite[][]{Arav12, Rogerson15}. The relationship between these variables and the density constraint is

\begin{equation}
n_e > \log{(\rm{EW_1}/\rm{EW_2})} / \alpha_{i-1} \Delta t_{\rm rest}
\label{eq:ne}
\end{equation}

A derivation of Equation~\ref{eq:ne} is given in Section~\ref{sec:density} of the Appendix. We assume an outflow temperature approximately equal to that of \civ-region broad line region gas, $\log{T}=4.3$ (e.g., \citealt{Krolik99}). For this temperature, the \ciii \ recombination coefficient quoted by the CHIANTI online database is $\alpha_{\rm CIII} = 2.45 \times 10^{-11} \rm{\ cm^{3} \ s^{-1}}$ \cite[][]{Dere97, Landi13}. We utilize this recombination coefficient for subsequent calculations.

With Equation~\ref{eq:ne}, we treat two cases of short-timescale variability to constrain density: the most significantly varying epoch pair (RM\,508 BAL[A], MJDs 57901--57918, $\sigma_{\Delta \rm{EW}} = -22.34$) and the most rapidly varying epoch pair of a C{\sc iv}$_{\rm 00}$ BAL exhibiting decreasing EW (RM\,257 BAL[B], MJDs 56717--56720, $\Delta t_{\rm rest} = 0.91 \rm{\ days}$). These epoch pairs return density constraints of $n_e \gtrsim 6.5 \times 10^{4} \rm{\ cm^{-3}}$ and $n_e \gtrsim 4.1 \times 10^{4} \rm{\ cm^{-3}}$, respectively. Using data on RM\,613 from \cite{Grier15} in Equation~\ref{eq:ne}, we calculate a density constraint of $n_e \gtrsim 3.4 \times 10^{4} \rm{\ cm^{-3}}$. Thus, the conservative approach we adopt in Equation~\ref{eq:ne} yields a lower-limit density from the large-amplitude variability in RM 508 BAL[A] which is about a factor of two larger than those from RM 257 BAL[B] in this 
work or from RM 613 reported by \cite{Grier15}. Previous studies (e.g., \citealt{Arav13}; \citealt{Borguet13}) have reported density constraints on the order of $n_e > 10^{3}-10^{4} \rm{\ cm^{-3}}$ for outflows at radii of kiloparsec scales --- our measured densities are consistent with such measurements. 

The radial distance from the quasar to the outflow for which the previously calculated density constraint remains valid can also be constrained via

\begin{equation}
r > \sqrt{\int_{\nu_i}^{\infty}{\frac{(L_\nu/h\nu)\sigma_\nu}{4\pi n_e\alpha_{i-1}}} d\nu}
\label{eq:dist}
\end{equation}

\noindent where $\sigma_\nu$ is the cross section of photons with energy $h\nu$ and $\nu_i$ is the ionization energy of ionization species $(i-1)$. A derivation of Equation~\ref{eq:dist} is given in Section~\ref{sec:distance} of the Appendix. Note that both the density and radius constraints are conditional; either the gas is more distant and has a greater $n_e$ than the respective lower limits above, or it is located closer than the radius limit with no constraint on its density (see Figure 14 of \citealt{Rogerson15}).

We determined $L_{\rm Bol}$ of RM\, 257 and RM\,508 using their respective observed flux densities at 3000\,\AA\ 1350\,\AA\ and adopting a bolometric correction factor of 5 and 4, respectively (bolometric corrections factors are from \citealt{Richards06}). We calculate $L_{\rm Bol}~=$~7.64~$\times$~10$^{46}$~erg~$s^{-1}$ for RM\,257 and $L_{\rm Bol}~=$~3.32~$\times$~10$^{47}$~erg~$s^{-1}$ for RM\,508.  Adopting the UV-soft spectral energy distribution of \cite{Dunn10}, we calculate $L_{\nu}$ and integrate Equation~\ref{eq:dist}, returning radial distance constraints of $r \gtrsim 248 \rm{\ pc}$ and $r \gtrsim 405 \rm{\ pc}$ for RMs\, 257 and 508, respectively. Additionally, we rescale the RM\,613 radial distance constraint $r \gtrsim 120 \rm{\ pc}$ reported by \cite{Grier15} to the improved density constraint calculated via Equation~\ref{eq:ne}, yielding a constraint of $r \gtrsim 406 \rm{\ pc}$. Our conditional distance constraints for all three quasars for which we have done the calculations (RM\,257, RM\,508, and RM\,613) are thus all on the order of hundreds of parsecs. 

Such large radii are inconsistent with the $\sim 10^{-3} \rm{\ pc}$ launching radius distance suggested by previous theoretical studies (e.g., \citealt{Murray95}; \citealt{Proga00}; \citealt{Higgenbottom13}). However, various studies have reported BAL outflows at greater distances (e.g., see Table 10 of \citealt{Dunn10}, with more recent works including, e.g., \citealt{Xu18}), and \citealt{Arav18} report evidence that more than half of all BALs are found at distances of greater than 100~pc. Our conditional distance constraints are consistent with this claim, though because they are conditional, they should not be viewed as direct support for it.

 \section{SUMMARY \& FUTURE WORK} 
 \label{sec:summary}
 We have systematically searched for short-timescale ($<10$-day) variability in a sample of 37 distinct \civ \ BALs in 27 unique quasars using data from the SDSS-RM project, which contains a large number of spectral epochs at a high cadence compared to previous studies of BAL variability. We further examine these spectra to evaluate models for variability mechanisms and compare our observed variability to previous sample-based studies (e.g., \citealt{Filizak13}). Our main findings are the following: 
 
	\begin{enumerate}
		\itemsep0em
    	\item We discover 54 cases of significant rapid variability in 19 unique BALs (see Section~\ref{sec:var_overview}) in 15 out of the 27 quasars in our sample ($55^{+18}_{-14}$\%), 19 of 37 \civ \ BAL troughs ($51^{+15}_{-12}$\%), and 54 of 1460 epoch pairs ($3.7 \pm 0.5$\%). This variability is on rest-frame timescales of 0.55--8.27 days, with a median of 2.55 days. This result demonstrates that short-term BAL variability is common; it has not been observed frequently in the past likely because high-SN observations intensively probing these short timescales did not exist. These quasars and their BAL troughs do not appear to be distinctive; their properties are similar to those of the general quasar population. See Section~\ref{sec:var_overview}. 
        \item We compare the overall variability properties of our sample ($\Delta$EW and fractional $\Delta$EW) to previous studies to determine whether our rapid variability is unusual or more extreme than previously reported. The amplitude of variability in our sample does not appear to differ qualitatively from that previously observed on comparable or slightly longer timescales (see Section~\ref{sec:samplechar}). 
        \item We searched for coordinated variability across our rapidly varying \civ \ BAL troughs and find that the velocities over which the variability is coordinated range from  $\approx$290 km~s$^{-1}$ to $\approx$7250 km~s$^{-1}$ within the troughs. More than half of our troughs show coordinated variations across velocities $>$1000~km~s$^{-1}$ (see Section~\ref{sec:coord}). This result is consistent with much of the observed variability being driven by changes in the ionization state of the outflow gas (e.g., \citealt{Filizak13}; \citealt{Grier15}), which naturally explains coordinated variability across a range of velocities. However, many of our troughs show much smaller ranges of coordinated variability, suggesting that cloud-crossing may be responsible for the variability in other cases. 
        \item We see evidence for coordinated variability between \civ \ troughs in quasars hosting more than one \civ \ BAL (see Section~\ref{sec:other_civ}). This observation again indicates that the variability cannot be explained solely by a cloud-crossing scenario, as coordinated variability at such a large spread of velocities is difficult to achieve in the context of cloud-crossing models.  
        \item We investigate the presence of other BAL species such as \siiv, \aliii, and \pv \ in our sample and find that many of our \civ \ BALs are saturated; in these cases, cloud-crossing scenarios and/or inhomogeneous partial covering are required to explain our observed rapid variability. See Section~\ref{sec:otherspecies}.  
        \item In many cases, we see variability consistent with models of ionization-state changes within the outflow; we thus adopt a simple model and employ the model to constrain various properties of the outflow in an example object (see Section~\ref{sec:physicalconstraints}). We find outflow density lower limits in agreement with those reported in previous work. 
    \end{enumerate}

Previous sample-based studies of BAL variability have found significant variability on all timescales explored; we similarly find significant BAL variability down to timescales of less than a day in the quasar rest frame. Our results show that future observational studies of BAL variability on longer timescales need to consider the likely possibility that the BALs have experienced significant variations, perhaps in many directions, within the time span between observations. In addition, models describing BAL formation and variability need to explain short-timescale variability to be viable, as such rapid variability is not a rare, extreme phenomenon, but appears to be common among quasars. 

Our study is the first to examine short BAL variability timescales (down to less than a day in the quasar rest frame) intensively for many different quasars. Future time-domain spectroscopic studies of quasars will be extremely useful expanding the sample of quasars searched and placing better constraints on the frequency of such variability. There are multiple upcoming industrial-scale spectroscopic programs that will provide data optimal for such efforts. The SDSS-V Black Hole Mapper program (e.g., \citealt{Kollmeier17}) will observe several quasar fields at a high cadence similar to the SDSS-RM, allowing us to explore hundreds of additional BAL quasars and largely expand our sample. Additionally, telescopes such as the 4-Metre Multi-Object Spectroscopic Telescope (4MOST; \citealt{dejong14}) are well-suited for high-cadence spectroscopic studies of quasars, which naturally produce data suitable for short-timescale BAL variability studies. These upcoming programs will be crucial in uncovering the drivers behind this variability and the implications for models of quasar outflows. 

\acknowledgments 
ZSH, CJG, WNB, DPS, and JRT acknowledge support from NSF grant AST-1517113. WNB and MV acknowledge support from NSF grant AST-1516784. PBH acknowledges support from the Natural Sciences and Engineering Research Council of Canada (NSERC), funding reference number 2017-05983. YS acknowledges support from an Alfred P. Sloan Research Fellowship and NSF grant AST-1715579. 

Funding for the Sloan Digital Sky Survey IV has been provided by
the Alfred P. Sloan Foundation, the U.S. Department of Energy Office of
Science, and the Participating Institutions. SDSS-IV acknowledges
support and resources from the Center for High-Performance Computing at
the University of Utah. The SDSS web site is www.sdss.org.

SDSS-IV is managed by the Astrophysical Research Consortium for the 
Participating Institutions of the SDSS Collaboration including the 
Brazilian Participation Group, the Carnegie Institution for Science, 
Carnegie Mellon University, the Chilean Participation Group, the French Participation Group, Harvard-Smithsonian Center for Astrophysics, 
Instituto de Astrof\'isica de Canarias, The Johns Hopkins University, 
Kavli Institute for the Physics and Mathematics of the Universe (IPMU) / 
University of Tokyo, Lawrence Berkeley National Laboratory, 
Leibniz Institut f\"ur Astrophysik Potsdam (AIP),  
Max-Planck-Institut f\"ur Astronomie (MPIA Heidelberg), 
Max-Planck-Institut f\"ur Astrophysik (MPA Garching), 
Max-Planck-Institut f\"ur Extraterrestrische Physik (MPE), 
National Astronomical Observatory of China, New Mexico State University, 
New York University, University of Notre Dame, 
Observat\'ario Nacional / MCTI, The Ohio State University, 
Pennsylvania State University, Shanghai Astronomical Observatory, 
United Kingdom Participation Group,
Universidad Nacional Aut\'onoma de M\'exico, University of Arizona, 
University of Colorado Boulder, University of Oxford, University of Portsmouth, 
University of Utah, University of Virginia, University of Washington, University of Wisconsin, 
Vanderbilt University, and Yale University.


\begin{thebibliography}{}
\expandafter\ifx\csname natexlab\endcsname\relax\def\natexlab#1{#1}\fi
\providecommand{\url}[1]{\href{#1}{#1}}
\providecommand{\dodoi}[1]{doi:~\href{http://doi.org/#1}{\nolinkurl{#1}}}
\providecommand{\doeprint}[1]{\href{http://ascl.net/#1}{\nolinkurl{http://ascl.net/#1}}}
\providecommand{\doarXiv}[1]{\href{https://arxiv.org/abs/#1}{\nolinkurl{https://arxiv.org/abs/#1}}}

\bibitem[{{Ahn} {et~al.}(2014){Ahn}, {Alexandroff}, {Allende Prieto}, {Anders},
  {Anderson}, {Anderton}, {Andrews}, {Aubourg}, {Bailey}, {Bastien}, \&
  et~al.}]{Ahn14}
{Ahn}, C.~P., {Alexandroff}, R., {Allende Prieto}, C., {et~al.} 2014, \apjs,
  211, 17, \dodoi{10.1088/0067-0049/211/2/17}

\bibitem[{{Arav} {et~al.}(2013){Arav}, {Borguet}, {Chamberlain}, {Edmonds}, \&
  {Danforth}}]{Arav13}
{Arav}, N., {Borguet}, B., {Chamberlain}, C., {Edmonds}, D., \& {Danforth}, C.
  2013, \mnras, 436, 3286, \dodoi{10.1093/mnras/stt1812}

\bibitem[{{Arav} {et~al.}(2005){Arav}, {Kaastra}, {Kriss}, {Korista}, {Gabel},
  \& {Proga}}]{Arav05}
{Arav}, N., {Kaastra}, J., {Kriss}, G.~A., {et~al.} 2005, \apj, 620, 665,
  \dodoi{10.1086/425560}

\bibitem[{{Arav} {et~al.}(2018){Arav}, {Liu}, {Xu}, {Stidham}, {Benn}, \&
  {Chamberlain}}]{Arav18}
{Arav}, N., {Liu}, G., {Xu}, X., {et~al.} 2018, \apj, 857, 60,
  \dodoi{10.3847/1538-4357/aab494}

\bibitem[{{Arav} {et~al.}(2008){Arav}, {Moe}, {Costantini}, {Korista}, {Benn},
  \& {Ellison}}]{Arav08}
{Arav}, N., {Moe}, M., {Costantini}, E., {et~al.} 2008, \apj, 681, 954,
  \dodoi{10.1086/588651}

\bibitem[{{Arav} {et~al.}(2012){Arav}, {Edmonds}, {Borguet}, {Kriss},
  {Kaastra}, {Behar}, {Bianchi}, {Cappi}, {Costantini}, {Detmers}, {Ebrero},
  {Mehdipour}, {Paltani}, {Petrucci}, {Pinto}, {Ponti}, {Steenbrugge}, \& {de
  Vries}}]{Arav12}
{Arav}, N., {Edmonds}, D., {Borguet}, B., {et~al.} 2012, \aap, 544, A33,
  \dodoi{10.1051/0004-6361/201118501}

\bibitem[{{Barlow}(1993)}]{Barlow93}
{Barlow}, T.~A. 1993, PhD thesis, California University

\bibitem[{{Baskin} {et~al.}(2014){Baskin}, {Laor}, \& {Stern}}]{Baskin14}
{Baskin}, A., {Laor}, A., \& {Stern}, J. 2014, \mnras, 445, 3025,
  \dodoi{10.1093/mnras/stu1732}

\bibitem[{{Blanton} {et~al.}(2017){Blanton}, {Bershady}, {Abolfathi},
  {Albareti}, {Allende Prieto}, {Almeida}, {Alonso-Garc{\'\i}a}, {Anders},
  {Anderson}, {Andrews}, {Aquino-Ort{\'\i}z}, {Arag{\'o}n-Salamanca},
  {Argudo-Fern{\'a}ndez}, {Armengaud}, {Aubourg}, {Avila-Reese}, {Badenes},
  {Bailey}, {Barger}, {Barrera-Ballesteros}, {Bartosz}, {Bates}, {Baumgarten},
  {Bautista}, {Beaton}, {Beers}, {Belfiore}, {Bender}, {Berlind}, {Bernardi},
  {Beutler}, {Bird}, {Bizyaev}, {Blanc}, {Blomqvist}, {Bolton}, {Boquien},
  {Borissova}, {van den Bosch}, {Bovy}, {Brandt}, {Brinkmann}, {Brownstein},
  {Bundy}, {Burgasser}, {Burtin}, {Busca}, {Cappellari}, {Delgado Carigi},
  {Carlberg}, {Carnero Rosell}, {Carrera}, {Chanover}, {Cherinka}, {Cheung},
  {G{\'o}mez Maqueo Chew}, {Chiappini}, {Doohyun Choi}, {Chojnowski}, {Chuang},
  {Chung}, {Cirolini}, {Clerc}, {Cohen}, {Comparat}, {da Costa}, {Cousinou},
  {Covey}, {Crane}, {Croft}, {Cruz-Gonzalez}, {Garrido Cuadra}, {Cunha},
  {Damke}, {Darling}, {Davies}, {Dawson}, {de la Macorra}, {Dell'Agli}, {De
  Lee}, {Delubac}, {Di Mille}, {Diamond-Stanic}, {Cano-D{\'\i}az}, {Donor},
  {Downes}, {Drory}, {du Mas des Bourboux}, {Duckworth}, {Dwelly}, {Dyer},
  {Ebelke}, {Eigenbrot}, {Eisenstein}, {Emsellem}, {Eracleous}, {Escoffier},
  {Evans}, {Fan}, {Fern{\'a}ndez-Alvar}, {Fernandez- Trincado}, {Feuillet},
  {Finoguenov}, {Fleming}, {Font-Ribera}, {Fredrickson}, {Freischlad},
  {Frinchaboy}, {Fuentes}, {Galbany}, {Garcia-Dias},
  {Garc{\'\i}a-Hern{\'a}ndez}, {Gaulme}, {Geisler}, {Gelfand},
  {Gil-Mar{\'\i}n}, {Gillespie}, {Goddard}, {Gonzalez-Perez}, {Grabowski},
  {Green}, {Grier}, {Gunn}, {Guo}, {Guy}, {Hagen}, {Hahn}, {Hall}, {Harding},
  {Hasselquist}, {Hawley}, {Hearty}, {Gonzalez Hern{\'a}ndez}, {Ho}, {Hogg},
  {Holley-Bockelmann}, {Holtzman}, {Holzer}, {Huehnerhoff}, {Hutchinson},
  {Hwang}, {Ibarra- Medel}, {da Silva Ilha}, {Ivans}, {Ivory}, {Jackson},
  {Jensen}, {Johnson}, {Jones}, {J{\"o}nsson}, {Jullo}, {Kamble}, {Kinemuchi},
  {Kirkby}, {Kitaura}, {Klaene}, {Knapp}, {Kneib}, {Kollmeier}, {Lacerna},
  {Lane}, {Lang}, {Law}, {Lazarz}, {Lee}, {Le Goff}, {Liang}, {Li}, {Li},
  {Lian}, {Lima}, {Lin}, {Lin}, {Bertran de Lis}, {Liu}, {de Icaza Lizaola},
  {Long}, {Lucatello}, {Lundgren}, {MacDonald}, {Deconto Machado}, {MacLeod},
  {Mahadevan}, {Geimba Maia}, {Maiolino}, {Majewski}, {Malanushenko},
  {Malanushenko}, {Manchado}, {Mao}, {Maraston}, {Marques- Chaves}, {Masseron},
  {Masters}, {McBride}, {McDermid}, {McGrath}, {McGreer}, {Medina Pe{\~n}a},
  {Melendez}, {Merloni}, {Merrifield}, {Meszaros}, {Meza}, {Minchev},
  {Minniti}, {Miyaji}, {More}, {Mulchaey}, {M{\"u}ller-S{\'a}nchez}, {Muna},
  {Munoz}, {Myers}, {Nair}, {Nandra}, {Correa do Nascimento}, {Negrete},
  {Ness}, {Newman}, {Nichol}, {Nidever}, {Nitschelm}, {Ntelis}, {O'Connell},
  {Oelkers}, {Oravetz}, {Oravetz}, {Pace}, {Padilla}, {Palanque- Delabrouille},
  {Alonso Palicio}, {Pan}, {Parejko}, {Parikh}, {P{\^a}ris}, {Park}, {Patten},
  {Peirani}, {Pellejero-Ibanez}, {Penny}, {Percival}, {Perez-Fournon},
  {Petitjean}, {Pieri}, {Pinsonneault}, {Pisani}, {Poleski}, {Prada},
  {Prakash}, {Queiroz}, {Raddick}, {Raichoor}, {Barboza Rembold}, {Richstein},
  {Riffel}, {Riffel}, {Rix}, {Robin}, {Rockosi}, {Rodr{\'\i}guez-Torres},
  {Roman-Lopes}, {Rom{\'a}n-Z{\'u}{\~n}iga}, {Rosado}, {Ross}, {Rossi}, {Ruan},
  {Ruggeri}, {Rykoff}, {Salazar-Albornoz}, {Salvato}, {S{\'a}nchez}, {Aguado},
  {S{\'a}nchez- Gallego}, {Santana}, {Santiago}, {Sayres}, {Schiavon}, {da
  Silva Schimoia}, {Schlafly}, {Schlegel}, {Schneider}, {Schultheis},
  {Schuster}, {Schwope}, {Seo}, {Shao}, {Shen}, {Shetrone}, {Shull}, {Simon},
  {Skinner}, {Skrutskie}, {Slosar}, {Smith}, {Sobeck}, {Sobreira}, {Somers},
  {Souto}, {Stark}, {Stassun}, {Stauffer}, {Steinmetz}, {Storchi-Bergmann},
  {Streblyanska}, {Stringfellow}, {Su{\'a}rez}, {Sun}, {Suzuki}, {Szigeti},
  {Taghizadeh-Popp}, {Tang}, {Tao}, {Tayar}, {Tembe}, {Teske}, {Thakar},
  {Thomas}, {Thompson}, {Tinker}, {Tissera}, {Tojeiro}, {Hernandez Toledo}, {de
  la Torre}, {Tremonti}, {Troup}, {Valenzuela}, {Martinez Valpuesta},
  {Vargas-Gonz{\'a}lez}, {Vargas- Maga{\~n}a}, {Vazquez}, {Villanova}, {Vivek},
  {Vogt}, {Wake}, {Walterbos}, {Wang}, {Weaver}, {Weijmans}, {Weinberg},
  {Westfall}, {Whelan}, {Wild}, {Wilson}, {Wood-Vasey}, {Wylezalek}, {Xiao},
  {Yan}, {Yang}, {Ybarra}, {Y{\`e}che}, {Zakamska}, {Zamora}, {Zarrouk},
  {Zasowski}, {Zhang}, {Zhao}, {Zheng}, {Zheng}, {Zhou}, {Zhou}, {Zhu},
  {Zoccali}, \& {Zou}}]{Blanton17}
{Blanton}, M.~R., {Bershady}, M.~A., {Abolfathi}, B., {et~al.} 2017, \aj, 154,
  28, \dodoi{10.3847/1538-3881/aa7567}

\bibitem[{{Borguet} {et~al.}(2013){Borguet}, {Arav}, {Edmonds}, {Chamberlain},
  \& {Benn}}]{Borguet13}
{Borguet}, B.~C.~J., {Arav}, N., {Edmonds}, D., {Chamberlain}, C., \& {Benn},
  C. 2013, \apj, 762, 49, \dodoi{10.1088/0004-637X/762/1/49}

\bibitem[{{Capellupo} {et~al.}(2013){Capellupo}, {Hamann}, {Shields},
  {Halpern}, \& {Barlow}}]{Capellupo13}
{Capellupo}, D.~M., {Hamann}, F., {Shields}, J.~C., {Halpern}, J.~P., \&
  {Barlow}, T.~A. 2013, \mnras, 429, 1872, \dodoi{10.1093/mnras/sts427}

\bibitem[{{Capellupo} {et~al.}(2011){Capellupo}, {Hamann}, {Shields},
  {Rodr{\'{\i}}guez Hidalgo}, \& {Barlow}}]{Capellupo11}
{Capellupo}, D.~M., {Hamann}, F., {Shields}, J.~C., {Rodr{\'{\i}}guez Hidalgo},
  P., \& {Barlow}, T.~A. 2011, \mnras, 413, 908,
  \dodoi{10.1111/j.1365-2966.2010.18185.x}

\bibitem[{{Capellupo} {et~al.}(2012){Capellupo}, {Hamann}, {Shields},
  {Rodr{\'{\i}}guez Hidalgo}, \& {Barlow}}]{Capellupo12}
---. 2012, \mnras, 422, 3249, \dodoi{10.1111/j.1365-2966.2012.20846.x}

\bibitem[{{Capellupo} {et~al.}(2017){Capellupo}, {Hamann}, {Herbst}, {Brandt},
  {Ge}, {P{\^a}ris}, {Petitjean}, {Schneider}, {Streblyanska}, \&
  {York}}]{Capellupo17}
{Capellupo}, D.~M., {Hamann}, F., {Herbst}, H., {et~al.} 2017, \mnras, 469,
  323, \dodoi{10.1093/mnras/stx870}

\bibitem[{{Cardelli} {et~al.}(1989){Cardelli}, {Clayton}, \&
  {Mathis}}]{Cardelli89}
{Cardelli}, J.~A., {Clayton}, G.~C., \& {Mathis}, J.~S. 1989, \apj, 345, 245,
  \dodoi{10.1086/167900}

\bibitem[{{Dawson} {et~al.}(2013){Dawson}, {Schlegel}, {Ahn}, {Anderson},
  {Aubourg}, {Bailey}, {Barkhouser}, {Bautista}, {Beifiori}, {Berlind},
  {Bhardwaj}, {Bizyaev}, {Blake}, {Blanton}, {Blomqvist}, {Bolton}, {Borde},
  {Bovy}, {Brandt}, {Brewington}, {Brinkmann}, {Brown}, {Brownstein}, {Bundy},
  {Busca}, {Carithers}, {Carnero}, {Carr}, {Chen}, {Comparat}, {Connolly},
  {Cope}, {Croft}, {Cuesta}, {da Costa}, {Davenport}, {Delubac}, {de Putter},
  {Dhital}, {Ealet}, {Ebelke}, {Eisenstein}, {Escoffier}, {Fan}, {Filiz Ak},
  {Finley}, {Font-Ribera}, {G{\'e}nova-Santos}, {Gunn}, {Guo}, {Haggard},
  {Hall}, {Hamilton}, {Harris}, {Harris}, {Ho}, {Hogg}, {Holder}, {Honscheid},
  {Huehnerhoff}, {Jordan}, {Jordan}, {Kauffmann}, {Kazin}, {Kirkby}, {Klaene},
  {Kneib}, {Le Goff}, {Lee}, {Long}, {Loomis}, {Lundgren}, {Lupton}, {Maia},
  {Makler}, {Malanushenko}, {Malanushenko}, {Mandelbaum}, {Manera}, {Maraston},
  {Margala}, {Masters}, {McBride}, {McDonald}, {McGreer}, {McMahon}, {Mena},
  {Miralda-Escud{\'e}}, {Montero-Dorta}, {Montesano}, {Muna}, {Myers},
  {Naugle}, {Nichol}, {Noterdaeme}, {Nuza}, {Olmstead}, {Oravetz}, {Oravetz},
  {Owen}, {Padmanabhan}, {Palanque-Delabrouille}, {Pan}, {Parejko},
  {P{\^a}ris}, {Percival}, {P{\'e}rez-Fournon}, {P{\'e}rez-R{\`a}fols},
  {Petitjean}, {Pfaffenberger}, {Pforr}, {Pieri}, {Prada}, {Price-Whelan},
  {Raddick}, {Rebolo}, {Rich}, {Richards}, {Rockosi}, {Roe}, {Ross}, {Ross},
  {Rossi}, {Rubi{\~n}o-Martin}, {Samushia}, {S{\'a}nchez}, {Sayres}, {Schmidt},
  {Schneider}, {Sc{\'o}ccola}, {Seo}, {Shelden}, {Sheldon}, {Shen}, {Shu},
  {Slosar}, {Smee}, {Snedden}, {Stauffer}, {Steele}, {Strauss}, {Streblyanska},
  {Suzuki}, {Swanson}, {Tal}, {Tanaka}, {Thomas}, {Tinker}, {Tojeiro},
  {Tremonti}, {Vargas Maga{\~n}a}, {Verde}, {Viel}, {Wake}, {Watson}, {Weaver},
  {Weinberg}, {Weiner}, {West}, {White}, {Wood-Vasey}, {Yeche}, {Zehavi},
  {Zhao}, \& {Zheng}}]{Dawson13}
{Dawson}, K.~S., {Schlegel}, D.~J., {Ahn}, C.~P., {et~al.} 2013, \aj, 145, 10,
  \dodoi{10.1088/0004-6256/145/1/10}

\bibitem[{{Dawson} {et~al.}(2016){Dawson}, {Kneib}, {Percival}, {Alam},
  {Albareti}, {Anderson}, {Armengaud}, {Aubourg}, {Bailey}, {Bautista},
  {Berlind}, {Bershady}, {Beutler}, {Bizyaev}, {Blanton}, {Blomqvist},
  {Bolton}, {Bovy}, {Brandt}, {Brinkmann}, {Brownstein}, {Burtin}, {Busca},
  {Cai}, {Chuang}, {Clerc}, {Comparat}, {Cope}, {Croft}, {Cruz-Gonzalez}, {da
  Costa}, {Cousinou}, {Darling}, {de la Macorra}, {de la Torre}, {Delubac}, {du
  Mas des Bourboux}, {Dwelly}, {Ealet}, {Eisenstein}, {Eracleous}, {Escoffier},
  {Fan}, {Finoguenov}, {Font-Ribera}, {Frinchaboy}, {Gaulme}, {Georgakakis},
  {Green}, {Guo}, {Guy}, {Ho}, {Holder}, {Huehnerhoff}, {Hutchinson}, {Jing},
  {Jullo}, {Kamble}, {Kinemuchi}, {Kirkby}, {Kitaura}, {Klaene}, {Laher},
  {Lang}, {Laurent}, {Le Goff}, {Li}, {Liang}, {Lima}, {Lin}, {Lin}, {Lin},
  {Long}, {Lundgren}, {MacDonald}, {Geimba Maia}, {Malanushenko},
  {Malanushenko}, {Mariappan}, {McBride}, {McGreer}, {M{\'e}nard}, {Merloni},
  {Meza}, {Montero-Dorta}, {Muna}, {Myers}, {Nandra}, {Naugle}, {Newman},
  {Noterdaeme}, {Nugent}, {Ogando}, {Olmstead}, {Oravetz}, {Oravetz},
  {Padmanabhan}, {Palanque-Delabrouille}, {Pan}, {Parejko}, {P{\^a}ris},
  {Peacock}, {Petitjean}, {Pieri}, {Pisani}, {Prada}, {Prakash}, {Raichoor},
  {Reid}, {Rich}, {Ridl}, {Rodriguez-Torres}, {Carnero Rosell}, {Ross},
  {Rossi}, {Ruan}, {Salvato}, {Sayres}, {Schneider}, {Schlegel}, {Seljak},
  {Seo}, {Sesar}, {Shandera}, {Shu}, {Slosar}, {Sobreira}, {Streblyanska},
  {Suzuki}, {Taylor}, {Tao}, {Tinker}, {Tojeiro}, {Vargas-Maga{\~n}a}, {Wang},
  {Weaver}, {Weinberg}, {White}, {Wood-Vasey}, {Yeche}, {Zhai}, {Zhao}, {Zhao},
  {Zheng}, {Ben Zhu}, \& {Zou}}]{Dawson16}
{Dawson}, K.~S., {Kneib}, J.-P., {Percival}, W.~J., {et~al.} 2016, \aj, 151,
  44, \dodoi{10.3847/0004-6256/151/2/44}

\bibitem[{{De Cicco} {et~al.}(2018){De Cicco}, {Brandt}, {Grier}, {Paolillo},
  {Filiz Ak}, {Schneider}, \& {Trump}}]{Decicco18}
{De Cicco}, D., {Brandt}, W.~N., {Grier}, C.~J., {et~al.} 2018, ArXiv e-prints,
  arXiv:1804.04666.
\newblock \doarXiv{1804.04666}

\bibitem[{{de Jong} {et~al.}(2014){de Jong}, {Barden}, {Bellido-Tirado},
  {Brynnel}, {Chiappini}, {Depagne}, {Haynes}, {Johl}, {Phillips}, {Schnurr},
  {Schwope}, {Walcher}, {Bauer}, {Cescutti}, {Cioni}, {Dionies}, {Enke},
  {Haynes}, {Kelz}, {Kitaura}, {Lamer}, {Minchev}, {M{\"u}ller}, {Nuza},
  {Olaya}, {Piffl}, {Popow}, {Saviauk}, {Steinmetz}, {Ural}, {Valentini},
  {Winkler}, {Wisotzki}, {Ansorge}, {Banerji}, {Gonzalez Solares}, {Irwin},
  {Kennicutt}, {King}, {McMahon}, {Koposov}, {Parry}, {Sun}, {Walton},
  {Finger}, {Iwert}, {Krumpe}, {Lizon}, {Mainieri}, {Amans}, {Bonifacio},
  {Cohen}, {Fran{\c{c}}ois}, {Jagourel}, {Mignot}, {Royer}, {Sartoretti},
  {Bender}, {Hess}, {Lang-Bardl}, {Muschielok}, {Schlichter}, {B{\"o}hringer},
  {Boller}, {Bongiorno}, {Brusa}, {Dwelly}, {Merloni}, {Nandra}, {Salvato},
  {Pragt}, {Navarro}, {Gerlofsma}, {Roelfsema}, {Dalton}, {Middleton}, {Tosh},
  {Boeche}, {Caffau}, {Christlieb}, {Grebel}, {Hansen}, {Koch}, {Ludwig},
  {Mandel}, {Quirrenbach}, {Sbordone}, {Seifert}, {Thimm}, {Helmi}, {trager},
  {Bensby}, {Feltzing}, {Ruchti}, {Edvardsson}, {Korn}, {Lind}, {Boland},
  {Colless}, {Frost}, {Gilbert}, {Gillingham}, {Lawrence}, {Legg}, {Saunders},
  {Sheinis}, {Driver}, {Robotham}, {Bacon}, {Caillier}, {Kosmalski}, {Laurent},
  \& {Richard}}]{dejong14}
{de Jong}, R.~S., {Barden}, S., {Bellido-Tirado}, O., {et~al.} 2014, in
  Ground-based and Airborne Instrumentation for Astronomy V, Vol. 9147, 91470M

\bibitem[{{de Kool} {et~al.}(2002){de Kool}, {Korista}, \& {Arav}}]{dekool02}
{de Kool}, M., {Korista}, K.~T., \& {Arav}, N. 2002, \apj, 580, 54,
  \dodoi{10.1086/343107}

\bibitem[{{Dere} {et~al.}(1997){Dere}, {Landi}, {Mason}, {Monsignori Fossi}, \&
  {Young}}]{Dere97}
{Dere}, K.~P., {Landi}, E., {Mason}, H.~E., {Monsignori Fossi}, B.~C., \&
  {Young}, P.~R. 1997, \aaps, 125, 149, \dodoi{10.1051/aas:1997368}

\bibitem[{{Di~Matteo} {et~al.}(2005){Di~Matteo}, {Springel}, \&
  {Hernquist}}]{Dimatteo05}
{Di~Matteo}, T., {Springel}, V., \& {Hernquist}, L. 2005, \nat, 433, 604,
  \dodoi{10.1038/nature03335}

\bibitem[{{Dunn} {et~al.}(2010){Dunn}, {Bautista}, {Arav}, {Moe}, {Korista},
  {Costantini}, {Benn}, {Ellison}, \& {Edmonds}}]{Dunn10}
{Dunn}, J.~P., {Bautista}, M., {Arav}, N., {et~al.} 2010, \apj, 709, 611,
  \dodoi{10.1088/0004-637X/709/2/611}

\bibitem[{{Eisenstein} {et~al.}(2011){Eisenstein}, {Weinberg}, {Agol},
  {Aihara}, {Allende Prieto}, {Anderson}, {Arns}, {Aubourg}, {Bailey},
  {Balbinot}, \& et~al.}]{Eisenstein11}
{Eisenstein}, D.~J., {Weinberg}, D.~H., {Agol}, E., {et~al.} 2011, \aj, 142,
  72, \dodoi{10.1088/0004-6256/142/3/72}

\bibitem[{{Filiz Ak} {et~al.}(2012){Filiz Ak}, {Brandt}, {Hall}, {Schneider},
  {Anderson}, {Gibson}, {Lundgren}, {Myers}, {Petitjean}, {Ross}, {Shen},
  {York}, {Bizyaev}, {Brinkmann}, {Malanushenko}, {Oravetz}, {Pan}, {Simmons},
  \& {Weaver}}]{Filizak12}
{Filiz Ak}, N., {Brandt}, W.~N., {Hall}, P.~B., {et~al.} 2012, \apj, 757, 114,
  \dodoi{10.1088/0004-637X/757/2/114}

\bibitem[{{Filiz Ak} {et~al.}(2013){Filiz Ak}, {Brandt}, {Hall}, {Schneider},
  {Anderson}, {Hamann}, {Lundgren}, {Myers}, {P{\^a}ris}, {Petitjean}, {Ross},
  {Shen}, \& {York}}]{Filizak13}
---. 2013, \apj, 777, 168, \dodoi{10.1088/0004-637X/777/2/168}

\bibitem[{{Filiz Ak} {et~al.}(2014){Filiz Ak}, {Brandt}, {Hall}, {Schneider},
  {Trump}, {Anderson}, {Hamann}, {Myers}, {P{\^a}ris}, {Petitjean}, {Ross},
  {Shen}, \& {York}}]{Filizak14}
---. 2014, \apj, 791, 88, \dodoi{10.1088/0004-637X/791/2/88}

\bibitem[{{Gehrels}(1986)}]{Gehrels86}
{Gehrels}, N. 1986, \apj, 303, 336, \dodoi{10.1086/164079}

\bibitem[{{Gibson} {et~al.}(2010){Gibson}, {Brandt}, {Gallagher}, {Hewett}, \&
  {Schneider}}]{Gibson10}
{Gibson}, R.~R., {Brandt}, W.~N., {Gallagher}, S.~C., {Hewett}, P.~C., \&
  {Schneider}, D.~P. 2010, \apj, 713, 220, \dodoi{10.1088/0004-637X/713/1/220}

\bibitem[{{Gibson} {et~al.}(2009{\natexlab{a}}){Gibson}, {Brandt}, {Gallagher},
  \& {Schneider}}]{Gibson09}
{Gibson}, R.~R., {Brandt}, W.~N., {Gallagher}, S.~C., \& {Schneider}, D.~P.
  2009{\natexlab{a}}, \apj, 696, 924, \dodoi{10.1088/0004-637X/696/1/924}

\bibitem[{{Gibson} {et~al.}(2008){Gibson}, {Brandt}, \& {Schneider}}]{Gibson08}
{Gibson}, R.~R., {Brandt}, W.~N., \& {Schneider}, D.~P. 2008, \apj, 685, 773,
  \dodoi{10.1086/590403}

\bibitem[{{Gibson} {et~al.}(2009{\natexlab{b}}){Gibson}, {Jiang}, {Brandt},
  {Hall}, {Shen}, {Wu}, {Anderson}, {Schneider}, {Vanden Berk}, {Gallagher},
  {Fan}, \& {York}}]{Gibson09b}
{Gibson}, R.~R., {Jiang}, L., {Brandt}, W.~N., {et~al.} 2009{\natexlab{b}},
  \apj, 692, 758, \dodoi{10.1088/0004-637X/692/1/758}

\bibitem[{{Grier} {et~al.}(2015){Grier}, {Hall}, {Brandt}, {Trump}, {Shen},
  {Vivek}, {Filiz Ak}, {Chen}, {Dawson}, {Denney}, {Green}, {Jiang},
  {Kochanek}, {McGreer}, {P{\^a}ris}, {Peterson}, {Schneider}, {Tao},
  {Wood-Vasey}, {Bizyaev}, {Ge}, {Kinemuchi}, {Oravetz}, {Pan}, \&
  {Simmons}}]{Grier15}
{Grier}, C.~J., {Hall}, P.~B., {Brandt}, W.~N., {et~al.} 2015, \apj, 806, 111,
  \dodoi{10.1088/0004-637X/806/1/111}

\bibitem[{{Grier} {et~al.}(2016){Grier}, {Brandt}, {Hall}, {Trump}, {Filiz Ak},
  {Anderson}, {Green}, {Schneider}, {Sun}, {Vivek}, {Beatty}, {Brownstein}, \&
  {Roman-Lopes}}]{Grier16}
{Grier}, C.~J., {Brandt}, W.~N., {Hall}, P.~B., {et~al.} 2016, \apj, 824, 130,
  \dodoi{10.3847/0004-637X/824/2/130}

\bibitem[{{Gunn} {et~al.}(2006){Gunn}, {Siegmund}, {Mannery}, {Owen}, {Hull},
  {Leger}, {Carey}, {Knapp}, {York}, {Boroski}, {Kent}, {Lupton}, {Rockosi},
  {Evans}, {Waddell}, {Anderson}, {Annis}, {Barentine}, {Bartoszek}, {Bastian},
  {Bracker}, {Brewington}, {Briegel}, {Brinkmann}, {Brown}, {Carr},
  {Czarapata}, {Drennan}, {Dombeck}, {Federwitz}, {Gillespie}, {Gonzales},
  {Hansen}, {Harvanek}, {Hayes}, {Jordan}, {Kinney}, {Klaene}, {Kleinman},
  {Kron}, {Kresinski}, {Lee}, {Limmongkol}, {Lindenmeyer}, {Long}, {Loomis},
  {McGehee}, {Mantsch}, {Neilsen}, {Neswold}, {Newman}, {Nitta}, {Peoples},
  {Pier}, {Prieto}, {Prosapio}, {Rivetta}, {Schneider}, {Snedden}, \&
  {Wang}}]{Gunn06}
{Gunn}, J.~E., {Siegmund}, W.~A., {Mannery}, E.~J., {et~al.} 2006, \aj, 131,
  2332, \dodoi{10.1086/500975}

\bibitem[{{Haggard} {et~al.}(2012){Haggard}, {Arraki}, {Green}, {Aldcroft}, \&
  {Anderson}}]{Haggard12}
{Haggard}, D., {Arraki}, K.~S., {Green}, P.~J., {Aldcroft}, T., \& {Anderson},
  S.~F. 2012, in Astronomical Society of the Pacific Conference Series, Vol.
  460, AGN Winds in Charleston, ed. G.~{Chartas}, F.~{Hamann}, \& K.~M.
  {Leighly}, 98

\bibitem[{{Hall} {et~al.}(2011){Hall}, {Anosov}, {White}, {Brandt}, {Gregg},
  {Gibson}, {Becker}, \& {Schneider}}]{Hall11}
{Hall}, P.~B., {Anosov}, K., {White}, R.~L., {et~al.} 2011, \mnras, 411, 2653,
  \dodoi{10.1111/j.1365-2966.2010.17870.x}

\bibitem[{{Hall} {et~al.}(2007){Hall}, {Sadavoy}, {Hutsemekers}, {Everett}, \&
  {Rafiee}}]{Hall07}
{Hall}, P.~B., {Sadavoy}, S.~I., {Hutsemekers}, D., {Everett}, J.~E., \&
  {Rafiee}, A. 2007, \apj, 665, 174, \dodoi{10.1086/519273}

\bibitem[{{Hamann} {et~al.}(2000){Hamann}, {Netzer}, \& {Shields}}]{Hamann00}
{Hamann}, F.~W., {Netzer}, H., \& {Shields}, J.~C. 2000, \apj, 536, 101,
  \dodoi{10.1086/308936}

\bibitem[{{Higginbottom} {et~al.}(2013){Higginbottom}, {Knigge}, {Long}, {Sim},
  \& {Matthews}}]{Higgenbottom13}
{Higginbottom}, N., {Knigge}, C., {Long}, K.~S., {Sim}, S.~A., \& {Matthews},
  J.~H. 2013, \mnras, 436, 1390, \dodoi{10.1093/mnras/stt1658}

\bibitem[{{Higginbottom} {et~al.}(2014){Higginbottom}, {Proga}, {Knigge},
  {Long}, {Matthews}, \& {Sim}}]{Higginbottom14}
{Higginbottom}, N., {Proga}, D., {Knigge}, C., {et~al.} 2014, \apj, 789, 19,
  \dodoi{10.1088/0004-637X/789/1/19}

\bibitem[{{King} \& {Pounds}(2015)}]{KingPounds15}
{King}, A., \& {Pounds}, K. 2015, Annual Review of Astronomy and Astrophysics,
  53, 115, \dodoi{10.1146/annurev-astro-082214-122316}

\bibitem[{{Kollmeier} {et~al.}(2017){Kollmeier}, {Zasowski}, {Rix}, {Johns},
  {Anderson}, {Drory}, {Johnson}, {Pogge}, {Bird}, {Blanc}, {Brownstein},
  {Crane}, {De Lee}, {Klaene}, {Kreckel}, {MacDonald}, {Merloni}, {Ness},
  {O'Brien}, {Sanchez-Gallego}, {Sayres}, {Shen}, {Thakar}, {Tkachenko},
  {Aerts}, {Blanton}, {Eisenstein}, {Holtzman}, {Maoz}, {Nandra}, {Rockosi},
  {Weinberg}, {Bovy}, {Casey}, {Chaname}, {Clerc}, {Conroy}, {Eracleous},
  {G{\"a}nsicke}, {Hekker}, {Horne}, {Kauffmann}, {McQuinn}, {Pellegrini},
  {Schinnerer}, {Schlafly}, {Schwope}, {Seibert}, {Teske}, \& {van
  Saders}}]{Kollmeier17}
{Kollmeier}, J.~A., {Zasowski}, G., {Rix}, H.-W., {et~al.} 2017, ArXiv
  e-prints, arXiv:1711.03234.
\newblock \doarXiv{1711.03234}

\bibitem[{{Krolik}(1999)}]{Krolik99}
{Krolik}, J.~H. 1999, {Active galactic nuclei : from the central black hole to
  the galactic environment}

\bibitem[{{Landi} {et~al.}(2013){Landi}, {Young}, {Dere}, {Del Zanna}, \&
  {Mason}}]{Landi13}
{Landi}, E., {Young}, P.~R., {Dere}, K.~P., {Del Zanna}, G., \& {Mason}, H.~E.
  2013, \apj, 763, 86, \dodoi{10.1088/0004-637X/763/2/86}

\bibitem[{{Lundgren} {et~al.}(2007){Lundgren}, {Wilhite}, {Brunner}, {Hall},
  {Schneider}, {York}, {Vanden Berk}, \& {Brinkmann}}]{Lundgren07}
{Lundgren}, B.~F., {Wilhite}, B.~C., {Brunner}, R.~J., {et~al.} 2007, \apj,
  656, 73, \dodoi{10.1086/510202}

\bibitem[{{Marshall} {et~al.}(1997){Marshall}, {Carone}, {Peterson}, {Clavel},
  {Crenshaw}, {Korista}, {Kriss}, {Krolik}, {Malkan}, {Morris}, \&
  {Reichert}}]{Marshall97}
{Marshall}, H.~L., {Carone}, T.~E., {Peterson}, B.~M., {et~al.} 1997, \apj,
  479, 222, \dodoi{10.1086/303850}

\bibitem[{{Misawa} {et~al.}(2010){Misawa}, {Kawabata}, {Eracleous}, {Charlton},
  \& {Kashikawa}}]{Misawa10}
{Misawa}, T., {Kawabata}, K.~S., {Eracleous}, M., {Charlton}, J.~C., \&
  {Kashikawa}, N. 2010, \apj, 719, 1890, \dodoi{10.1088/0004-637X/719/2/1890}

\bibitem[{{Moll} {et~al.}(2007){Moll}, {Schindler}, {Domainko}, {Kapferer},
  {Mair}, {van Kampen}, {Kronberger}, {Kimeswenger}, \& {Ruffert}}]{Moll07}
{Moll}, R., {Schindler}, S., {Domainko}, W., {et~al.} 2007, \aap, 463, 513,
  \dodoi{10.1051/0004-6361:20066386}

\bibitem[{{Murray} {et~al.}(1995){Murray}, {Chiang}, {Grossman}, \&
  {Voit}}]{Murray95}
{Murray}, N., {Chiang}, J., {Grossman}, S.~A., \& {Voit}, G.~M. 1995, \apj,
  451, 498, \dodoi{10.1086/176238}

\bibitem[{{P{\^a}ris} {et~al.}(2018){P{\^a}ris}, {Petitjean}, {Aubourg},
  {Myers}, {Streblyanska}, {Lyke}, {Anderson}, {Armengaud}, {Bautista},
  {Blanton}, {Blomqvist}, {Brinkmann}, {Brownstein}, {Brandt}, {Burtin},
  {Dawson}, {de la Torre}, {Georgakakis}, {Gil-Mar{\'\i}n}, {Green}, {Hall},
  {Kneib}, {LaMassa}, {Le Goff}, {MacLeod}, {Mariappan}, {McGreer}, {Merloni},
  {Noterdaeme}, {Palanque-Delabrouille}, {Percival}, {Ross}, {Rossi},
  {Schneider}, {Seo}, {Tojeiro}, {Weaver}, {Weijmans}, {Y{\`e}che}, {Zarrouk},
  \& {Zhao}}]{Paris18}
{P{\^a}ris}, I., {Petitjean}, P., {Aubourg}, {\'E}., {et~al.} 2018, \aap, 613,
  A51, \dodoi{10.1051/0004-6361/201732445}

\bibitem[{{Park} {et~al.}(2013){Park}, {Woo}, {Denney}, \& {Shin}}]{Park13}
{Park}, D., {Woo}, J.-H., {Denney}, K.~D., \& {Shin}, J. 2013, \apj, 770, 87,
  \dodoi{10.1088/0004-637X/770/2/87}

\bibitem[{{Pei}(1992)}]{Pei92}
{Pei}, Y.~C. 1992, \apj, 395, 130, \dodoi{10.1086/171637}

\bibitem[{{Proga}(2000)}]{Proga00}
{Proga}, D. 2000, \apj, 538, 684, \dodoi{10.1086/309154}

\bibitem[{{Proga} {et~al.}(2012){Proga}, {Rodriguez-Hidalgo}, \&
  {Hamann}}]{Proga12}
{Proga}, D., {Rodriguez-Hidalgo}, P., \& {Hamann}, F. 2012, in Astronomical
  Society of the Pacific Conference Series, Vol. 460, AGN Winds in Charleston,
  ed. G.~{Chartas}, F.~{Hamann}, \& K.~M. {Leighly}, 171

\bibitem[{{Richards} {et~al.}(2006){Richards}, {Strauss}, {Fan}, {Hall},
  {Jester}, {Schneider}, {Vanden Berk}, {Stoughton}, {Anderson}, {Brunner},
  {Gray}, {Gunn}, {Ivezi{\'c}}, {Kirkland}, {Knapp}, {Loveday}, {Meiksin},
  {Pope}, {Szalay}, {Thakar}, {Yanny}, {York}, {Barentine}, {Brewington},
  {Brinkmann}, {Fukugita}, {Harvanek}, {Kent}, {Kleinman}, {Krzesi{\'n}ski},
  {Long}, {Lupton}, {Nash}, {Neilsen}, {Nitta}, {Schlegel}, \&
  {Snedden}}]{Richards06}
{Richards}, G.~T., {Strauss}, M.~A., {Fan}, X., {et~al.} 2006, \aj, 131, 2766,
  \dodoi{10.1086/503559}

\bibitem[{{Rogerson} {et~al.}(2018){Rogerson}, {Hall}, {Ahmed}, {Rodr{\'\i}guez
  Hidalgo}, {Brandt}, \& {Filiz Ak}}]{Rogerson18}
{Rogerson}, J.~A., {Hall}, P.~B., {Ahmed}, N.~S., {et~al.} 2018, \apj, 862, 22,
  \dodoi{10.3847/1538-4357/aabfe5}

\bibitem[{{Rogerson} {et~al.}(2015){Rogerson}, {Hall}, {Rodr{\'{\i}}guez
  Hidalgo}, {Pirkola}, {Brandt}, \& {Filiz Ak}}]{Rogerson15}
{Rogerson}, J.~A., {Hall}, P.~B., {Rodr{\'{\i}}guez Hidalgo}, P., {et~al.}
  2015, ArXiv e-prints.
\newblock \doarXiv{1509.02842}

\bibitem[{{Sabra} \& {Hamann}(2005)}]{Sabra05}
{Sabra}, B.~M., \& {Hamann}, F. 2005, ArXiv Astrophysics e-prints

\bibitem[{{Schlafly} \& {Finkbeiner}(2011)}]{SchaflyFinkbeiner11}
{Schlafly}, E.~F., \& {Finkbeiner}, D.~P. 2011, \apj, 737, 103,
  \dodoi{10.1088/0004-637X/737/2/103}

\bibitem[{{Shen} {et~al.}(2015){Shen}, {Brandt}, {Dawson}, {Hall}, {McGreer},
  {Anderson}, {Chen}, {Denney}, {Eftekharzadeh}, {Fan}, {Gao}, {Green},
  {Greene}, {Ho}, {Horne}, {Jiang}, {Kelly}, {Kinemuchi}, {Kochanek},
  {P{\^a}ris}, {Peters}, {Peterson}, {Petitjean}, {Ponder}, {Richards},
  {Schneider}, {Seth}, {Smith}, {Strauss}, {Tao}, {Trump}, {Wood-Vasey}, {Zu},
  {Eisenstein}, {Pan}, {Bizyaev}, {Malanushenko}, {Malanushenko}, \&
  {Oravetz}}]{Shen15}
{Shen}, Y., {Brandt}, W.~N., {Dawson}, K.~S., {et~al.} 2015, \apjs, 216, 4,
  \dodoi{10.1088/0067-0049/216/1/4}

\bibitem[{{Shen} {et~al.}(2016){Shen}, {Horne}, {Grier}, {Peterson}, {Denney},
  {Trump}, {Sun}, {Brandt}, {Kochanek}, {Dawson}, {Green}, {Greene}, {Hall},
  {Ho}, {Jiang}, {Kinemuchi}, {McGreer}, {Petitjean}, {Richards}, {Schneider},
  {Strauss}, {Tao}, {Wood-Vasey}, {Zu}, {Pan}, {Bizyaev}, {Ge}, {Oravetz}, \&
  {Simmons}}]{Shen16}
{Shen}, Y., {Horne}, K., {Grier}, C.~J., {et~al.} 2016, \apj, 818, 30,
  \dodoi{10.3847/0004-637X/818/1/30}

\bibitem[{{Shen} {et~al.}(2018){Shen}, {Hall}, {Horne}, {Zhu}, {McGreer},
  {Simm}, {Trump}, {Kinemuchi}, {Brandt}, {Green}, {Grier}, {Guo}, {Ho},
  {Homayouni}, {Jiang}, {I-Hsiu Li}, {Morganson}, {Petitjean}, {Richards},
  {Schneider}, {Starkey}, {Wang}, {Chambers}, {Kaiser}, {Kudritzki}, {Magnier},
  \& {Waters}}]{Shen18}
{Shen}, Y., {Hall}, P.~B., {Horne}, K., {et~al.} 2018, ArXiv e-prints,
  arXiv:1810.01447.
\newblock \doarXiv{1810.01447}

\bibitem[{{Smee} {et~al.}(2013){Smee}, {Gunn}, {Uomoto}, {Roe}, {Schlegel},
  {Rockosi}, {Carr}, {Leger}, {Dawson}, {Olmstead}, {Brinkmann}, {Owen},
  {Barkhouser}, {Honscheid}, {Harding}, {Long}, {Lupton}, {Loomis}, {Anderson},
  {Annis}, {Bernardi}, {Bhardwaj}, {Bizyaev}, {Bolton}, {Brewington}, {Briggs},
  {Burles}, {Burns}, {Castander}, {Connolly}, {Davenport}, {Ebelke}, {Epps},
  {Feldman}, {Friedman}, {Frieman}, {Heckman}, {Hull}, {Knapp}, {Lawrence},
  {Loveday}, {Mannery}, {Malanushenko}, {Malanushenko}, {Merrelli}, {Muna},
  {Newman}, {Nichol}, {Oravetz}, {Pan}, {Pope}, {Ricketts}, {Shelden},
  {Sandford}, {Siegmund}, {Simmons}, {Smith}, {Snedden}, {Schneider},
  {SubbaRao}, {Tremonti}, {Waddell}, \& {York}}]{Smee13}
{Smee}, S.~A., {Gunn}, J.~E., {Uomoto}, A., {et~al.} 2013, \aj, 146, 32,
  \dodoi{10.1088/0004-6256/146/2/32}

\bibitem[{{van der Marel} \& {Franx}(1993)}]{vanderMarel93}
{van der Marel}, R.~P., \& {Franx}, M. 1993, \apj, 407, 525,
  \dodoi{10.1086/172534}

\bibitem[{{Vanden Berk} {et~al.}(2001){Vanden Berk}, {Richards}, {Bauer},
  {Strauss}, {Schneider}, {Heckman}, {York}, {Hall}, {Fan}, {Knapp},
  {Anderson}, {Annis}, {Bahcall}, {Bernardi}, {Briggs}, {Brinkmann}, {Brunner},
  {Burles}, {Carey}, {Castander}, {Connolly}, {Crocker}, {Csabai}, {Doi},
  {Finkbeiner}, {Friedman}, {Frieman}, {Fukugita}, {Gunn}, {Hennessy},
  {Ivezi{\'c}}, {Kent}, {Kunszt}, {Lamb}, {Leger}, {Long}, {Loveday}, {Lupton},
  {Meiksin}, {Merelli}, {Munn}, {Newberg}, {Newcomb}, {Nichol}, {Owen}, {Pier},
  {Pope}, {Rockosi}, {Schlegel}, {Siegmund}, {Smee}, {Snir}, {Stoughton},
  {Stubbs}, {SubbaRao}, {Szalay}, {Szokoly}, {Tremonti}, {Uomoto}, {Waddell},
  {Yanny}, \& {Zheng}}]{Vandenberk01}
{Vanden Berk}, D.~E., {Richards}, G.~T., {Bauer}, A., {et~al.} 2001, \aj, 122,
  549, \dodoi{10.1086/321167}

\bibitem[{{Vivek} {et~al.}(2016){Vivek}, {Srianand}, \& {Gupta}}]{Vivek16}
{Vivek}, M., {Srianand}, R., \& {Gupta}, N. 2016, \mnras, 455, 136,
  \dodoi{10.1093/mnras/stv2240}

\bibitem[{{Vivek} {et~al.}(2012){Vivek}, {Srianand}, {Mahabal}, \&
  {Kuriakose}}]{Vivek12}
{Vivek}, M., {Srianand}, R., {Mahabal}, A., \& {Kuriakose}, V.~C. 2012, \mnras,
  421, L107, \dodoi{10.1111/j.1745-3933.2012.01216.x}

\bibitem[{{Vivek} {et~al.}(2014){Vivek}, {Srianand}, {Petitjean}, {Mohan},
  {Mahabal}, \& {Samui}}]{Vivek14}
{Vivek}, M., {Srianand}, R., {Petitjean}, P., {et~al.} 2014, \mnras, 440, 799,
  \dodoi{10.1093/mnras/stu288}

\bibitem[{{Welling} {et~al.}(2014){Welling}, {Miller}, {Brandt}, {Capellupo},
  \& {Gibson}}]{Welling14}
{Welling}, C.~A., {Miller}, B.~P., {Brandt}, W.~N., {Capellupo}, D.~M., \&
  {Gibson}, R.~R. 2014, \mnras, 440, 2474, \dodoi{10.1093/mnras/stu402}

\bibitem[{{Weymann} {et~al.}(1991){Weymann}, {Morris}, {Foltz}, \&
  {Hewett}}]{Weymann91}
{Weymann}, R.~J., {Morris}, S.~L., {Foltz}, C.~B., \& {Hewett}, P.~C. 1991,
  \apj, 373, 23, \dodoi{10.1086/170020}

\bibitem[{{White} {et~al.}(1997){White}, {Becker}, {Helfand}, \&
  {Gregg}}]{White97}
{White}, R.~L., {Becker}, R.~H., {Helfand}, D.~J., \& {Gregg}, M.~D. 1997,
  \apj, 475, 479, \dodoi{10.1086/303564}

\bibitem[{{Xu} {et~al.}(2018){Xu}, {Arav}, {Miller}, \& {Benn}}]{Xu18}
{Xu}, X., {Arav}, N., {Miller}, T., \& {Benn}, C. 2018, \apj, 858, 39,
  \dodoi{10.3847/1538-4357/aab7ea}

\end{thebibliography}

\appendix 
\section{Notes on Individual Objects}
We here provide additional information and notes on the individual sources in our quasar sample.\\ 

\subsection{RM 039 (SDSS\,J141607.12+531904.8)} 
RM 039 hosts an extremely broad, deep \civ \ BAL, and it also contains deep \siiv \ and \aliii \ BALs, placing it into the \civ$_{\rm SA}$ category. This object has a P\,{\sc v} BAL at corresponding velocities, indicating that the \civ \ BAL is highly saturated. 

\subsection{RM 073 (SDSS\,J141741.72+530519.0)} 
RM\,073 contains two \civ \ BALs, neither of which is identified to be significantly variable on short timescales. There is no evidence for corresponding \siiv \ or \aliii \ BALs or mini-BALs. 

\subsection{RM 116 (SDSS\,J141432.46+523154.5)} 
RM\,116 formally has two distinct \civ \ BALs identified; however, visual inspection of the mean spectrum suggests that these two BALs may belong to the same absorption system. There also exists a \siiv \ absorption feature, and a few accompanying narrow \aliii \ troughs in the region. 

\subsection{RM 128 (SDSS\,J141103.17+531551.3)}
RM 128 has two BALs, both identified as rapidly varying but on different dates. In all four epochs exhibiting short-timescale variability, the other BAL also experiences an EW shift in the same direction, but at a lower significance. We interpret this as tentative evidence for coordinated variability, although the low significance of the secondary BAL variability prohibits a definite conclusion. The higher-velocity BAL in RM\,128 (BAL[A]) also has a corresponding \siiv \ BAL as well as slight evidence for shallow \aliii \ absorption at the same velocities. The lower-velocity BAL[B] does not appear to have any accompanying \siiv \ or \aliii \ absorption. 

\subsection{RM 155 (SDSS\,J141123.68+532845.7)} 
Our spectra for RM\,155 do not have complete coverage of the \siiv \ region; however, we see evidence for the beginning of a \siiv \ trough at corresponding velocities to the very deep \civ \ BAL, as well as a strong \aliii \ absorption feature. Thus, this is a C\,{\sc iv}$_{\rm SA}$ quasar.

\subsection{RM 195 (SDSS\,J141935.58+525710.7)} 
RM\,195 hosts two formal \civ \ BALs; however, the BAL at higher outflow velocities was excluded from our sample due to the presence of sky line contamination (see Section~\ref{sec:finalsample}). There exists evidence for an accompanying \siiv \ feature to this higher-velocity trough; however, the BAL that is included in our study shows no accompanying \siiv \ or \aliii \ absorption.  

\subsection{RM 217 (SDSS\,J141000.68+532156.1)}
This quasar has two identified \civ \ BALs, but only one is identified as rapidly varying. The high-velocity BAL[A] shows less-significant variability --- in two epoch pairs, however, the less-significant variability is in the same direction as the significant variability. In one epoch pair, the EW of BAL[A] changes in the opposite direction, and it has disappeared completely in the last epoch pair. This quasar is one of two that harbor a high-velocity BAL (BAL[A]) that disappears during the campaign. Additionally, this quasar has \siiv \ BALs present at velocities corresponding to both \civ \ BALs, although no \aliii \ absorption is visible. 

\subsection{RM 257 (SDSS\,J140931.90+532302.2)} 
RM\,257 has two BALs, both identified as rapidly varying. BAL[A] only has two epoch pairs identified as rapidly-varying (MJDs 57510--57518 and 57518--57544). BAL[B] is also significantly variable in the same direction during these epoch pairs. In the additional three epoch pairs that BAL[B] is identified as rapidly variable, BAL[A] also varies in the same direction, but the variability is significant at less than 4$\sigma$. Thus, we again find tentative evidence for coordinated variability. RM\,257 BAL[A] has accompanying \siiv \ and \aliii \ troughs. The \pv \ region corresponding to these velocities is cut off at the blue end of the wavelength region, but the spectrum does appear to have  \pv \ absorption.

\subsection{RM 284 (SDSS\,J141927.35+533727.7)} 
RM\,284 shows broad \civ \ absorption within our velocity search range --- however, this quasar also has a very high-velocity \civ \ trough (not studied in this work) extending from approximately 35,000 to 59,000 km s$^{-1}$ with a maximum depth of approximately 50\% of the continuum flux. The absorption EW of the very high-velocity trough has generally decreased from 2014 to 2017, but the maximum depth has not. We see the red edge of an accompanying very high-velocity N\,{\sc v} absorption trough at the extreme blue edge of the SDSS-RM spectrum, but we lack coverage of the \pv \ region at these velocities. The strength of any \siiv \ absorption  in BAL[A] in this quasar is impossible to determine due to confusion with the very high-velocity \civ \ trough, so we assign the C\,{\sc iv}$_{\rm N0}$ designation to this target to indicate this uncertainty. There does not appear to be substantial \aliii \ absorption, although the region blueward of \aliii \ has considerable blended Fe emission, making it difficult to interpret the spectrum.

\subsection{RM 339 (SDSS\,J142014.84+533609.0)}
RM\,339 contains BAL troughs in both \siiv \ and \aliii \ at similar velocities to the \civ \ trough. Thus, this is a C\,{\sc iv}$_{\rm SA}$ trough. 

\subsection{RM 357 (SDSS\,J141955.27+522741.1)} 
RM\,357 has two \civ \ BAL troughs. BAL[A], the higher-velocity BAL, displays rapid variability on two separate occasions. On both of those occasions, BAL[B] also exhibits rapid variability in the same direction. BAL[B] shows rapid variability in five additional epoch pairs as well, but in two of these pairs, BAL[A] has disappeared entirely. In the epoch pairs where BAL[A] remains present, it varies in the same direction as BAL[B], but at a lower significance. Thus, we yet again see tentative evidence for coordinated variability. RM\,357 BAL[B] has accompanying \siiv \ absorption, but no signs of \aliii \ features. The \pv \ region is not covered by our spectra for this quasar. 

\subsection{RM 361 (SDSS\,J142100.22+524342.3)} 
RM\,361 has a single, deep BAL trough. The \siiv \ region is not covered by our spectra, so we assign this BAL to the C\,{\sc iv}$_{\rm N0}$ category. There is not strong evidence for \aliii \ absorption.  

\subsection{RM 408 (SDSS\,J141409.85+520137.2)} 
RM\,408 has visible narrow \siiv \ and \aliii \ absorption at velocities corresponding to the \civ \ BAL[A]; however, both of these features are too narrow to meet our formal BAL definition, so we assign this BAL to the \civ$_{00}$ category. We do not have coverage of the \pv \ region in this object.

\subsection{RM 508 (SDSS\,J142129.40+522752.0)} 
This quasar exhibits \civ \ absorption with no accompanying \siiv, \aliii, or \pv \ features. Additional \civ \ absorption is visible at lower velocities that is both slightly too shallow and narrow to meet the formal BAL definition. The \civ \ BAL[A] in this target varies significantly on short timescales, sometimes with exceptionally high significance (i.e., greater than 20$\sigma_{\Delta \rm{EW}}$), making this quasar especially noteworthy. There also may be a weak, broad, very high-velocity \civ \ absorption feature at rest-frame 1350\,\AA. 

\subsection{RM 509 (SDSS\,J142233.74+525219.8)}
RM\,509 has two \civ \ BALs, but only the lower-velocity BAL[B] is rapidly variable. This variability only occurs in a single pair of epochs. Between this rapidly varying pair of epochs, BAL[A] varies in the same direction but at lower significance. The higher-velocity BAL[A] has an accompanying \siiv \ trough, but no \aliii \ feature; the lower-velocity BAL[B] shows no BALs or mini-BALs of either species. There is no evidence for a \pv \ BAL in this target. 

\subsection{RM 564 (SDSS\,J142306.05+531529.0)}
This source has a high-velocity \civ \ BAL with accompanying \siiv \ absorption, but no \aliii \ feature. We also see a significant feature near 1320\,\AA \ in the \siiv \ region that must be an extremely high-velocity \civ \ trough.

\subsection{RM 565 (SDSS\,J140634.14+525407.8)}
The spectra for RM\,565 barely clear our SN threshold; it is thus difficult to identify different species throughout the spectra. However, we identify two BALs in this quasar. The velocities covered by BAL[A] are not fully covered by the spectra in the \siiv \ region, so we are unsure if \siiv \ is present at these velocities. \aliii \ does not appear to be present in the spectra. The lower-velocity BAL[B] appears to have accompanying \siiv \ absorption, but no \aliii \ absorption. 

\subsection{RM 613 (SDSS\,J141007.73+541203.4)} 
RM\,613 has one \civ \ BAL that meets the formal BAL definition, although there is an additional lower-velocity mini-BAL. \siiv \ absorption is likely present at velocities corresponding to the BAL, but, at $<$10\% of the continuum flux, it does not formally meet the definition of a BAL. There is no evidence for accompanying \aliii \ absorption. This quasar was investigated by \cite{Grier15}, hereafter G15, who reported the first detection of significant BAL variability on timescales shorter than five days in the quasar rest frame. Their investigation revealed four different pairs of spectra separated by less than five rest-frame days that exhibited $> 4\sigma_{\Delta \rm{EW}}$ BAL variability. We find qualitatively similar results in our study (i.e., the BAL varies on short timescales in several epoch pairs); however, we identify fewer (and different) pairs of epochs as significantly variable.  

The identified variable epoch pairs from G15 are MJDs 56697--56715 (Pair 1), MJDs 56751--56755 (Pair 2), MJDs 56768--56772 (Pair 3), and MJDs 56783--56795 (Pair 4). Pair 2 was also identified in our study as significantly variable. Pairs 1 and 4 were not identified in our study due to the implementation of the $G$ criterion, and Pair 3 was not identified due to the re-binning of the spectra causing the $\Delta$EW value to fall just below our 4$\sigma$ significance threshold. In addition, we identify two additional significantly-variable epoch pairs that were not noted by G15; prior to the spectral binning, these epoch pairs also lay just below the 4$\sigma$ significance cutoff. We do not believe the epochs identified by G15 are in error; in fact, we find it likely that all of these epoch pairs show real BAL variability. 
We again stress our goal of assembling a clean, conservative collection of significant BAL variations in this study, and our strict significance criteria is likely to eliminate some epoch pairs that exhibit true variability.

\subsection{RM 631 (SDSS\,J140554.87+530323.5)} 
This quasar has a \civ \ BAL accompanied by \siiv \ absorption. Additionally, a \aliii \ absorption feature exists, but it does not drop down below a normalized flux density of 0.9. Thus, this is a \civ$_{\rm S0}$ BAL. However, strong evidence for \aliii \ exists, and evidence for \pv \ is apparent as well. 

\subsection{RM 717 (SDSS\,J141648.26+542900.9)}
RM\,717 exhibits a broad \civ \ BAL with multiple components; we see a similar structure in the accompanying \siiv \ BAL at corresponding velocities. Hints of possible similarly structured \aliii \ absorption are visible, but the flux does not drop below 0.9 at any point. 

\subsection{RM 722 (SDSS\,J142419.18+531750.6)}
This object displays a single \civ \ BAL with no accompanying broad \siiv \ or \aliii \ absorption, though there is narrow \siiv \ absorption superimposed on the \siiv \ emission line. 

\subsection{RM 729 (SDSS\,J142404.67+532949.3)}
RM\,729 contains a fairly deep \civ \ BAL trough with accompanying \siiv \ absorption and traces of accompanying \aliii \ absorption. The \aliii \ absorption is not quite deep enough to be considered a formal BAL or mini-BAL. There is potential evidence for a \pv \ BAL feature, though the low SN in the \pv \ region makes this uncertain. 
RM\,729 also has a second formal \civ \ BAL present in its spectrum at higher velocity; however, it was excluded from our study due to contamination by bad pixels (see Section~\ref{sec:finalsample}). 

\subsection{RM 730 (SDSS\,142225.03+535901.7)} 
RM\,730 possesses two \civ \ BALs that are very close together (see Figure~\ref{fig:meanspec}), though they are detected as two distinct BALs separated by about 500~km~s$^{-1}$ (see their $v_{\rm min}$ and $v_{\rm max}$ measurements in Table~\ref{Table:baltbl}). The higher-velocity BAL[A] shows significant short-term variability in two pairs of epochs; BAL[B] in this object does not vary significantly over these time periods. BAL[A] shows coordinated variability on velocity scales of 750 and 1250 km~s$^{-1}$ (only about 5\% and 9\% of the entire trough width) during the two identified epoch pairs. There is a \siiv \ BAL present at velocities corresponding to BAL[A], and traces of \aliii \ absorption, but the \aliii \ features are too shallow to be considered BALs; we thus place this trough into the \civ$_{\rm S0}$ category. There is no evidence for \siiv, \aliii, or \pv \ BALs or mini-BALs at velocities corresponding to the lower-velocity BAL[B]. 

\subsection{RM 743 (SDSS\,J142405.10+533206.3)} 
RM\,743 has strong \civ, \siiv, and \aliii \ BAL features present. In addition, Al\,{\sc ii}\,$\lambda 1670$ (and probably Mg\,{\sc ii}) absorption also appears in this quasar. 

\subsection{RM 770 (SDSS\,J142106.86+533745.2)}
This quasar shows no \siiv \ or \aliii \ absorption. 

\subsection{RM 785 (SDSS\,J141322.43+523249.7)} 
This source has a single \civ \ BAL, accompanied by \siiv \ absorption. There are potential \pv \ BALs at corresponding velocities, although the low SN of the region makes this a tentative interpretation.

\subsection{RM 786 (SDSS\,J141421.53+522940.1)} 
RM\,786 has a very shallow high-velocity BAL[A] that varies rapidly for one epoch pair. For the same epoch pair, the low-velocity, deeper BAL[B] varies in the same direction, suggesting coordinated variability between the two. BAL[B] varies significantly in many more epochs, however --- sometimes BAL[A] varies in the same direction, but for two epoch pairs, the troughs vary in opposite directions. Again, we note that the BAL[A] is varying at low significance in these cases, so the disagreement is tentative rather than conclusive.

\section{Derivation of Physical Constraint Equations} 
\label{sec:derivations}

We below provide details on the model and calculations discussed in Section~\ref{sec:physicalconstraints}. 
\subsection{Ionization Variability Model}
In a simple ionization-variability scenario, a step-function change in ionizing flux by some factor $A$ alters the ionization balance in the absorber, resulting in a change in the observed optical depth of a given tracer ion by a factor $B$ on a density-dependent equilibration timescale.
If the electron density in the absorber is constant over its velocity range, then the equilibration timescale and the factor $B$ will be 
constant with velocity as well.\footnote{A step function is just one member of a family of ionizing-flux light curves that will create an 
optical-depth change of a factor $B$ between two given epochs. We are not concerned here with reconstructing the true ionizing flux light 
curve, merely with testing if the resulting optical depth change is constant with velocity.}

We can test for this predicted optical-depth change by a constant factor by assuming a uniform partial-covering scenario (e.g., \cite{dekool02}).
We perform this test for the two epochs in which RM\,508 shows the greatest change in its absorption depth (MJDs 57901 and 57918). For these two epochs, hereafter denoted 1 and 2, we take the normalized flux profile in the 2900 ~km~s$^{-1}$ wide trough and bin by three pixels to reduce correlations between pixels. We assume a value of the covering factor $C$ and calculate
$$\tau_1(\lambda) = -\ln[(F_1(\lambda)+1-C)/C]$$
$$\tau_2(\lambda) = -\ln[(F_2(\lambda)+1-C)/C]$$
and the ratio $B(\lambda)=\tau_1(\lambda)/\tau_2(\lambda)$, as well as the accompanying uncertainties (we are approximating the \civ \ doublet as a singlet, since the trough is broader than the doublet separation). We then find the weighted average $\bar{B}$ over the entire trough. We simultaneously fit the trough profiles with the above partial-covering model with those values of $C$ and $\bar{B}$ and a 14-point optical depth profile (one for each velocity), determined using both $F_1$ and $F_2$ at each velocity.
Mathematically, what we fit is $C$, $\bar{B}$, and 14 values of 
$\tau_{\rm 1,model}$ (one at each pixel), where
$$ \tau_{\rm 1,model}=-(\ln[(F_1(\lambda)+1-C)/C] + $$
$$\bar{B}\ln[(F_2(\lambda)+1-C)/C])/2 $$
$$ {\rm and~} \tau_{\rm 2,model}=\tau_{\rm 1,model}/\bar{B}. $$
We repeat this exercise for all possible values of $C\leq 1$. We find a minimum $\chi^2=22.3$ for 12 degrees of freedom for $C=1$, which is a marginally acceptable model ($p~=~3.48\times~10^{-2}$, 1.82$\sigma$).
The relatively large minimum $\chi^2$ may indicate that the adopted model is an oversimplification, or it may arise from the use of only two epochs of data to fit the optical-depth profile. Our results indicate that the observed variability is consistent with a model of an absorber that fully covers the continuum source, is optically thin in \civ \ (with a maximum optical depth $\tau_{\rm max}=0.847\pm 0.024$), has constant electron density over its velocity range, and responds to a variable ionizing flux by a drop in optical depth of a factor of $\bar{B}=2.04 \pm 0.07$ in 4.03 rest-frame days.

\subsection{Density Constraint}
\label{sec:density} 
Following \cite{Grier15}, we further use the observed changes in EW and the timescale of these changes to measure a lower limit on the density of the outflow material. For ionization state $i$ of some element which begins in photoionization equilibrium, the density of species $n_i$ can be modeled as varying with a characteristic timescale $t_i^*$ as 

\begin{equation}
n_i(t)=n_i(0)\exp{(t/t_i^*)}
\label{eq:apndnsty}
\end{equation} 

\cite[][]{Arav12, Rogerson15}. Assuming the outflow is optically thin, number density is linearly proportional to column density. Thus, for the case of two relevant quasar observations, $n_i(t)/n_i(0) = \rm{EW_2}/\rm{EW_1}$, and $t = \Delta t_{\rm rest}$. Equation~\ref{eq:apndnsty} can therefore be rewritten and solved for the characteristic timescale $t_i^*$, giving

\begin{equation}
	t_i^* = \Delta t_{\rm rest} \log{(\rm{EW_2}/\rm{EW_1})}^{-1}
    \label{eq:apchartm1}
\end{equation}

Using the ionization rate per ion of state $i$ ($I_i$), the ion's recombination coefficient to ionization state $i-1$ ($\alpha_{i-1}$), and the fractional change in $I_i$ ($f_i$, for which $f_i = I_i(t)/I_i(0) - 1$ and $-1 < f_i < +\infty$), the characteristic timescale $t_i^*$ can also be calculated as

\begin{equation}
	t_i^* = [-f_i(I_i - n_e\alpha_{i-1})]^{-1}
    \label{eq:apchartm2}
\end{equation}

assuming photoionization and recombination are the only processes relevant to ionization state change (i.e., collisional ionization is ignored). A negative $t_i^*$ implies $n_i$ decreasing with time.

For an outflow sufficiently separated from the quasar, recombination dominates photoionization ($n_e\alpha_{i-1} \gg I_i$). Consider the scenario in which ionizing flux vanishes, implying $f_i = -1$ (e.g., \citealt{Capellupo13}). For this limiting case, $t_i^*$ is the recombination time of the ion, and Equation~\ref{eq:apchartm2} reduces to $t_i^* = -1 /n_e\alpha_{i-1}$. Holding firm our previous assumption of optical thinness, this result can be combined with Equation~\ref{eq:apchartm1} to derive the lower density limit

\begin{equation}
n_e > \log{(\rm{EW_1}/\rm{EW_2})} / \alpha_{i-1} \Delta t_{\rm rest}
\label{eq:apne}
\end{equation}

\noindent which is Equation~\ref{eq:ne}, used to constrain outflow density in Section~\ref{sec:physicalconstraints}

\subsection{Radial Distance Constraint}
\label{sec:distance}

For an optically thin cloud containing some element in ionization state $i$ at radial distance $r$ from a quasar with luminosity $L_{\nu}$, the ionization rate per ion in state $i$ is calculated as

\begin{equation}
	I_i = \int_{\nu_i}^{\infty}{\frac{(L_\nu/h\nu)\sigma_\nu}{4\pi r^2} d\nu}
	\label{eq:Ii}
\end{equation}

\noindent for which $\sigma_\nu$ is the ionization cross section for photons of energy $h\nu$ and $\nu_i$ is the ionization energy of ionization species $(i-1)$.

In the derivation of our density constraint (see Section~\ref{sec:density} of the Appendix), we assume that the outflow is sufficiently separated from the quasar such that recombination (with recombination coefficient $\alpha_{i-1}$) dominates photoionization ($I_i \ll n_e\alpha_{i-1}$). As radial distance from the quasar decreases, ionizing flux increases and photoionization comes to instead dominate recombination ($I_i \gg n_e\alpha_{i-1}$). Between photoionization-dominated distances and recombination-dominated distances exists an equilibrium distance ($r_{\rm equal}$) for which the rate of photoionization and the rate of recombination are equivalent ($I_i = n_e\alpha_{i-1}$). Due to the aforementioned assumptions of our density constraint derivation, this constraint is only valid at distances larger than the equilibrium distance ($r > r_{\rm equal}$). From these arguments, we substitute $I_i = n_e\alpha_{i-1}$ and $r > r_{\rm equal}$ into Equation~\ref{eq:Ii}, obtaining the lower radial distance limit

\begin{equation}
	r > r_{\rm equal} = \sqrt{\int_{\nu_i}^{\infty}{\frac{(L_\nu/h\nu)\sigma_\nu}{4\pi n_e\alpha_{i-1}}} d\nu}
\end{equation}

which is Equation~\ref{eq:dist}, used to constrain the radial distance of outflows in Section~\ref{sec:physicalconstraints}.

\end{document}